# Divergent Perturbation Series

## I. M. Suslov

*Kapitza Institute for Physical Problems, Moscow, 117334 Russia*
*e-mail: suslov@kapitza.ras.ru*

*Divergent series are the devil's invention...*
N.Abel, 1828   *

**Abstract**—Various perturbation series are factorially divergent. The behavior of their high-order terms can be found by Lipatov's method, according to which they are determined by instanton configurations of appropriate functional integrals. When the Lipatov asymptotics is known and several lowest order terms of the perturbation series are found by direct calculation of diagrams, one can gain insight into the behavior of the remaining terms of the series. Summing it, one can solve (in a certain approximation) various strong-coupling problems. This approach is demonstrated by determining the Gell-Mann–Low functions in $\varphi^4$ theory, QED, and QCD for arbitrary coupling constants. An overview of the mathematical theory of divergent series is presented, and interpretation of perturbation series is discussed. Explicit derivations of the Lipatov asymptotics are presented for some basic problems in theoretical physics. A solution is proposed to the problem of renormalon contributions, which hampered progress in this field in the late 1970s. Practical perturbation-series summation schemes are described for a coupling constant of order unity and in the strong-coupling limit. An interpretation of the Borel integral is given for "non-Borel-summable" series. High-order corrections to the Lipatov asymptotics are discussed.

## CONTENTS

* Cited following the book [1]

## 1. DYSON'S ARGUMENT: IMPORTANT PERTURBATIVE SERIES HAVE ZERO RADIUS OF CONVERGENCE

Classical books on diagrammatic techniques [2–4] describe the construction of diagram series as if they were well defined. However, almost all important perturbation series are hopelessly divergent since they have zero radii of convergence. The first argument to this effect was given by Dyson [5] with regard to quantum electrodynamics. Here, it is reiterated by using simpler examples.

Consider a Fermi gas with a delta-function interaction $g\delta(\mathbf{r} - \mathbf{r}')$ and the corresponding perturbation series in terms of the coupling constant $g$. Its radius of convergence is determined by the distance from the origin to the nearest singular point in the complex plane and can be found as follows. In the case of a repulsive interaction ($g > 0$), the ground state of the system is a Fermi liquid. When the interaction is attractive ($g < 0$), the Cooper instability leads to superconductivity (see Fig. 1a). As $g$ is varied, the ground state qualitatively changes at $g = 0$. Thus, the nearest singular point is located at the origin, and the convergence radius of the series is zero.

An even simpler example is the energy spectrum of a quantum particle in the one-dimensional anharmonic potential

$$U(x) = x^2 + gx^4. \tag{1.1}$$

Whereas the system has well-defined energy levels when $g > 0$, these levels are metastable when $g < 0$ since the particle can escape to infinity (see Fig. 1b). Therefore, the perturbation series in terms of $g$ is divergent for any finite $g$ as it can be tested by direct calculation of its coefficients. The calculation of the first 150 coefficients of this series in [6] was the first demonstration of its divergence and gave possibility of its detailed study.

Zero radius of convergence looks "accidental" in quantum-mechanical problems: it takes a place when a potential of special form is taken and a "bad" definition of coupling constant is chosen. However, zero radius of convergence is encountered in all fundamental quantum field theories with a single coupling constant.

Even though Dyson's argument is unquestionable, it was hushed up or decried for many years: the scientific community was not ready to face the problem of the hopeless divergency of perturbation series.

## 2. LIPATOV'S METHOD: QUANTITATIVE ESTIMATION OF DIVERGENCY OF SERIES

A further step was made in 1977, when Lipatov's method was proposed [7] as a tool for calculating high-order terms in perturbation series and making quantitative estimates for its divergence. The idea of the method is as follows. If a function $F(g)$ can be expanded into a power series of $g$,

$$F(g) = \sum_{N=0}^{\infty} F_N g^N, \qquad (2.1)$$

then the coefficients $F_N$ in the expansion can be determined as

$$F_N = \int_C \frac{dg}{2\pi i} \frac{F(g)}{g^{N+1}}, \qquad (2.2)$$

where the contour $C$ goes around the point $g = 0$ in the complex plane. Rewriting the denominator as $\exp\{-(N+1)\ln g\}$, for large $N$ one may hope that the saddle-point method can be applied to the resulting exponential with a large exponent.

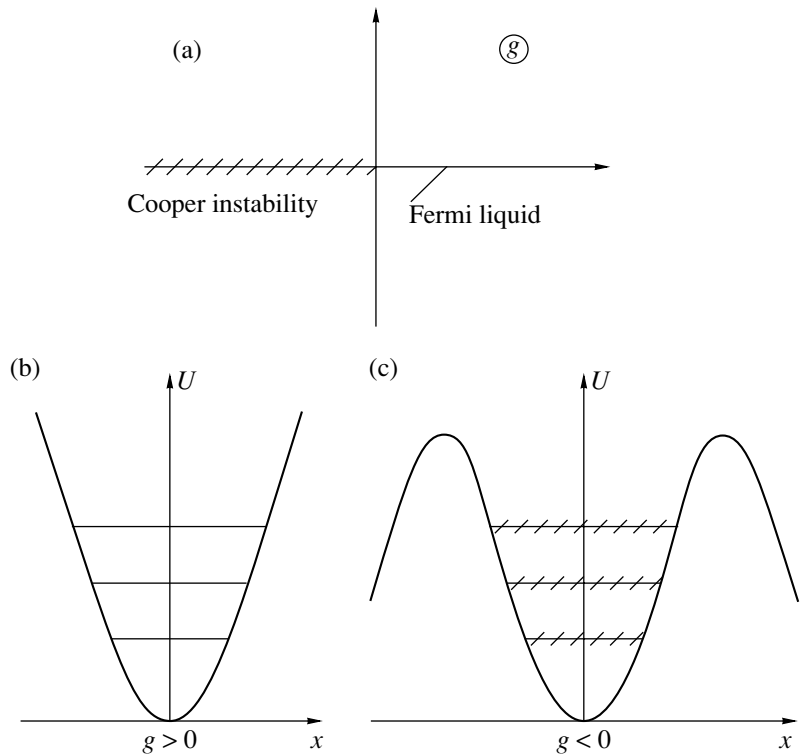

**Fig. 1.** Graphic illustration of Dyson's argument.

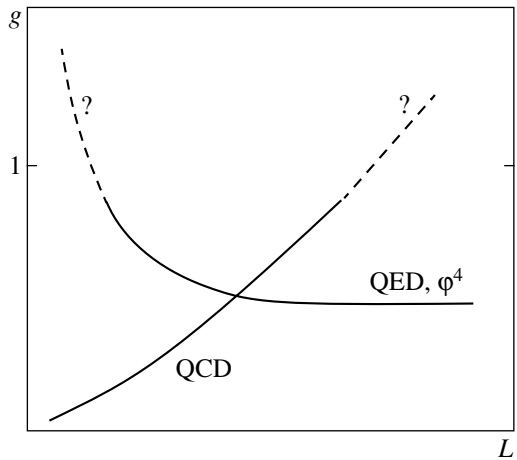

**Fig. 2.** Effective coupling versus length scale in $\varphi^4$ theory, QED, and QCD.

It is well known that the problems tractable by diagrammatic technique can be reformulated in terms of functional integrals of the form

$$Z(g) = \int D\varphi \exp(-S_0\{\varphi\} - g S_{\text{int}}\{\varphi\}), \qquad (2.3)$$

with the expansion coefficients

$$Z_N = \int_C \frac{dg}{2\pi i g} \\ \times \int D\varphi \exp(-S_0\{\varphi\} - g S_{\text{int}}\{\varphi\} - N \ln g). \qquad (2.4)$$

Lipatov suggested seeking the saddle-point configuration of (2.4) in $g$ and $\varphi$ simultaneously, rather than with respect to $g$ only. The desired configuration exists in all cases of interest and is realized on a localized function $\varphi(x)$ called *instanton*. It turns out that the saddle-point approximation is always applicable when $N$ is large irrespective of its applicability to integral (2.3). This finding has important consequences: whereas functional integrals cannot generally be calculated exactly, they can always be calculated in the saddle-point approximation.

Once expansion coefficients are known for a functional integral, expansions of Green functions, vertices, tc., can be found, because factorial series can be treated as easily as finite expressions, due to existence of simple algebra (see [8, Section 5.3]). Generally, the Lipatov asymtotics of the expansion coefficients for any quantity $F(g)$ is

$$F_N = c \Gamma(N + b) a^N, \qquad (2.5)$$

where $\Gamma(x)$ is the gamma function and $a$, $b$, and $c$ are parameters depending on the specific problem under analysis. In the framework of a particular theory, $a$ is a universal constant, $b$ is a parameter depending on $F(g)$, and $c$ depends on external coordinates or momenta.

When the Lipatov asymptotic form is known and a few lowest order terms of a perturbation series are found by direct calculation of diagrams, one can gain insight into the behavior of the remaining terms of the series and perform their summation to solve various strong-coupling problems in a certain approximation. The most important consequence is the possibility of finding the Gell-Mann–Low function $\beta(g)$, which determines the effective coupling constant $g(L)$ as a function of length scale:

$$-\frac{dg}{d\ln L^2} = \beta(g). \qquad (2.6)$$

In relativistic theories, the first term in the expansion of $\beta(g)$ is quadratic, $\beta(g) = \beta_2 g^2 + \ldots$. For a small $g$, Eq. (2.6) yields the well-known result [3, 9, 10]

$$g(L) = \frac{g_0}{1 - \beta_2 g_0 \ln(L^2/L_0^2)}, \qquad (2.7)$$

where $g_0$ is the value of $g(L)$ on a length scale $L_0$. In both quantum electrodynamics (QED) and $\varphi^4$ theory, the constant $\beta_2$ is positive, and $g(L)$ is an increasing function at small $L$ (see Fig. 2). In quantum chromodynamics (QCD), the sign of $\beta_2$ is negative. Accordingly, the interaction between quarks and gluons is weak at small $L$ (asymptotic freedom), while its increase with $L$ demonstrates a tendency toward confinement (see Fig. 2). One problem of primary importance is extension of (2.7) to intermediate and strong coupling. According to the classification put forward in [2] (for \beta_2>0), the function $g(L)$ tends to a constant if $\beta(g)$ has a zero at a finite $g$ and continues to increase ad infinitum, as $L \longrightarrow 0$, if $\beta(g) \propto g^\alpha$ with $\alpha \le 1$. If $\beta(g) \propto g^\alpha$ with $\alpha > 1$, then two interpretations are plausible. On the one hand, assuming finite interaction at long distances, one would have a self-contradictory theory: the effective charge $g(L)$ goes to infinity at a finite $L_c$ (Landau pole), while the function $g(L)$ is undefined at $L < L_c$. On the other hand, a field theory interpreted as a continuum limit of lattice models is "trivial" since the interaction vanishes as $L \longrightarrow \infty$ ("zero-charge" property). The first attempts to determine the Gell-Mann–Low function in $\varphi^4$ theory were made in [11–13].

Originally developed for scalar theories (such as $\varphi^4$ [7]), Lipatov's method was extended to vector fields [14], fermion problems [15], scalar electrodynamics [16], and the Yang–Mills theories [17, 18], as well as to a variety of problems in quantum mechanics (see [19] and reviews in [20, 21]). Next in order were its applications to theories of practical interest, QED [22, 23] and QCD [24–26].

In all theories mentioned above, factorially divergent series were obtained. Assuming that divergent series are "the devil's invention," one should admit that the Creator has also taken part: description of physical reality leads to divergent series expansions with striking regularity.

## 3. INTERPRETATION OF PERTURBATION SERIES: A SURVEY OF THE MATHEMATICAL THEORY OF DIVERGENT SERIES

The modern status of divergent series suggests that techniques for manipulating them should be included in a minimum syllabus for graduate students in theoretical physics. However, the theory of divergent series is almost unknown to physicists, because the corresponding parts of standard university courses in calculus date back to the mid-nineteenth century, when divergent series were virtually banished from mathematics. The discussion that follows provides a brief review of the mathematical theory of divergent series [27].

### 3.1. Can We Deal with Divergent Series?

Dealing with series of the form

$$a_0 + a_1 + a_2 + a_3 + \ldots + a_N + \ldots \qquad (3.1)$$

for the first time, one may be tempted to treat them as if they were finite sums. However, this is incorrect in the general case, because a series can be treated as a finite sum only if it is absolutely convergent [28], i.e.,

$$|a_0| + |a_1| + |a_2| + |a_3| + \ldots + |a_N| + \ldots < \infty. \qquad (3.2)$$

If the series is convergent, but no absolutely, as the the alternating harmonic series,

$$1 - \frac{1}{2} + \frac{1}{3} - \frac{1}{4} + \frac{1}{5} - \ldots, \qquad (3.3)$$

one cannot permute its terms in an arbitrary manner: by Riemann's theorem, not absolutely convergent series can be rearranged to converge to any specified sum [28]. Indeed, the sum of a convergent series is defined as the limit of its partial sums, and any result can be obtained by shifting negative terms rightwards and positive terms leftwards, or vice versa.

Expectably, the analysis of divergent series is even more complicated because of a greater number of forbidden operations on them:

(a) obviously, terms cannot be permuted;

(b) terms cannot be grouped either, e.g.,

$$\begin{aligned} 1 - 1 + 1 - 1 + 1 - 1 + \ldots &\neq (1 - 1) + (1 - 1) \\ + (1 - 1) + \ldots &\neq 1 + (-1 + 1) + (-1 + 1) \\ &\quad + (-1 + 1) + \ldots; \end{aligned} \qquad (3.4)$$

(c) a series cannot be "padded" by inserting zero terms,

$$\begin{aligned} &a_0 + a_1 + a_2 + a_3 + \ldots \\ &\neq a_0 + 0 + a_1 + 0 + a_2 + 0 + a_3 + 0 + \ldots. \end{aligned} \qquad (3.5)$$

Now we can formulate the basic idea of the theory of divergent series: in principle, they can be consistently manipulated if one follows rules that are much more stringent than those for operations on finite sums or convergent series.

### 3.2. Euler's Principle

What are the new rules to be followed? A preliminary answer to this question was given by L. Euler, who was the true pioneer in developing the theory of divergent series. Euler ruled out the use of number series (3.1) and expansions over arbitrary basis functions,[1]

$$a_0 f_0(x) + a_1 f_1(x) + a_2 f_2(x) + \ldots + a_N f_N(x) + \ldots, \qquad (3.6)$$

and emphasized a special role played by power series

$$a_0 + a_1 x + a_2 x^2 + a_3 x^3 + \ldots + a_N x^N + \ldots. \qquad (3.7)$$

Power series expansions are special due to existence of the natural numbering of terms, information of which is preserved under permutations or other operations. As a result, power series can be treated as finite sums. It is clear that forbidden operations are ruled out automatically: if number series (3.1) is interpreted as the limit of power series (3.7) as $x \longrightarrow 1$, then any permutation, pad-

---

[1] This discussion concerns to divergent series only. Convergent expansions such as (3.6) (e.g., over an orthogonal basis) are obviously admissible.

ding with zero terms, or association leads to a series different from the initial one:

$$a_1 + a_0 + a_3 + a_2 + \ldots$$
$$\longrightarrow a_1 + a_0 x + a_3 x^2 + a_2 x^3 \ldots, \quad (3.8)$$

$$a_0 + 0 + a_1 + 0 + a_2 + 0 + a_3 + \ldots$$
$$\longrightarrow a_0 + 0 \cdot x + a_1 x^2 + 0 \cdot x^3 + a_2 x^4 \quad (3.9)$$
$$+ 0 \cdot x^5 + a_3 x^6 \ldots,$$

$$(a_0 + a_1) + (a_2 + a_3) + (a_4 + a_5) + \ldots$$
$$\longrightarrow (a_0 + a_1) + (a_2 + a_3)x + (a_4 + a_5)x^2 + \ldots . \quad (3.10)$$

The fundamental reason for validity of Euler's principle lies in fact that a power series is absolutely convergent within its circle of convergence and defines an analytic function that can be continued outside its domain of convergence. Accordingly, free manipulations of power series are admissible either as operations on absolutely convergent series or by the principle of analytic continuation. However, an analytic function may have several branches, and information about them is lost when divergent series are employed. Therefore, Euler's approach is not complete, and its application may lead to poorly defined expressions requiring correct interpretation. As a consequence, its rigorous mathematical substantiation is hampered by difficult problems. Generally, constructive results in the theory of divergent series correspond to the partial proofs of Euler's principle under different restrictive assumptions [27]. It looks that this principle is valid in the entire parameter space spanned by the coefficients of a series except for a set of measure zero, where it is valid only when the definition of the sum of the series is appropriately generalized. The excluded set has a complicated structure and is difficult to specify by proving a finite number of theorems. For this reason, Euler's principle cannot be adopted in modern mathematics without reservation. However, it is not rejected either, because it does not seem to be disproved by any known fact.

Basically, Euler's principle is consistent with common practice in theoretical physics. It is commonly believed that formal manipulations of power series on a "symbolic" level cannot lead to results that are definitely incorrect even if divergent series are used in intermediate calculations. Moreover, ill-defined expressions do not present principal problem, since their correct interpretation can be found from physical considerations by applying various rules for avoiding singularities, which are so skillfully devised by physicists. When applying this approach, one should follow two rules: never substitute the numerical values of $x$ before the series is transformed into a convergent one and never perform Taylor expansions in the clearly singular points.

With regard to the latter requirement, note that the series used in quantum field theories have zero radii of convergence, but arise from functional integral (2.3) as a result of a regular expansion of the exponential in terms of $g$ and a subsequent (incorrect) interchange of summation and integration. In essence, an aim of the summation theory is a reverse permutation, which can be performed "in another place", providing a freedom of formal manipulations.

It may seem that the restriction to power series expansions is very stringent. Actually, this is not so, because a number series may arise in a physical application only when some particular values are assigned to parameters of the model. Usually, a power series in at least one parameter can be obtained by returning to the general formulation of the problem or by generalizing the model. This is frequently done by using relatively simple tricks. For example, if the potential energy in the Schrödinger equation is treated as a perturbation, then the resulting expansion is not a power series. However, if $U(x)$ is replaced with $gU(x)$ before performing the series expansion (with a view to setting $g \longrightarrow 1$ as a final step), then a power series in $g$ will be obtained.

Now, a few words should be said about number series of type (3.1). In principle, they can be consistently manipulated [27] if (3.1) is considered as a symbolic representation that cannot be identified with a conventional sum (otherwise, one is led to paradoxes commonly discussed in textbooks [28]). Manipulations of this kind are performed according to ad hoc rules known only to specialists. In their constructive part, these rules can be derived from Euler's principle if number series (3.1) is identified with power series (3.7) in the limit of $x \longrightarrow 1$. Such identification can be always done formally, but one should be sure that series (3.1) has not been modified by rearranging, discarding zero terms, etc. Since the fulfillment of this requirement cannot be reliably checked unless the number series is derived from a known power series, number series per se are of no practical importance.

As an implementation of Euler's approach, consider the well-known Borel transformation: dividing and multiplying each term of a series by $N!$, introducing the integral representation of the gamma function, and interchanging summation and integration, one obtains

$$F(g) = \sum_{N=0}^{\infty} F_N g^N = \sum_{N=0}^{\infty} \frac{F_N}{N!} \int_0^{\infty} dx\, x^N e^{-x} g^N$$
$$= \int_0^{\infty} dx\, e^{-x} \sum_{N=0}^{\infty} \frac{F_N}{N!} (gx)^N. \quad (3.11)$$

The power series on the right-hand side has factorially improved convergence and defines the Borel transform $B(z)$ of $F(g)$:

$$F(g) = \int_0^{\infty} dx\, e^{-x} B(gx), \quad B(z) = \sum_{N=0}^{\infty} \frac{F_N}{N!} z^N. \quad (3.12)$$

The Borel transformation provides a natural procedure for summing factorially divergent series in quantum field theories.

### 3.3. How Should We Define the Sum of a Series?

Now, let us discuss the modern approach to the problem. An "ideal" program of formalization can be represented as follows.

1. Formulate a definition of the sum $S$ of a series that is equivalent to the conventional definition in the case of a convergent series.

2. Consider a class $L$ of transformations of one series into another that leaves the sum invariant:

$$a_0 + a_1 + a_2 + a_3 + \ldots + a_N + \ldots = S$$
$$\longrightarrow b_0 + b_1 + b_2 + b_3 + \ldots + b_N + \ldots = S. \quad (3.13)$$

3. Verify that class $L$ is a sufficiently wide and can be efficiently used to transform convergent series into divergent ones and vice versa.

4. Specify the class $L$; it sets rules for manipulating series expansions without checking their convergence.

Has this program ever been implemented in modern mathematics? In fact, it has been, to the extent that a subclass of $L$, sufficiently wide to solve practical problems, has been specified. However, the theory cannot be presented in the elegant form outlined above. Indeed, difficulties arise even in implementing the first step: no definition of a sum equally suited to any particular problem is available. Summation methods for strongly divergent series are not instrumental as applied to weakly divergent ones, and vice versa. For this reason, a laissez-faire approach is adopted: any definition of a sum is formally admissible, and mutual consistency of different definitions is the only subject to be analyzed on an abstract mathematical level. The choice of a particular definition is left to the user. This attitude of mathematicians is somewhat doubtful: if the user knows the definition of the sum, he can make the rest without problem. However, this attitude is sufficiently grounded (see Section 7).

In principle, it is known how these difficulties may be resolved. Recall how the temperature standard is introduced in physics. Since no temperature measurement method is universally applicable, several temperature standards are introduced (for high, low, ultralow etc. temperatures) that lead to identical results in the temperature regions where they overlap. An analogous approach can be adopted in the theory of divergent series, where a variety of "good" (mutually consistent) summation methods are available:[2] a combined definition of sum can be accepted by using these methods. Since good methods usually based on Euler's principle, this approach reverts us to this principle, but on a higher formal level and with certain restrictions.

As examples, consider the following possible definitions of sum.

**Euler's definition.** If power series (3.7) is convergent at small $x$, then it defines a regular function $f(x)$ whose analytic continuation is the sum of series (3.7) outside its circle of convergence.

In physical applications, this definition is adopted without reservation. As noted above, it is not complete, because the choice of a branch of the analytic function remains an open question. However, when this definition is meaningful, all calculations can be performed by using only convergent series, and the uncertainty is thus eliminated. A theory of divergent series is really necessary when the radius of convergence is zero, i.e., when Euler's definition is meaningless.

**Borel's definition** (applicable in the latter case as well). The sum of series (3.11) is given by (3.12). This definition agrees with other definitions based on Euler's principle and satisfies all necessary requirements.

### 3.4. Asymptotic Interpretation of Divergent Series

Modern theory of divergent series has "two sources and two parts." The foregoing discussion deals with the essentials of the summation theory presented in its complete form by Borel [29]. More widely known is the asymptotic interpretation of divergent series proposed by Poincaré [30]. A power series expansion of a function $f(x)$ is asymptotic if

$$f(x) = a_0 + a_1 x + a_2 x^2 + \ldots + a_N x^N + R_N(x), \quad (3.14)$$

where

$$R_N(x) = O(x^{N+1}), \quad x \longrightarrow 0, \quad (3.15)$$

i.e., if $f(x)$ is accurately approximated by a truncated series for sufficiently small $x$. The asymptotic interpretation is constructive only if the problem at hand involves a small parameter. However, in such case, it is the most convinient: one need not sum any high-order terms and even should not be interested in their behavior. Another advantage lies in the possibility of constructing asympotic expansion (3.6) over arbitrary basis functions, provided that each $f_n(x)$ approaches zero faster than does $f_{n-1}(x)$.[3]

There is no one-to-one correspondence between functions and asymptotic power series expansions, because $f(x)$ can be modified by adding a function for which all coefficients in (3.14) vanish, such as $\exp(-a/x)$. When a series is convergent at small $x$, this uncertainty is eliminated by imposing the condition of analyticity at $x = 0$. However, this cannot be done for series with zero radius of convergence. At first glance, this would imply that a divergent series cannot be assigned any particular sum.

---

[2] Note that there are also a lot of "bad" methods, which contradict to each other.

[3] If $x$ is not small, then expansion (3.6), in contrast to (3.7), does not admit any meaningful interpretation at all. However, this is not very actual: what sense is in the use of expansion that is not regular and does not involve any small parameter?

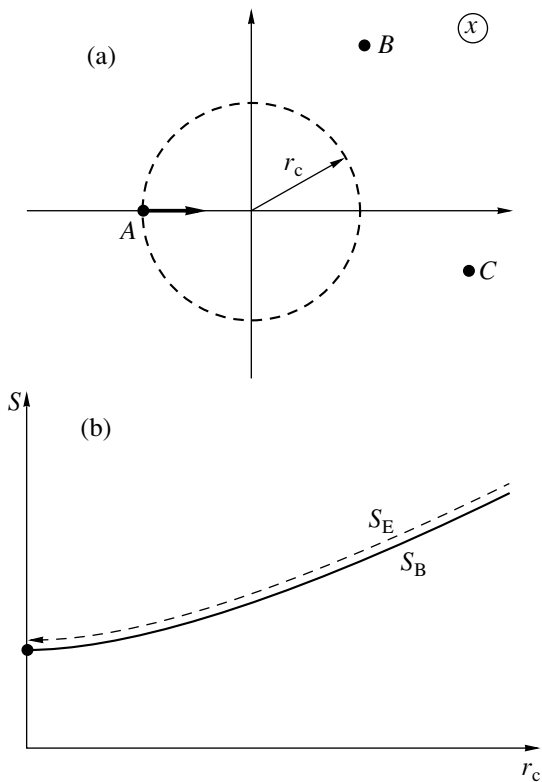

**Fig. 3.** (a) Convergence radius $r_c$ decreases as singular point $A$ approaches the origin. (b) Euler sum $S_E$ equals Borel sum $S_B$ when $r_c$ is finite; when $r_c = 0$, the former is meaningless, whereas the latter extends the former by continuity.

Actually, this is not true. Asymptotic equality means that the partial sum $a_0 + \ldots + a_N x^N$ "resembles" the function $f(x)$ up to some remainder term $R_N(x)$. It is no surprise that many functions meet this requirement within the prescribed accuracy. Their variety can be reduced by diminishing $R_N(x)$, so that ultimately a single function remains. In fact, this is exactly what should be done: when the remainder term is subject to certain constraints, the functions and the asymptotic series expansions are in one-to-one correspondence. Namely, it suffices to replace standard condition (3.15) with the so-called strong asymptotic condition

$$|R_N(x)| < C_N |x|^{N+1}, \quad x \in G, \qquad (3.16)$$

where $G$ is a region containing the point $x = 0$ and $C_N$ are the specially chosen coefficients. The single function that satisfies the strong asymptotic condition and "resembles" an asymptotic series "most closely" can be naturally accepted as its sum. By Watson's theorem on Borel summability [27], this unique function is given by the Borel integral in (3.12) for a broad class of divergent series. Thus, both summation theory and asymptotic theory naturally lead one to adopt Borel's sum as the natural sum of a divergent series.

The discussion above clearly solves the problem of nonperturbative contributions, such as $\exp(-a/x)$, which is frequently brought up as an argument against the use of perturbation series. When Borel's definition is adopted, such terms should not be added to Borel integral (3.12). Formal manipulations of power series expansions will not lead to terms of this kind if Taylor expansions are not performed in the clearly singular points.

### 3.5. Physical Arguments

Now, let us discuss physical arguments in support of the interpretation of perturbation series in the Borel sense.

Suppose that power series (3.7) has a finite radius of convergence. Then, the corresponding analytic function $f(x)$ has singular points $A$, $B$, $C$, ... at finite distances from the origin (see Fig. 3a). In this case, it can be shown that Borel's definition of a sum is equivalent to Euler's, which is definitely suitable for physical applications.

Let the singular point A approaches to the origin. If the point $A$ is a pole or power-like singularity, the coefficients of the series diverge, and the expansion becomes meaningless when the convergence radius tends to zero. However, there exist singularities that can be moved to the origin without causing divergence of the coefficients: these are branch points with exponentially decreasing jump at the cut. In particular, Lipatov asymptotic form (2.5) is associated with the following jump in $F(g)$ at the cut [31, 32]:[4]

$$\Delta F(g) = 2\pi i c \left(\frac{1}{ag}\right)^b \exp\left(-\frac{1}{ag}\right). \qquad (3.17)$$

Both Borel's and Euler's sums vary with radius of convergence $r_c$, remaining equal. When $r_c = 0$, Euler's sum is meaningless, whereas Borel's sum corresponds to Euler's definition extended by continuity (see Fig. 3b).

The limit of $r_c \longrightarrow 0$ is amenable to a straightforward physical interpretation. Consider the Fermi gas with a delta-function interaction discussed in Section 1. The Cooper instability occurs for arbitrary $g < 0$ only at zero temperature. As the temperature $T$ is raised to a finite value, the instability domain shifts to negative $g$ by an amount $g_c$ determined by a Bardeen–Cooper–

---

[4] Correspondence between (3.17) and (2.5) can be established by calculating the jump at the cut for Borel's sum of a series with expansion coefficients having asymptotic form (2.5). Alternatively, one can write Cauchy's integral formula for a point $g$ lying in the domain of analyticity and deform the integration contour so that it goes around the cut. Then, the jump at the cut given by (3.17) will lead to asymptotic expression (2.5).

Schrieffer-like relation, $T \propto \exp\{-\text{const}/g_c\}$. The corresponding perturbation series has a finite convergence radius $g_c$, which tends to zero with decreasing temperature. Generally, the value of some quantity calculated at strictly zero temperature differs from its value obtained in the limit $T \longrightarrow 0$. However, the physically meaningful result is that obtained for $T \longrightarrow 0$. Thus, the value at $T = 0$ should be always defined by continuous extension and can be obtained by Borel summation.

## 4. LIPATOV ASYMPTOTIC FORMS FOR SPECIFIC MODELS

The calculation of Lipatov asymptotic form (2.5) is tedious if all parameters $a$, $b$, and $c$ should be found. As for its functional form, it can be easily found by performing a formal saddle-point expansion and separating the dependence on $N$. In what follows, calculations of this kind are performed for several fundamental models in theoretical physics.

### 4.1. $\varphi^4$ Theory

To begin with, consider the $n$-component $\varphi^4$ theory. The corresponding action is

$$S\{g, \varphi\} = \int d^d x \left\{ \frac{1}{2} \sum_{\alpha=1}^n [\partial_\mu \varphi_\alpha(x)]^2 + \frac{1}{2} m^2 \sum_{\alpha=1}^n \varphi_\alpha^2(x) \right.$$
$$\left. + \frac{1}{4} g \left( \sum_{\alpha=1}^n \varphi_\alpha^2(x) \right)^2 \right\} \quad (4.1.1)$$

($d$ is the space dimension). Functional integrals of the form

$$Z_M(g) = \int D\varphi \, \varphi_{\alpha_1}(x_1) \varphi_{\alpha_2}(x_2) \ldots \varphi_{\alpha_M}(x_M) \quad (4.1.2)$$
$$\times \exp(-S\{g, \varphi\})$$

define $M$-point Green functions,

$$G_M(g) = \frac{Z_M(g)}{Z_0(g)}, \quad (4.1.3)$$

which are diagrammatically represented by $M$-legged graphs. Hereinafter, integral (4.1.2) is written in compact form as

$$Z(g) = \int D\varphi \, \varphi^{(1)} \ldots \varphi^{(M)} \exp(-S\{g, \varphi\}), \quad (4.1.4)$$

and normalized to an analogous integral with $M = 0$ and $g = 0$, with the factor $Z_0^{-1}(0)$ included into $D\varphi$. Actually, the explicit form of the action is not essential in the present analysis, and only its homogeneity properties are used to write

$$S\{g, \varphi\} = \frac{S\{\phi\}}{g}, \quad \text{where} \quad \phi = \frac{\varphi}{\sqrt{g}}. \quad (4.1.5)$$

First, consider a finite-dimensional integral having the form of (4.1.4) with $D\varphi = d\varphi_1 d\varphi_2 \ldots d\varphi_m$ and define

$$\phi = \begin{vmatrix} \phi_1 \\ \phi_2 \\ \cdot \\ \phi_m \end{vmatrix}, \quad S'\{\phi\} = \begin{vmatrix} \partial S/\partial \phi_1 \\ \partial S/\partial \phi_2 \\ \ldots \\ \partial S/\partial \phi_m \end{vmatrix}, \quad (4.1.6)$$

$$S''\{\phi\} = \left\| \frac{\partial^2 S}{\partial \phi_i \partial \phi_j} \right\|.$$

This notation makes it possible to write any expression in a form analogous to the corresponding one-dimensional integral. In the infinite-dimensional limit, $S'\{\phi\}$ and $S''\{\phi\}$ become the first and second functional derivatives, interpreted as a vector and linear operator, respectively.

According to Section 2, the expansion coefficients are

$$Z_N = \oint_C \frac{dg}{2\pi i g} \int D\varphi \, \varphi^{(1)} \ldots \varphi^{(M)}$$
$$\times \exp\left( -\frac{S\{\phi\}}{g} - N \ln g \right), \quad (4.1.7)$$

and the saddle-point conditions have the form

$$S'\{\phi_c\} = 0, \quad g_c = \frac{S\{\phi_c\}}{N}. \quad (4.1.8)$$

The expansion of the exponent in (4.1.7) to quadratic terms in $\delta\phi = \phi - \phi_c$ and $\delta g = g - g_c$ is

$$-N - N \ln g_c - \frac{N(\delta\phi, S''\{\phi_c\}\delta\phi)}{2 S\{\phi_c\}} - \frac{N}{2g_c^2}(\delta g)^2. \quad (4.1.9)$$

Since

$$\delta\phi = \sqrt{g_c}\left( \delta\varphi + \frac{\delta g}{2g_c}\varphi_c \right), \quad \delta\varphi = \varphi - \varphi_c, \quad (4.1.10)$$

the origin of $\delta\varphi$ can be shifted to obtain

$$Z_N = e^{-N} g_c^{-N-M/2} \int_{-\infty}^{\infty} \frac{dt}{2\pi} \int D\varphi \, \phi_c^{(1)} \ldots \phi_c^{(M)}$$
$$\times \exp\left( -\frac{1}{2}(\delta\varphi, S''\{\phi_c\}\delta\varphi) + \frac{N}{2}t^2 \right), \quad (4.1.11)$$

where $\delta g = ig_c t$, because the saddle point is passed in the vertical direction. The Gaussian integration yields

$$Z_N = \frac{\text{const}}{\sqrt{-\det[S''\{\phi_c\}]}} S\{\phi_c\}^{-N} \Gamma\left( N + \frac{M}{2} \right); \quad (4.1.12)$$

i.e., Lipatov asymptotic form (2.5) is recovered.

The result given by (4.1.12) is independent of $m$, remaining valid as $m \longrightarrow \infty$, i.e., in the functional-integral limit. However, any realistic functional integral contains zero modes associated with the symmetry of action under a continuous group defined by the operator $\hat{L}$, $S\{\phi\} = S\{\hat{L}\phi\}$. If $\phi_c$ is an instanton (i.e., $S'\{\phi_c\} = 0$), then so is $\hat{L}\phi_c$ ($S'\{\hat{L}\phi_c\} = 0$). By the continuity of the group operation, there exists an operator $\hat{L}$ that is arbitrarily close to the identity operator, $\hat{L}_\epsilon = 1 + \epsilon\hat{T}$. It is easy to see that $\hat{T}$\phi_c is the eigenvector of the operator $S''\{\phi\}$ associated with its zero eigenvalue. Therefore, $\det[S''\{\phi_c\}] = 0$, and expression (4.1.12) is divergent. However, its divergence is spurious, being related with inapplicability of the Gaussian approximation to integrals over zero modes.

To calculate integrals of this kind correctly, collective variables $\lambda_i$ are formally defined as functionals of an arbitrary field configuration, $\lambda_i = f_i\{\phi\}$. Thus, the "center" $x_0$ of an instanton can be defined by the relation

$$\int d^d x \phi^4(x)(x - x_0) = 0, \qquad (4.1.13)$$

giving an example for definition of collective variable

$$x_0 = \frac{\int d^d x \phi^4(x) x}{\int d^d x \phi^4(x)}. \qquad (4.1.14)$$

The integration over collective variables is introduced by inserting the following partition of unity into the integrand in (4.1.11):

$$1 = \prod_{i=1}^{r} \int d\lambda_i \delta(\lambda_i - f_i\{\phi\}), \qquad (4.1.15)$$

where $f_i\{\phi\}$ can be defined as homogeneous functionals of $\phi$ of degree zero[5] (cf. (4.1.14)). If the arguments of the delta functions in (4.1.15) are linearized in the neighborhood of a saddle-point configuration,

$$1 = \prod_{i=1}^{r} \int d\lambda_i \delta(\lambda_i - f_i\{\phi_c\} - (f_i'\{\phi_c\}, \delta\phi))$$

$$= \prod_{i=1}^{r} \int d\lambda_i \delta(\lambda_i - f_i\{\phi_c\} - \sqrt{g_c}(f_i'\{\phi_c\}, \delta\phi)), \qquad (4.1.16)$$

and the instanton is chosen so that $\lambda_i - f_i\{\phi_c\} = 0$ (e.g., using a solution that is symmetric about the point $x = x_0$ in (4.1.14)), then $\phi_c$ is a function of $\lambda_i$, i.e., $\phi_c \equiv \phi_\lambda$. Per-form a linear change of variables $\delta\phi \longrightarrow \hat{S}\delta\phi$ with $\det\hat{S} = 1$ to diagonalize the matrix $S''\{\phi_c\}$ and set

$$D\phi = D'\phi \prod_{i=1}^{r} d\tilde{\phi}_i, \qquad (4.1.17)$$

where the $r$ variables (denoted by the tilde) that correspond to the zero eigenvalues of $\hat{S}''\{\phi_c\}$ and actually do not contribute to the exponential in (4.1.11) are factored out. Substituting (4.1.16) and (4.1.17) into (4.1.11), removing the delta functions by performing integration in $\delta\tilde{\phi}_i$, and calculating the integral in $D'\phi$, we obtain

$$Z_N = c S_0^{-N} \Gamma\left(N + \frac{M+r}{2}\right), \qquad (4.1.18)$$

$$S_0 = S\{\phi_c\},$$

$$c = \frac{S_0^{-(M+r)/2}}{(2\pi)^{1+r/2}} \sqrt{-\frac{\det S''\{0\}}{\det[S''\{\phi_c\}]_{P'} \det[f'\{\phi_c\}]_P}}$$

$$\times \int \prod_{i=1}^{r} d\lambda_i \phi_\lambda^{(1)} \ldots \phi_\lambda^{(M)}, \qquad (4.1.19)$$

where $f'\{\phi_c\}$ is the operator defined by the matrix consisting of the columns $f_i'\{\phi_c\}$, and the subscripts $P$ and $P'$ denote projections onto the subspace spanned by the zero modes and its complement, respectively.[6] The ultraviolet divergences that arise when the constant $c$ is calculated are eliminated by conventional renormalization of mass and charge. A general renormalization scheme of this kind was developed by Brezin and Parisi (see [8, 34]). Specific values of the parameters in (4.1.18) can be found in [7, 14, 34, 35], and the most general formal results were presented in [36–38].

According to (4.1.18), each degree of freedom associated with a zero mode contributes 1/2 to the argument of the gamma function. This resembles the classical equipartition law and a more careful consideration reveals a direct analogy. Indeed, the conventional partition function $Z$ is a configuration-space integral of $\exp(-H/T)$. As the number $r_{osc}$ of oscillatory degrees of freedom increases by unity, $Z$ changes to $ZT^{1/2}$ and a corresponding 1/2 is added to specific heat [39]. Integral (4.1.4) is dominated by the exponential $\exp(-S\{\phi\}/g)$, and the coupling constant $g$ plays the role of temperature. An increase by unity in the number $r$ of zero modes corresponds to a decrease by unity in $r_{osc}$ and change from $Z$ to $Zg^{-1/2}$. To calculate the Lipatov asymptotic form, the factor $g^{-1/2}$ is estimated at the saddle point $g_c \sim 1/N$ (see (4.1.8)), $Z_N$ is replaced by $Z_N N^{1/2}$, and 1/2 is added to the argument of the gamma function. In $\varphi^4$ theory with $d < 4$, the total number of zero modes is $r = d + n - 1$, including $d$ instanton translations and $n - 1$ instanton rotations in a vector space. In a four-dimensional massless theory, there also exists a dilatation mode, corresponding to variation of the instanton radius and related with scale invariance.

---

[5] The result is actually independent of the particular form of the functionals [33], and only their linear independence is essential.

[6] In some cases, when $\det[f'\{\phi_c\}]_P$ depends on collective variables, it should be factored into the integral in $d\lambda_i$.

The equipartition law may be violated in the presence of soft modes related with approximate symmetries: some degrees of freedom resemble zero modes in the first approximation (see Fig. 4a), but a more accurate analysis shows that they correspond to motion in a slowly varying potential (Fig. 4b), which may have a nonanalytic minimum (Fig. 4c). Evidently, the contribution of such a mode to the argument of the gamma-function is neither zero nor 1/2.

The problem arising in the presence of soft modes is that the instanton $\phi_c$ is only an approximate solution of the equation $S'\{\phi\} = 0$, while the exact solution may not exist at all. Accordingly, the linear terms in the expansion of $S\{\phi\}$ in powers of $\delta\phi$ should be accurately eliminated. The collective variable characterizing the location in a slowly varying potential (see Fig. 4b) can be formally defined as a functional of an arbitrary field configuration: $z = f\{\phi\}$. An idea is to seek extremum of the action under the constraint $f\{\phi\} = $ const (i.e., for a constant $z$) and then to integrate over $z$. Correspondingly, the instanton is determined by the equation

$$S'\{\phi_c\} - \mu f'\{\phi_c\} = 0, \quad (4.1.20)$$

where $\mu$ is a Lagrange multiplier, and the integration over $z$ is introduced by inserting the following partition of unity into the functional integral:

$$\begin{aligned} 1 &= \int dz \delta(z - f\{\phi\}) \\ &= \int dz \delta(z - f\{\phi_c\} - (f'\{\phi_c\}, \delta\phi)). \end{aligned} \quad (4.1.21)$$

Using the condition $z = f\{\phi_c\}$ to fix an arbitrariness in the choice of an instanton, one obtains

$$Z(g) = \int D\varphi \varphi^{(1)} \ldots \varphi^{(M)}$$

$$\times \exp\left\{-\frac{S\{\phi_c\} + (S'\{\phi_c\}, \delta\phi) + \frac{1}{2}(\delta\phi, S''\{\phi_c\}\delta\phi)}{g}\right\}$$

$$\times \int dz \delta(-(f'\{\phi_c\}, \delta\phi)), \quad (4.1.22)$$

and the linear in $\delta\phi$ terms in the exponential are eliminated by a delta function due to condition (4.1.20). Since $\phi_c$ is a function of $z$, integration with respect to $D\varphi$ results in a nontrivial integral in $z$, which corresponds to the motion in a slowly varying potential (Fig. 4b). Note that the transformations performed in (4.1.20)–(4.1.22) are not restrictive, and any degree of freedom can be treated as a soft mode. However, this has a sense only if the validity of the Gaussian approximation is questionable.

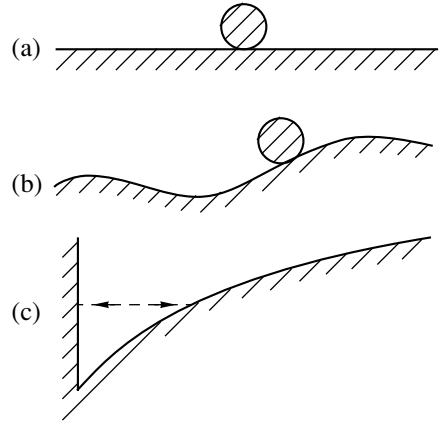

**Fig. 4.** Soft mode looks as zero mode in the first approximation (a) and corresponds to motion in a slowly varying potential in a more detailed consideration (b); (c) the potential may have a nonanalytic minimum.

Examples of soft modes are the dilatations in the massive four-dimensional or $(4 - \epsilon)$-dimensional $\varphi^4$ theory [37, 38] and the variation of the distance between elementary instantons in a two-instanton configuration (see Sections 4.4 and 9). The analysis above shows that, generally, the shift $b$ in the argument of the gamma function in (2.5) includes contributions of external lines ($M/2$), zero modes ($r/2$), and the additional contribution $\nu$ related to soft modes.

### 4.2. Quantum Electrodynamics

The simplest functional integral in quantum electrodynamics (vacuum integral) has the form

$$Z = \int DA D\bar{\psi} D\psi \exp\left\{-\int d^4x \left[\frac{1}{4}(\partial_\mu A_\nu - \partial_\nu A_\mu)^2 \right.\right.$$

$$\left.\left. + \bar{\psi}(i\gamma_\nu \partial_\nu - m + e\gamma_\nu A_\nu)\psi\right]\right\}, \quad (4.2.1)$$

where $A_\nu$ is the vector potential, and $\bar{\psi}$, $\psi$ denote fermionic fields constructed with the use of Grassmanian variables. The latter are abstract quantities for which formal algebraic operations are defined [40, 41], and the standard Lipatov method cannot be applied directly when they enter to an exponential. A remedy can be found [15] by noting that the action is quadratic in the fermionic fields, while calculation of the Gaussian integral (being one of the standard operations in the Grassmanian algebra) gives the determinant of the corresponding quadratic form:

$$Z = \int DA \det(i\gamma_\nu \partial_\nu - m + e\gamma_\nu A_\nu)$$
$$\times \exp\left\{-\frac{1}{4}\int d^4x (\partial_\mu A_\nu - \partial_\nu A_\mu)^2\right\}. \quad (4.2.2)$$

If det(…) is represented as exp{logdet(…)}, then the resulting effective action contains only the vector potential $A_\nu$ and can be treated by Lipatov's method.

The determinant of an operator is too difficult to be used constructively, and considerable effort has been applied to reduce it to a tractable form (see [22, 23]). The difficulty consisted in establishing the general properties of a saddle-point configuration when no tractable expression for effective action was available [22]. It was found that the saddle-point value of $eA_\nu(x)$ is large. Accordingly, one can make use of the asymptotic form of the determinant as $e \longrightarrow i\infty$ since the fastest growth corresponds to a pure imaginary $e$ [20]:

$$\det(i\gamma_\nu \partial_\nu - m + e\gamma_\nu A_\nu)$$
$$= \exp\left\{\frac{e^4}{12\pi^2}\int d^4x (A_\nu^2)^2\right\}. \quad (4.2.3)$$

Expression (4.2.3) is not gauge invariant. It is valid only in a restricted set of gauges for which the length scale of vector-potential variation is comparable to that of the physical electromagnetic field, which is treated as semiclassical.[7] Actually, these gauges are close to the Lorentz gauge, as can be shown by considering configurations characterized by high symmetry [20, 23].

Substituting (4.2.3) into (4.2.2), one obtains the functional integral containing the effective action

$$S_{\text{eff}}\{A\} = \int d^4x \left\{\frac{1}{4}(\partial_\mu A_\nu - \partial_\nu A_\mu)^2 - \frac{4}{3}g^2(A_\nu^2)^2\right\}, (4.2.4)$$

for which asymptotic behavior of perturbation theory can be found in the saddle-point approximation. The structure of the asymptotics is determined by the homogeneity properties of the action, which are analogous to those in $\varphi^4$ theory with $g^2$ used as a coupling constant. According to Section 4.1, the general term of the asymptotics has the form $cS_0^{-N}\Gamma(N+b)g^{2N}$, where $S_0$ is an instanton action. In fact, the series expansion is developed in arbitrary (and not only even) powers of $g$, and the substitution $N \longrightarrow N/2$ leads to $cS_0^{-N/2}\Gamma(N/2 + b)g^N$ as the $N$th-order contribution. To justify this formal substitution, we note that the direct expansion in powers of the last term in (4.2.4) is not correct, because the functional integration would involve configurations for which (4.2.4) is not valid. Calculation should be performed by the saddle-point method, which yields a continuous function of $N$, and the fact that this function should be taken at integer or half-integer points is considered as an external condition.

---

[7] The general scheme for deriving expressions of type (4.2.3) is illustrated in Section 4.3 by using a simpler example.

By using the value of the instanton action, the coefficients of high-order terms in the expansion of (4.2.1) are expressed as follows [14, 15]:

$$Z_N = \text{const}\, S_0^{-N/2}\Gamma\left(\frac{N+r}{2}\right), \quad S_0 = \frac{4\pi^3}{3^{3/2}}. \quad (4.2.5)$$

Here, the total number $r$ of zero modes is 11, including four translations, a scale transformation, and six four-dimensional rotations, since the symmetry of an instanton is similar to that of an irregular solid.

The scheme developed above can be applied to calculate other quantities [42]. The most general vertex in QED contains $M$ photon legs and $2L$ electron legs and corresponds to the functional integral

$$Z_{M,L} = \int DAD\overline{\psi}D\psi A(x_1)$$
$$...A(x_M)\psi(y_1)\overline{\psi}(z_1)...\psi(y_L)\overline{\psi}(z_L)$$
$$\times \exp\left\{-\int d^4x\left[\frac{1}{4}(\partial_\mu A_\nu - \partial_\nu A_\mu)^2\right.\right. \quad (4.2.6)$$
$$\left.\left.+ \overline{\psi}(i\gamma_\nu\partial_\nu - m + e\gamma_\nu A_\nu)\psi\right]\right\}.$$

Integration over the fermionic fields results in

$$Z_{M,L} = \int DAA(x_1)$$
$$...A(x_M)G(y_1, z_1)...G(y_L, z_L)$$
$$\times \det(i\gamma_\nu\partial_\nu - m + e\gamma_\nu A_\nu) \quad (4.2.7)$$
$$\times \exp\left\{-\frac{1}{4}\int d^4x(\partial_\mu A_\nu - \partial_\nu A_\mu)^2\right\} + ....$$

Here, $G(x, x')$ is the Green function of the Dirac operator,

$$(i\gamma_\nu\partial_\nu - m + e\gamma_\nu A_\nu)G(x, x') = \delta(x - x'), \quad (4.2.8)$$

and the ellipsis stands for terms with different pairings of $\psi(y_i)$ and $\overline{\psi}(z_k)$. The structure of the result can be found by performing calculations as demonstrated above, i.e., essentially by dimensional analysis. It can readily be shown that $e_c \sim N^{-1/4}$ and $A_c(x) \sim N^{1/2}$ for a saddle-point configuration. To determine the dimension of $G(x, x')$, write out the Dyson equation that follows from (4.2.8):

$$G(x, x') = G_0(x - x') - \int d^4 y G_0(x - y) e \gamma_\nu A_\nu(y) G(y, x'). \quad (4.2.9)$$

To elucidate the structure of the solution, consider the scalar counterpart of (4.2.9) and assume that the function $A_\nu(x)$ is localized within a small neighborhood of $x = 0$. Then, the equation is easily solved by setting $G(y, x') \approx G(0, x')$:

$$G(x, x') = G_0(x - x') - \frac{G_0(-x') \int d^4 y G_0(x - y) e \gamma_\nu A_\nu(y)}{1 + \int d^4 y G_0(-y) e \gamma_\nu A_\nu(y)}. \quad (4.2.10)$$

Since $eA_\nu(x) \sim N^{1/4}$ and (4.2.10) tends to a finite limit as $e \longrightarrow \infty$, the result is $G(x, x') \sim N^0$. It is reasonable to expect that its validity is independent of the assumptions used in its derivation. The $N$th-order contribution to integral (4.2.6) has the form

$$\text{const} \left( \frac{3^{3/2}}{4\pi^3} \right)^{N/2} \Gamma\left( \frac{N + r + M}{2} \right) (-g)^N \quad (4.2.11)$$

for even $M$ and a similar form multiplied by $eN^{1/4}$ for odd $M$.

### 4.3. Other Fermionic Models

Another example is given by Yukawa model [15], describing effective fermion interaction due to exchange by bosons (decription of electron–phonon interaction in metals is reduced to this model with minor changes):

$$Z = \int D\varphi D\bar\psi D\psi \exp\left\{-\int d^d x \left[\frac{1}{2}(\partial_\mu \varphi)^2 + \frac{1}{2} m^2 \varphi^2 + \bar\psi(i\gamma_\nu \partial_\nu + M)\psi + \lambda \bar\psi \varphi \psi \right]\right\}. \quad (4.3.1)$$

Integration over the fermionic fields results in

$$Z = \int D\varphi \det(i\gamma_\nu \partial_\nu + M + \lambda\varphi) \\ \times \exp\left\{-\int d^d x \left[\frac{1}{2}(\partial_\mu \varphi)^2 + \frac{1}{2} m^2 \varphi^2\right]\right\}. \quad (4.3.2)$$

The transformation of the fermion determinant begins with the solution of an analogous problem for the determinant of the Schrödinger operator normalized to the determinant of the unperturbed problem (normalization of this kind always arises due to the normalization of a functional integral to the vacuum integral of the interaction-free theory):

$$D(z) = \frac{\det[-\Delta - E + zV(x)]}{\det[-\Delta - E]} \\ = \det\left[1 + z\frac{V(x)}{-\Delta - E}\right]. \quad (4.3.3)$$

It can easily be shown that

$$D(z) = \prod_s \left(1 + \frac{z}{\mu_s}\right), \quad (4.3.4)$$

where $\mu_s$ denotes the eigenvalues of the problem

$$\{-\Delta - E - \mu_s V(x)\} e_s(x) = 0. \quad (4.3.5)$$

The number $s$ of energy states below $E$ for an electron moving in a semiclassical potential $-\mu V(x)$ can be found in the Thomas–Fermi approximation. By introducing the local Fermi momentum

$$p(x) = \sqrt{E + \mu V(x)} \approx [\mu V(x)]^{1/2}, \quad (4.3.6)$$

it is expressed as

$$s = \int n(x) d^d x = \int \frac{K_d}{d} p^d(x) d^d x \\ \approx \int \frac{K_d}{d} [\mu V(x)]^{d/2} d^d x, \quad (4.3.7)$$

where $n(x)$ is the local electron density and $K_d$ is the area of a $d$-dimensional unit sphere divided by $(2\pi)^d$. Since the value of $\mu_s$ in (4.3.5) corresponds to the condition that exactly $s$ electron energy states lie below $E$, expression (4.3.7) describes the asymptotic behavior of $\mu_s$ for large $s$:

$$s = A\mu_s^{d/2}, \quad A = \frac{K_d}{d} \int [V(x)]^{d/2} d^d x. \quad (4.3.8)$$

Now, the value of (4.3.4) can easily be estimated at large $z$:

$$\ln D(z) = \sum_{s \sim 1}^\infty \ln\left(1 + \frac{z}{\mu_s}\right) \\ \approx A\frac{d}{2} \int_{\sim A^{-2/d}}^\infty d\mu_s \mu_s^{d/2 - 1} \ln\left(1 + \frac{z}{\mu_s}\right) \quad (4.3.9) \\ \approx A\frac{d}{2} z^{d/2} \int_0^{\sim zA^{2/d}} \frac{dx}{x^{1 + d/2}} \ln(1 + x) \approx Az^{d/2} \frac{\pi}{\sin(\pi d/2)}.$$

When $d < 2$, the upper limit in the integral can be set to be infinite, and the integral is calculated by changing from $x$ to $ax$ in the logarithm and differentiating the result

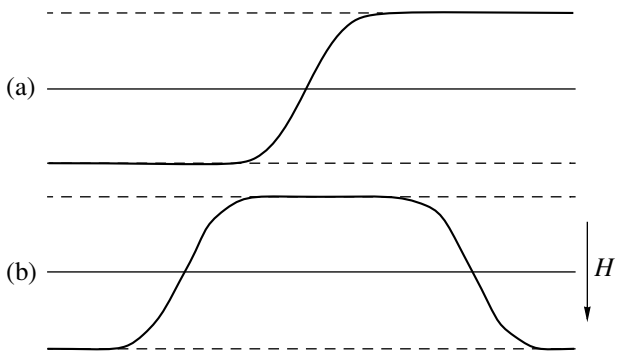

**Fig. 5.** Domain wall as example of topological instanton for the problems with degenerate vacuum (a). If degeneracy of vacuum states is removed, a saddle-point configuration corresponds to an instanton-anti-instanton pair (b).

with respect to $a$. The substitution of $K_d = 2^{1-d}\pi^{-d/2}/\Gamma(d/2)$ yields the final result

$$D(z) = \exp\left\{-\frac{\Gamma(-d/2)}{(4\pi)^{d/2}}\int d^d x [zV(x)]^{d/2}\right\}. \quad (4.3.10)$$

It is well known that the Dirac operator can be obtained by factorization of the Klein–Gordon operator, which is transformed into the Schrödinger operator by Wick rotation:

$$-\Delta + m^2 = (i\gamma_\nu\partial_\nu + m)(-i\gamma_\nu\partial_\nu + m). \quad (4.3.11)$$

An analogous relation holds after $m$ is replaced with $\lambda\varphi(x)$, where $\varphi(x)$ is a slowly varying function. For large $\lambda$, it can readily be shown that

$$\det\left(\frac{i\gamma_\nu\partial_\nu + M + \lambda\varphi}{i\gamma_\nu\partial_\nu + M}\right) \approx \sqrt{\det\left(\frac{-\Delta + \lambda^2\varphi^2(x)}{-\Delta}\right)}$$

$$\approx \exp\left\{-\frac{\Gamma(-d/2)}{2(4\pi)^{d/2}}\int d^d x [\lambda\varphi(x)]^d\right\} \quad (4.3.12)$$

and (4.3.2) acquire the effective action

$$S_{\text{eff}} = \int d^d x\left[\frac{1}{2}(\partial_\mu\varphi)^2 + \frac{1}{2}m^2\varphi^2\right.$$
$$\left.+ \frac{\Gamma(-d/2)}{2(4\pi)^{d/2}}g^{d/2}\varphi^d(x)\right], \quad (4.3.13)$$

where $g = \lambda^2$ is as an effective coupling constant, in terms of which the integral (4.3.1) is expanded.

When $2 \le d < 4$ or $d \ge 4$, the divergence of the integral in (4.3.9) is eliminated by renormalization of mass and charge [34] so that $\ln(1 + x)$ is replaced with $\ln(1 + x) - x$ and $\ln(1 + x) - x + x^2/2$, respectively. The integral is calculated by changing from $x$ to $ax$ and differentiating the result with respect to $a$. The result obtained for $d \ne 2, 4$ is formally identical to (4.3.13). If $d = 2$ or 4, then the calculation can be performed to logarithmic accuracy by taking into account the finite upper limit in the last integral in (4.3.9).

Effective action (4.3.13) can be rewritten as

$$S_{\text{eff}}\{g, \varphi\} = \frac{S\{\phi\}}{g^{1/\alpha}}, \quad (4.3.14)$$

where

$$\phi = \varphi g^{1/2\alpha}, \quad \alpha = \frac{d-2}{d},$$

and the change to $\tilde{g} = g^{1/\alpha}$ and $\tilde{N} = \alpha N$ can be performed to reduce the expression for the expansion coefficients to a form analogous to that in $\varphi^4$ theory. The final result is similar to $\varphi^4$ theory up to the change $N \longrightarrow \alpha N$:

$$Z_N = cS_0^{-\alpha N}\Gamma(\alpha N + b). \quad (4.3.15)$$

### 4.4. Degenerate Vacuum

Calculations of the Lipatov asymptotic forms for theories with degenerate vacuum require special analysis.

Consider, for example, the one-dimensional Ising ferromagnet with a doubly degenerate ground state with all spins either up or down. In addition to these vacuums, there exists a classical domain-wall solution (an example of *topological instanton*), which corresponds to transition between the two degenerate vacuums (Fig. 5a). The issues to be resolved in such problems arise from the following: (a) the contribution of topological instantons to asymptotic expression for expansion coefficients is pure imaginary, which implies that they are insignificant in some sense; (b) generally, the absence of other nontrivial classical solutions in problems of this kind is established by special theorems.

To elucidate these issues, suppose that the degeneracy of the vacuum states is eliminated by applying a magnetic field aligned with the ferromagnet's axis. Then, domain-wall-like excitations cannot exist, because they are associated with an infinitely large additional energy (in the infinite-volume limit). However, excitations can exist in the form of instanton–anti-instanton pairs (Fig. 5b). If the interaction between the components of such a pair is repulsive, then there exists an equilibrium distance between the components. If magnetic field is increased, the equilibrium distance decreases and a localized instanton typical for Lipatov's method arises. If magnetic field is decreased, the equilibrium distance increases, with simultaneous growth of its fluctuations. As a result, the pair breaks up into free instanton and anti-instanton in the limit of strictly degenerate vacuum states.

Because of strong fluctuations in the distance between the components, the saddle-point approximation can be applied in Lipatov's method only if $N \gg 1/\epsilon$ (rather than $N \gg 1$), where $\epsilon$ is a small parameter characterizing the difference between the vacuums. However, there exists an intermediate asymptotic regime, which is virtually independent of $\epsilon$, for $1 \ll N \ll 1/\epsilon$. As $\epsilon \longrightarrow 0$, this intermediate asymptotic behavior transfers into the true asymptotics of the degenerate problem. This implies that the latter asymptotics can be found without analyzing the case of strictly zero $\epsilon$. For finite $\epsilon$, single instantons do not exist and therefore do not contribute to the asymptotics of perturbation theory, which is determined by the instanton–anti-instanton pair.

The physical picture described above was developed in an analysis of the quantum mechanical problem of double-well oscillator [43]. Analogous picture can be expected for other theories with degenerate vacuum, among which Yang–Mills theories are of particular interest.

### 4.5. Yang–Mills Theory and QCD

The topological instanton found in [44] for the Yang–Mills theory was the earliest evidence of the existence of degenerate vacuum in QCD. In [45], the saddle-point calculation of a functional integral was performed for the one-instanton configuration of $SU(2)$ Yang–Mills fields coupled to fermions and scalar particles. This result was extended to arbitrary $SU(N_c)$ symmetry in [46]. In an analysis of saddle-point configurations performed for the Yang–Mills field coupled to a scalar field in [18], a continuous transformation of the $\varphi^4$ theory instanton into a saddle-point configuration for the pure Yang–Mills theory was found. The latter configuration was shown to correspond to an instanton–anti-instanton pair; i.e., a physical picture described above was confimed for the Yang-Mills theory. The result of [45] for a single instanton was used in [17] to calculate the contribution of the instanton–anti-instanton configuration to asymptotic behavior of perturbation theory for $SU(2)$ Yang–Mills fields. The Lipatov asymptotic forms for realistic QCD were calculated in [24–26]. A general scheme of these calculations is presented below.

As a first step, we formulate a rule for combining of instantons [33] on example of the functional integral

$$Z_M(g) = \int DA A^{(1)} A^{(2)} \ldots A^{(M)} \exp(-S\{A, g\}), \quad (4.5.1)$$

where $A^{(i)}$ is a bosonic field, and the superscript $i$ stands for both coordinate and internal degrees of freedom. Suppose that the action $S\{A, g\}$ is rewritten as $S\{B\}/g^2$ by changing from $A$ to $B/g$ and the equation $S'\{B\} = 0$ has an instanton solution $B_c$. By following the scheme developed in Section 4.1, it can readily be shown that the one-instanton contribution to $Z_M(g)$ has the form

$$Z_M^{(1)}(g) = c_0 g^{-M-r} e^{-S_0/g^2}$$
$$\times \int \prod_{i=1}^{r} d\lambda_i B_\lambda^{(1)} B_\lambda^{(2)} \ldots B_\lambda^{(M)}, \quad (4.5.2)$$

where $S_0 = S\{B_c\}$, $r$ is the number of zero modes, $\lambda_i$ denotes the corresponding collective variables, and $B_\lambda$ is the instanton configuration depending on these variables.

If $B_c$ is the combination $B_\lambda + B_{\lambda'}$ of elementary instantons, then the corresponding two-instanton contribution can be represented as the sum of terms, having the form

$$Z_{LL'}(g) = c_0^2 g^{-M-2r} e^{-2S_0/g^2}$$
$$\times \int \prod_{i=1}^{r} d\lambda_i d\lambda_i' B_\lambda^{(1)} \ldots B_\lambda^{(L)} B_{\lambda'}^{(1)} \ldots B_{\lambda'}^{(L')} \quad (4.5.3)$$
$$\times \exp\left(-\frac{S_{\text{int}}(B_\lambda, B_{\lambda'})}{g^2}\right)$$

with $L + L' = M$. The instanton–instanton interaction $S_{\text{int}}(B_\lambda, B_{\lambda'})$ is defined by the relation

$$S\{B_\lambda + B_{\lambda'}\}$$
$$\equiv S\{B_\lambda\} + S\{B_{\lambda'}\} + S_{\text{int}}(B_\lambda, B_{\lambda'}). \quad (4.5.4)$$

When the interaction is neglected, the right-hand side of (4.5.3) reduces to the product of two expressions having the form of (4.5.2), with $M = L$ and $M = L'$. Due to the exponential factor, the instanton–instanton interaction is limited by the condition $S_{\text{int}}(B_\lambda, B_{\lambda'}) \lesssim g^2$. When $g$ is small, the interaction is insignificant, and the overlap of $B_\lambda$ and $B_{\lambda'}$ can be neglected. The resulting sum in $L$ and $L'$ contains only the terms with $L = M$, $L' = 0$ and $L = 0$, $L' = M$, which are obviously equal. The ensuing factor 2 is canceled by the combinatorial factor $1/2!$ introduced to preclude double counting of configurations. The resulting two-instanton contribution,

$$Z_M^{(2)}(g) = c_0^2 g^{-M-2r} e^{-2S_0/g^2}$$
$$\times \int \prod_{i=1}^{r} d\lambda_i d\lambda_i' B_\lambda^{(1)} \ldots B_\lambda^{(M)} \exp\left(-\frac{S_{\text{int}}(B_\lambda, B_{\lambda'})}{g^2}\right), \quad (4.5.5)$$

entails a rule for combining instantons: in addition to the information contained in (4.5.2), it is necessary to know the instanton–instanton interaction in the domain where the interaction is weak.

In relativistic scale-invariant theories, it is convinient to single out integrations over $\rho$, $x_0$, the radius and a center of instanton [41], understanding under $\lambda_i$ only internal degrees of freedom. Then (4.5.2) takes a form:

$$Z_M^{(1)}(g) = c_H g^{-M-r} e^{-S_0/g^2} \int \prod_i d\lambda_i \int d^4 x_0 \qquad (4.5.6)$$
$$\times \int d\rho \rho^{-M-5} e^{\nu \ln \mu \rho} B_\lambda(y_1) \ldots B_\lambda(y_M),$$

where $y_i = (x_i - x_0)/\rho$, $\nu = -\beta_2 S_0$, $\beta_2$ is the lowest order nonvanishing expansion coefficient for the Gell-Mann–Low function, $\mu$ is momentum at the normalization point, and $\exp(\nu \ln \mu \rho)$ is uniquely factored out by virtue of the renormalizability condition [41]. Formula (4.5.6) is consistent with 't Hooft's result for $SU(N_c)$ Yang–Mills fields ($S_0 = 8\pi^2$, $r = 4N_c$, $\nu = (11N_c - 2N_f)/3$) [45, 46] and (with $g^2$ replaced by $g$) with the corresponding result in $\varphi^4$ theory [7, 14, 37].

The QCD Lagrangian has the form

$$L = -\frac{1}{4}(F^a_{\mu\nu})^2 - \frac{1}{2\alpha}(\partial_\mu A^a_\mu)^2$$
$$+ \sum_f \bar\psi_f \hat D \psi_f + \partial_\mu \bar\omega^a (\partial_\mu \omega^a - \bar g f^{abc} \omega^b A^c_\mu), \qquad (4.5.7)$$
$$F^a_{\mu\nu} = \partial_\mu A^a_\nu - \partial_\nu A^a_\mu + \bar g f^{abc} A^b_\mu A^c_\nu,$$
$$\hat D = i\gamma_\mu(\partial_\mu - i\bar g A^a_\mu T^a),$$

where $A^a_\nu$, $\psi_f$, and $\omega^a$ denote gluon, quark, and ghost fields, respectively; $T^a$ and $f^{abc}$ are the generators of the fundamental representation and structure constants of the Lie algebra, respectively; $\alpha$ is the gauge parameter; and the subscript "f" denotes the types of quarks, whose total number is $N_f$. The preexponential factor in the most general functional integral for QCD contains $M$ gluon fields, $2L$ ghost fields, and $2K$ quark fields:

$$Z_{MLK} = \int DAD\bar\omega D\omega D\bar\psi D\psi A(x_1)$$
$$\ldots A(x_M) \omega(y_1)\bar\omega(\bar y_1) \ldots \omega(y_L)\bar\omega(\bar y_L) \qquad (4.5.8)$$
$$\times \psi(z_1)\bar\psi(\bar z_1) \ldots \psi(z_K)\bar\psi(\bar z_K)$$
$$\times \exp(-S\{A, \bar\omega, \omega, \bar\psi, \psi\}),$$

where the vector indices that are not essential for the present analysis are omitted. By replacing $A$ with $B/\bar g$, the Euclidean action is rewritten as

$$S\{A, \bar\omega, \omega, \bar\psi, \psi\} \longrightarrow \frac{S\{B\}}{\bar g^2}$$
$$+ \int d^4 x \left[ \bar\omega \hat Q \omega + \sum_f \bar\psi_f \hat D \psi_f \right]. \qquad (4.5.9)$$

Integration over the fermionic fields results in

$$Z_{MLK} = (1/\bar g)^M \int DAB(x_1)$$
$$\ldots B(x_M) G(y_1, \bar y_1) \ldots G(y_L, \bar y_L) \qquad (4.5.10)$$
$$\times \tilde G(z_1, \bar z_1) \ldots \tilde G(z_K, \bar z_K) \det\hat Q (\det\hat D)^{N_f}$$
$$\times \exp\{-S\{B\}/\bar g^2\} + \ldots,$$

where $G$ and $\tilde G$ are the Green functions of the operators $\hat Q$ and $\hat D$, and the ellipsis stands for terms with different pairings. It is important here that $S\{B\}$, $G$, and $\tilde G$ are independent of $\bar g$. Functional integral (4.5.10) is dominated by the Yang–Mills action, and the corresponding one-instanton contribution can be written out by analogy with (4.5.2). The asymptotic behavior of perturbation theory is determined by an instanton–antiinstanton contribution calculated by analogy with (4.5.5). The instanton–instanton interaction is specified by introducing a conformal parameter $\xi$:

$$S_{int} = -h\xi, \quad \xi = \frac{\rho_I^2 \rho_A^2}{(R^2 + \rho_I^2 + \rho_A^2)^2}, \qquad (4.5.11)$$

where $\rho_I$ and $\rho_A$ denote the instanton and anti-instanton radii, $R$ is the distance between their centers, and $h = h(\lambda, \lambda')$ depends on their mutual orientation in the isotopic space [24]. Next, it should be noted that $\det\hat D \{B_\lambda + B_{\lambda'}\} \neq \det\hat D \{B_\lambda\} \det\hat D \{B_{\lambda'}\}$ since $\det\hat D\{B_\lambda\}$ does not vanish only if the finite quark mass is taken into account, whereas $\det\hat D \{B_\lambda + B_{\lambda'}\}$ is determined by the instanton–instanton interaction and is finite in the massless limit (see [24]),

$$\det\hat D\{B_\lambda + B_{\lambda'}\} = \text{const}\, \xi^{3/2}. \qquad (4.5.12)$$

By factoring the integrals over the instanton radii and centers and changing to the momentum representation, the instanton–anti-instanton contribution is rewritten as

$$Z_{MLL'}^{(IA)} = \frac{\text{const}}{\bar g^{M+2r}} e^{-2S_0/g^2} \int \prod_i d\lambda_i d\lambda_i'$$
$$\times \int d\rho \rho^{3M+6L+5L'-5} e^{2\nu \ln \mu \rho} \langle B_\lambda \rangle_{\rho p_1} \ldots \langle B_\lambda \rangle_{\rho p_M} \qquad (4.5.13)$$
$$\times \langle G_\lambda \rangle_{\rho k_1, \rho k_1'} \ldots \langle G_\lambda \rangle_{\rho k_L, \rho k_L'} \langle \tilde G_\lambda \rangle_{\rho q_1, \rho q_1'}$$
$$\ldots \langle \tilde G_\lambda \rangle_{\rho q_L, \rho q_L'} \int_0^{\sim 1} \frac{d\xi}{\xi^{1+\nu/2-3N_f/2}} e^{-h(\lambda,\lambda')\xi/\bar g^2} + \ldots,$$

where $\langle B \rangle_k$, $\langle G \rangle_{k,k'}$, and $\langle \tilde G \rangle_{k,k'}$ denote Fourier components of $B(x)$, $G(x, x')$, and $\tilde G(x, x')$, respectively; $\rho \equiv$

$\rho_I$; use is made of the fact that the dominant contribution is due to the region where $R \sim \rho_A \gg \rho_I$; and relation (18) from [24] is taken into account.

By following [17] (see also [33]), the last integral is replaced by the corresponding jump at the cut. Then, (4.5.13) yields a jump in the total value of $Z_{MLL'}$ at the cut:

$$\Delta Z_{MLL'}(\bar{g}) = i\,\mathrm{const}\left(\frac{1}{\bar{g}}\right)^{M+2r+v-3N_f} \times \exp\left(-\frac{2S_0}{\bar{g}^2}\right), \qquad (4.5.14)$$

where the independence of $B_\lambda$, $G_\lambda$, and $\tilde{G}_\lambda$ of $\bar{g}$ is used and all integrals are assumed to be convergent.[8] By virtue of relation (3.17) between the jump at the cut and the asymptotic form of expansion coefficients, the $N$th-order contribution to $Z_{MLL'}$ is

$$[Z_{MLK}]_N \bar{g}^{2N} = \mathrm{const}(16\pi^2)^{-N}$$
$$\times \Gamma\left(N + \frac{M}{2} + 4N_c + \frac{11(N_c - N_f)}{6}\right)\bar{g}^{2N} \qquad (4.5.15)$$

for even $M$ and with additional multiplier $\bar{g}N^{1/2}$ for odd $M$ [17, 24, 47]. This result is analogous to that discussed in Section 4.1: the term $M/2$ in the argument of the gamma function is determined by the number of external lines, $4N_c$ is half the number of zero modes, and $11(N_c - N_f)/6$ is the additional contribution of the soft mode corresponding to variation of the instanton–anti-instanton distance. Specific values of the constant factor were calculated in [17, 24, 25].

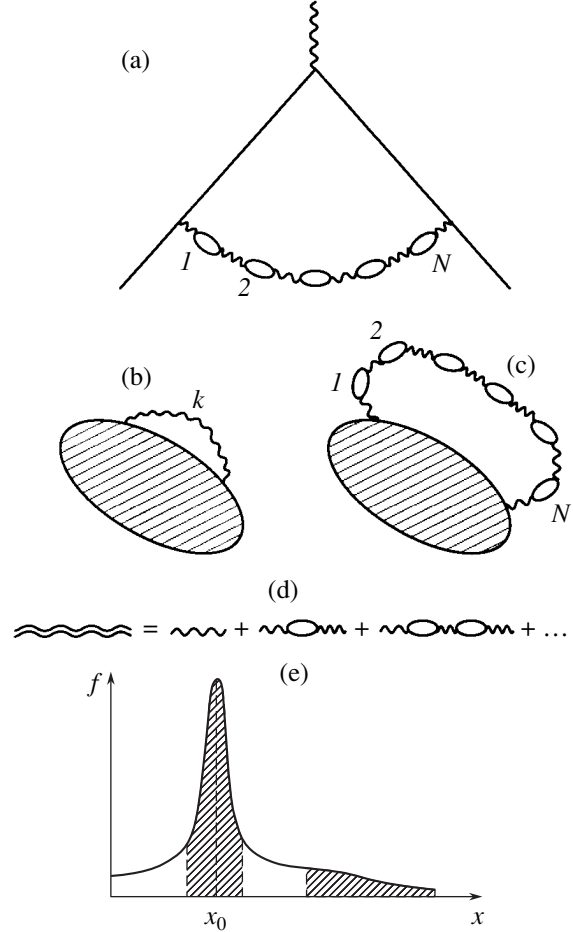

**Fig. 6.** (a) Example of single QED diagram of $N$th-order giving contribution $N!$ [49]. More wide class of renormalon diagrams is obtained by distinguishing the internal photon line (b) and inserting a chain of electron "bubbles" in it (c). (d) Insertions into the photon line correspond to "dressing" of interaction. (e) Example, when saddle-point method is formally applicable, but leads to incorrect results.

## 5. RENORMALONS. PROBLEM OF MATHEMATICAL SUBSTANTIATION OF LIPATOV'S METHOD

### 5.1. 't Hooft's Argumentation

The invention of Lipatov's method was widely recognized, and it was immediately applied to almost every topical problem in theoretical physics (see [19]). However, the validity of Lipatov's method was questioned as early as in 1977. The criticism dates back to [49], where the following interesting remark was made. Lipatov's result (2.5) is usually interpreted as a contribution of the factorial number of diagrams of order $(ag)^N$. However, this interpretation is incorrect in the general case: there exist individual $N$th-order diagrams (with long chains of "bubbles") whose contributions are proportional to $N!$ (see Fig. 6a); they were called renormalons since they arise only in renormalizable theories:[9] Though the example discussed in [49] and illustrated by Fig. 6a was taken from quantum electrodynamics, analogous diagrams arise in QCD and four-dimensional $\varphi^4$ theory. Strictly speaking, Lautrup's remark is inconsequential, since Lipatov's method relies on formal calculation of functional integral (2.4) and does not involve any statistical analysis of diagrams. Therefore, it should be expected that the renormalon contributions are already included in (2.5).

However, 't Hooft claimed in [50] that renormalons provide an independent mechanism of divergence of perturbation series and their contribution is not contained in the Lipatov asymptotics. The argumentation put forward by 't Hooft relies on an analysis of the analyticity properties of Borel transforms. Indeed, the

---

[8] In the quark–quark correlation function, the integral in $\rho$ involves divergences. The method for eliminating them proposed in [24, 26] evokes doubts [48]. For $M \geq 1$, the integral is convergent.

[9] We have in mind the theories with running coupling, where logarithmic situation takes place. No renormalons arise in super-renormalizable theories.

Borel transform of the function represented by a series with expansion coefficients $ca^N\Gamma(N+b)$ has a singular point at $z = 1/a$:

$$B(z) = \sum_N ca^N N^{b-1} z^N \sim (1-az)^{-b}, \quad (5.1)$$
$$za \longrightarrow 1.$$

Thus, the value of $a$ in (2.5) determines the location of a singular point in the Borel plane. This conclusion was obtained by 't Hooft without reference to Lipatov's method. Representing action as $S\{\phi\}/g$ (see Section 4), one can rewrite a general functional integral and the definition of Borel transform (3.12) as follows:

$$Z(g) = \int D\phi \exp\left(-\frac{S\{\phi\}}{g}\right), \quad (5.2)$$

$$Z(g) = \int_0^\infty dx\, e^{-x/g} B(x), \quad (5.3)$$

where the factors $g^n$ are omitted since they cancel out when Green functions are calculated as ratios of two functional integrals. Then, the Borel transform of functional integral (5.2) is

$$B(z) = \int D\phi\, \delta(z - S\{\phi\}) = \oint_{z = S\{\phi\}} \frac{d\sigma}{|S'\{\phi\}|}, \quad (5.4)$$

where $|S'\{\phi\}|$ is the modulus of the vector defined in (4.1.6), and the last integral is calculated over the hypersurface $z = S\{\phi\}$. If $\phi_c(x)$ is an instanton, then $S'\{\phi_c\} = 0$ and (5.4) has a singular point at $z = S\{\phi_c\}$, which coincides with $1/a$ for the instanton having the minimum action $S_0$. Furthermore, there exist singular points at the points $mS_0$, which correspond to $m$ remote instantons, and singularities corresponding to instantons of other form. If $z = S_0$ is the singular point nearest to the origin, then Lipatov asymptotic form (2.5) is valid. However, 't Hooft hypothesized that singularities other than instantons may exist, in which case the asymptotic behavior of expansion coefficients may be determined by the non-instanton singular point nearest to the origin.

Renormalons were considered by 't Hooft as a possible new mechanism of arising of singularities. The virtual photon line with momentum $k$ in an arbitrary QED diagram (see Fig. 6b) represents an large-momentum integral of the form

$$\int d^4k\, k^{-2n}, \quad (5.5)$$

where $n$ is integer. After all renormalizations are performed, the integral is convergent and $n \geq 3$. When $N$ electron bubbles are inserted into the photon line (see Fig. 6c), the integrand is multiplied by $\ln^N(k^2/m^2)$ ($m$ is electron mass), and the resulting integral is proportional to $N!$. Insertions in the photon line correspond to "dressing" of coupling. Accordingly, $g_0$ is replaced by a running coupling $g(k^2)$ arising in the integrand of (5.5). Summation of diagrams of the form shown in Fig. 6c is equivalent to the use of the one-loop approximation $\beta(g) = \beta_2 g^2$ for the Gell-Mann–Low function and leads to a well-known result:

$$g(k^2) = \frac{g_0}{1 - \beta_2 g_0 \ln(k^2/m^2)}. \quad (5.6)$$

Performing the integration over $k^2 \gtrsim m^2$, we obtain

$$\int d^4k\, k^{-2n} g(k^2) = g_0 \sum_N \int d^4k\, k^{-2n} \left(\beta_2 g_0 \ln\frac{k^2}{m^2}\right)^N$$
$$\sim g_0 \sum_N N! \left(\frac{\beta_2}{n-2}\right)^N g_0^N. \quad (5.7)$$

After the Borel summation, this yilds renormalon singularities at the points[10]

$$z_n = \frac{n-2}{\beta_2}, \quad n = 3, 4, 5, \ldots, \quad (5.8)$$

in the Borel plane $z$. In $\varphi^4$ theory and QED, instanton and renormalon singularities lie on the negative and positive half-axes, respectively (see Fig. 7a); in QCD, the converse is true. The analysis presented above shows that factorial contributions due to individual diagrams arise in any field theory where the leading term in the expansion of $\beta(g)$ is quadratic.

It is obvious that 't Hooft's argumentation with regard to renormalons leaves unanswered the following basic questions: Why should certain sequences of diagrams be considered particularly important even though they comprise only a small fraction of all diagrams? How should we deal with double counting? (In other words, how do we know that renormalons are not taken into account in instanton contribution (2.5)?) However, setting of the problem on the possibility of non-instanton contributions to the asymptotic behavior of expansion coefficients is of essential interest: it brings to light a shortcoming in the mathematical substantiation of Lipatov's method. Indeed, consider a function $f(x)$ that has a sharp peak at $x_0$ and a slowly decaying "tail" at large $x$ (Fig. 6e), so that the contributions of the peak and tail regions to the integral $\int f(x)\, dx$ are comparable. An analysis of the integral would reveal the existence of a saddle point at $x_0$ and (if it is sufficiently sharp) show that the saddle-point method is formally applicable. However, the calculation of the integral in the saddle-point approximation would be incorrect, because the contribution of the tail

---

[10] Analogous singularities with $n = 0, -1, -2, \ldots$ (known as *infrared renormalons*) arise in the integral over the small-momentum region.

would be lost. If tails of this kind contribute to (2.4), then Lipatov's method fails.

Since there is hardly any alternative to the saddle-point method in calculations of functional integrals, direct analysis of possible tail contributions cannot be performed, and 't Hooft's argumentation is difficult to disprove. Nevertheless, it is "unnatural" in a certain sense: for any finite-dimensional integral (5.2), it can be shown that (a) its value for $g \longrightarrow 0$ is determined by saddle-point configurations [51] (according to (4.1.8), $g_c \longrightarrow 0$ as $N \longrightarrow \infty$) and (b) all singularities in the Borel plane are associated with action extrema ('t Hooft's argumentation based on (5.4) is necessary and sufficient). Therefore, renormalon singularities may arise only in the limit of an infinite-dimensional integral. However, the constructive argumentation in support of their existence is rather weak and can easily be disproved by a careful analysis [48]. Further studies showed that summation of a more complicated sequences of diagrams leads to substantial modification of the renormalon contribution. The common coefficient before it becomes totally indeterminate [52], and the possibility that it vanishes cannot be ruled out. Thus, the existence of renormalon singularities is not an established fact, and this is admitted even by the most enthusiastic advocates of this hypothesis [53].

Nevertheless, 't Hooft's view immediately became popular [54–60]. Probanly, this is explained by the use of the convinient diagrammatic technique. One can easily insert a chain of bubbles into any diagram and explore the qualitative consequences of divergency of perturbation series for any phenomenon under study. As for instanton approach, it can be combined with diagrammatic calculations (see [8]), but the procedure is very cumbersome. Of course, the use of renormalons as a "model" does not arouse objections. Moreover, one can formulate conditions when such model is justified (see Section 5.3). Regrettably further investigation of high-order behavior of perturbation theory was hampered after 't Hooft's lecture [50], which had thrown doubt on the validity of Lipatov's method. As a consequence of the drop in its popularity, the complete perturbation-theory asymptotics in both QED and QCD remain uncalculated to this day, though all fundamental issues were resolved in the late 1970s.

### 5.2. Absence of Renormalon Singularities in $\varphi^4$ Theory

In the analysis of the Borel transforms arising in $\varphi^4$ theory presented in [48], it was shown that they are analytic in the complex plane with a cut extending from the nearest instanton singularity to infinity (see Fig. 7b), in agreement with a hypothesis (put forward by Le Guillou and Zinn-Justin [35]) that underlies an extremely efficient summation method, conformal–Borel technique (see Section 6.1). A comparison with 't Hooft's picture (Fig. 7a) shows that all instanton singularities are absorbed by the cut, while renormalon singularities are absent.

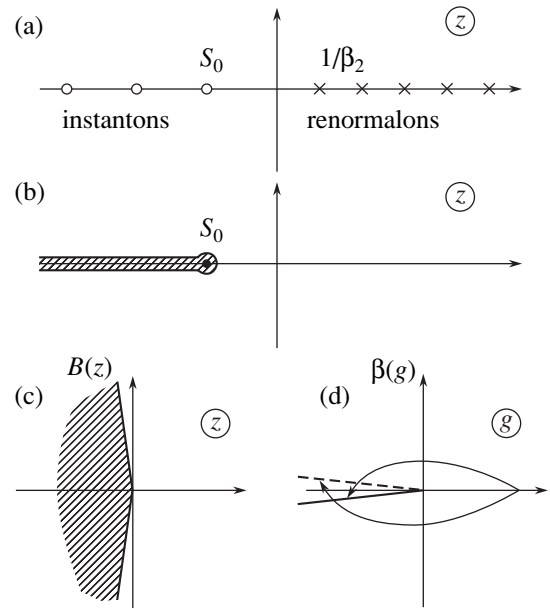

**Fig. 7.** (a) Singularities in $\varphi^4$ theory according to 't Hooft [50]. (b) Domain of analyticity according to [48]. Analyticity of $B(z)$ for $|\arg z| < \pi/2 + \delta$ (c) entails analyticity of $\beta(g)$ for $|\arg g| < \pi + \delta$ (d), i.e., on the entire physical sheet of the Riemann surface.

The approach developed in [48] relies on the use of the modified Borel transform in which $N!$ is replaced by $\Gamma(N + b_0)$ with an arbitrary $b_0$:

$$F(g) = \int_0^\infty dx\, e^{-x} x^{b_0 - 1} B(gx),$$
$$B(g) = \sum_{N=0}^\infty \frac{F_N}{\Gamma(N + b_0)} g^N \quad (5.9)$$

(Borel–Leroy transform). It can readily be shown that all Borel transforms are analytic in the same domain, which is easy to find by setting $b_0 = 1/2$, since the corresponding Borel transform preserves exponential form:

$$F(g) = e^{-g} \longrightarrow B(z) = \frac{1}{2\sqrt{\pi}} \{e^{2i\sqrt{z}} + \text{c.c.}\}. \quad (5.10)$$

Accordingly, the Borel transform of (2.3) can be expressed as

$$B(z) = \frac{1}{2\sqrt{\pi}} \int D\varphi \exp(-S_0\{\varphi\}) \\ \times [\exp(2i\sqrt{z}S_{\text{int}}\{\varphi\}) + \text{c.c.}], \quad (5.11)$$

and used to find the domain of analyticity for any finite-dimensional integral (2.3) (i.e. functional integral, determined on a finite-size lattice) for $m^2 > 0$.

The infinite-volume limit can be taken after Green functions are constructed as ratios of two integrals of the type (2.3). This limit is not singular, if the system under study is not at a phase-transition point: then partition into quasi-independent subsystems is possible due to finiteness of the correlation length; in $\varphi^4$ theory, this possibility is guaranteed by the condition $m^2 > 0$.

Transition to the continuum limit does not present any problem in the absence of ultraviolet divergences, which corresponds to $d < 2$. When the theory is divergent in the ultraviolet limit, the proof consists of the following steps:

(a) the domain of analyticity of $B(z)$ is determined for a finite cutoff parameter $\Lambda$ by using Feynman regularization;

(b) the domain of analyticity is found for the Borel transform of the Gell-Mann–Low function and the anomalous dimensions defined in the cutoff scheme, whose dependence on $\Lambda$ fades out as $\Lambda \longrightarrow \infty$;

(c) the invariance of the domain of analyticity under charge renormalization is proved;

(d) the domain of analyticity is determined for renormalized vertices and renormalization-group functions in other renormalization schemes.

Let us discuss a subtle point of the proof that was not elucidated in [48]. Any quantity calculated perturbatively is a function of the bare charge $g_B$ and the cutoff parameter $\Lambda$. Changing to a renormalized charge $g$, one obtains a function $F(g, \Lambda)$ that have weak dependence on $\Lambda$, but approaches a finite limit as $\Lambda \longrightarrow \infty$ due to renormalizability. Similarly, its Borel transform $B(z, \Lambda)$ tends to a finite limit $B(z)$. In [48], it was rigorously proved that $B(z, \Lambda)$ is analytic in the complex $z$ plane with a cut extending from the nearest instanton singularity to infinity when $\Lambda$ is finite. The function $B(z)$ is analytic in the same domain if the series is uniformly convergent (by the Weierstrass theorem [61]), which is the case when $B(z, \Lambda)$ is bounded (by compactness principle [62]). Therefore, regularity of $B(z)$ is guaranteed if the limit with respect to $\Lambda$ is finite. However, renormalizability has been rigorously proved only in the framework of perturbation theory, i.e., for the coefficients of expansions in $g$ and $z$, rather than directly for the functions $F(g, \Lambda)$ and $B(z, \Lambda)$. The proof presented in [48] assumes the existence of finite limits on the level of functions and is incomplete in this respect. However, the existence of these finite limits should be considered as a necessary physical condition for true renormalizability. This condition is directly related to the necessity of redefinition of functional integrals discussed below (see Section 5.3).

### 5.3. General Criterion for the Absence of Renormalon Singularities

The absence of renormalon singularities in the four-dimensional $\varphi^4$ theory, which is a typical "renormalon" theory, puts to question the general concept of renormalon. The problem of renormalons in an arbitrary field theory was elucidated in [63]. Returning to quantum electrodynamics, consider the simplest possible class of renormalon diagrams corresponding to all kinds of insertions into a photon line (see Figs. 6b and 6c). When the function $\beta(g)$ is known, all of these diagrams can easily be summed by solving the Gell-Mann–Low equation

$$\frac{dg}{d\ln k^2} = \beta(g) = \beta_2 g^2 + \beta_3 g^3 + \dots \qquad (5.12)$$

under the initial condition $g(k^2) = g_0$ for $k^2 = m^2$; then one can judge on existence of renormalon singularities analyzing the expansion in $g_0$ for an integral of type (5.7).

The solution to Eq. (5.12) is

$$F(g) = F(g_0) + \ln\frac{k^2}{m^2},$$
$$\text{where } F(g) = \int\frac{dg}{\beta(g)}. \qquad (5.13)$$

In view of the behavior of $F(g)$ for small g, the following expression can be used:

$$F(g) = -\frac{1}{\beta_2 g} + f(g),$$
$$\lim_{g \to 0} gf(g) = 0. \qquad (5.14)$$

The formal solution of (5.13) for $g$ is

$$g(k^2) = F^{-1}\left\{-\frac{1}{\beta_2 g_0} + f(g_0) + \ln\frac{k^2}{m^2}\right\}. \qquad (5.15)$$

Here, the right-hand side is regular at $g_0 = 0$; i.e., it can be represented as a series in powers of $g_0$ of the form

$$g = \sum_{N=1}^{\infty} A_N \left\{\frac{g_0}{r(x)}\right\}^N, \quad x = \beta_2 \ln\frac{k^2}{m^2}, \qquad (5.16)$$

where $r(x)$ is the radius of convergence and $A_N$ behaves as a power of $N$. The radius of convergence is determined by the distance to the singular point nearest to the origin.

If $z_c$ is a singular point of the function $F^{-1}(z)$, then the singular points in $g_0$ of (5.15) satisfy the equation

$$z_c = -\frac{1}{\beta_2 g_0} + f(g_0) + \ln\frac{k^2}{m^2} \quad (5.17)$$

or

$$g_0 x - 1 = \beta_2 g_0 [z_c - f(g_0)]. \quad (5.18)$$

If $z_c$ is finite, then Eq. (5.18) at large x has a small root $g_0 \approx 1/x$, since its right-hand side is negligible by virtue of (5.14). As a result, there exists a singular point at $g_c \approx 1/x$, and the series in (5.16) is

$$g(k^2) = \sum_{N=1}^{\infty} A_N (g_0 x)^N$$
$$= \sum_{N=1}^{\infty} A_N \left(\beta_2 \ln\frac{k^2}{m^2}\right)^N g_0^N. \quad (5.19)$$

Sustitution (5.19) into integral (5.7) leads to renormalon singularities at the points (5.8) (note that the integral is dominated by the contributions of large $k$, which correspond to large $x$). If $z_c = \infty$, then Eq. (5.17) has no solution for $g_0 \sim 1/x$, and the expansion of type (5.19) is possible with coefficients, decreasing faster than any exponential. Thus, the renormalon contribution is definitely smaller than the instanton one and the Borel plane does not contain renormalon singularities.

If the function $z = F(g)$ is regular at $g_0$ and $F'(g_0) \neq 0$, then its inverse $g = F^{-1}(z)$ is regular, existing in the neighborhood of $g_0$. Therefore, the singular points of $F^{-1}(z)$ are $z_c = F(g_c)$, where $g_c$ is any point determined by condition

$$F'(g_c) = 0 \text{ or } F'(g_c) \text{ does not exist.} \quad (5.20)$$

In summary, renormalon singularities exist if there is at least one point $g_c$ (including $g_c = \infty$) satisfying condition (5.20) and $z_c = F(g_c) < \infty$. Otherwise, renormalon singularities do not exist.

In terms of the $\beta$ function, this result imply that renormalon singularities do not exist if $\beta(g) \sim g^\alpha$ with $\alpha \leq 1$ at infinity and its singularities at finite $g_c$ are so weak that the function $1/\beta(g)$ is nonintegrable at $g_c$ (e.g., $\beta(g) \sim (g - g_c)^\gamma$ with $\gamma \geq 1$). When either condition is violated, there exist renormalon singularities in the points (5.8).

An analysis of more complicated classes of renormalon diagrams relying on the general Callan–Symanzik renormalization-group equation [63] leads to similar conclusions: necessary and sufficient conditions for the existence of renormalon singularities can be established, but no definite assertions can be made by using only results of renormalization-group analysis.

Now, recall that the perturbation series expansion of $\beta(g)$ is factorially divergent because there exists a cut emanating from the origin in the complex $g$ plane. Therefore, both $g = 0$ and $g = \infty$ are branch points, and $\beta(g)$ can be represented by a Borel integral:

$$\beta(g) = \int_0^\infty dz\, e^{-z} B(gz) = g^{-1} \int_0^\infty dz\, e^{-z/g} B(z). \quad (5.21)$$

Suppose that the Borel transform $B(z)$ behaves as $z^\alpha$ as $z$ goes to infinity[11] (in which case $\beta(g) \propto g^\alpha$) and is a regular function for $|\arg z| < \pi/2 + \delta$, where $\delta > 0$ (see Fig. 7c). Then, $\beta(g)$ is a regular function for $|\arg g| < \pi + \delta$ (see Fig. 7d), which implies the absence of singularities at finite $g$ in the physical sheet of the Riemann surface. Then the behavior of $\beta$ at infinity ($\beta(g) \sim g^\alpha$ with $\alpha \leq 1$) gives the condition for absence of renormalon singularities.

This criterion can be constructively used as follows. According to 't Hooft's picture (see Fig. 7a), instanton and (possible) renormalon singularities lie, respectively, on the negative and positive half-axes in both $\varphi^4$ theory and QED. Let us assume that renormalon singularities do not exist; then (a) the regularity condition for $\beta(g)$ at finite $g$ (see Figs. 7c and 7d) holds, (b) the asymptotic behavior of the expansion coefficients $\beta_N$ is determined by the nearest instanton singularity and can be found by Lipatov's method, and (c) the behavior of $\beta(g)$ at infinity can be uniquely determined by summing the corresponding perturbation series expansion since the Borel integral is well defined. If $\beta(g)$ grows faster than $g^\alpha$ with $\alpha > 1$, then the initial assumption is incorrect, and the existence of renormalon singularities is proved by contradiction. If $\beta(g) \sim g^\alpha$ with $\alpha \leq 1$, then the assumption on the absense of renormalon singularities is self-consistent. These results are extended to QCD by changing the signs of $g$ and $z$.

The program of determination of Gell-Mann–Low functions outlined here was implemented in [42, 47, 64, 65] (see discussion in Section 8). The exponent $\alpha$ is close to unity in both $\varphi^4$ theory and QED and essentially smaller than unity in QCD. Therefore, renormalon singularities can be self-consistently eliminated (up to uncertainty of results). Moreover, it can be argued that $\alpha = 1$ in both $\varphi^4$ theory and QED. Anyway, since $\beta(g)$ is nonalternating in both theories, the condition for the absence of renormalon singularities in them is identical to the condition for their internal consistency.

---

[11] Exponential and logarithmic behavior can be included in these considerations and correspond to \alpha=\infty and \alpha=0.

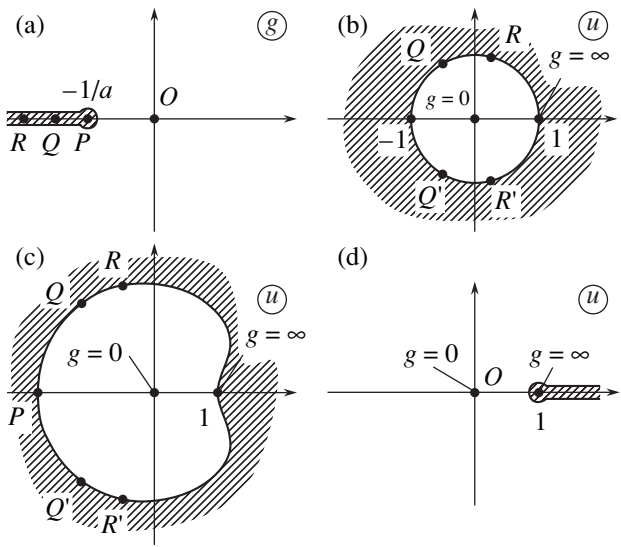

**Fig. 8.** (a) Analiticity domain for Borel transform is the complex plane with the cut $(-\infty, -1/a)$; (b) it can be conformally mapped to a unit circle. If analytic extension is restricted to the positive half-axis, then (c) conformal mapping can be performed to any domain such that $u = 1$ is its boundary point nearest to the origin; (d) an exreme case of such domain is the plane with the cut $(1, \infty)$.

The only field theory in which the existence of renormalons is considered as firmly established is the $O(n)$-symmetric sigma model in the limit of $n \longrightarrow \infty$ [53]. In this theory, the single-loop $\beta$ function is exact, and $\beta(g) \propto g^2$ at any $g$. Since $\alpha = 2$, renormalons cannot be eliminated self-consistently. However, this theory is self-contradictory in the four-dimensional case. It should be noted that truncation of the series for \beta-function at any finite number of terms immediately creates renormalon singularities. This shows that the renormalon problem cannot be solved in the framework of loop expansion [56, 57].

Note that possibilty of the existence of renormalon singularities makes the functional integrals ill-defined. The classical definition of the functional integral via the perturbation theory is unsatisfactory because of the divergence of the expansion in terms of the coupling constant. Its constructive summation requires knowledge of analyticity properties in the Borel plane (see Section 6), which are uncertain untill it is established whether the renormalon singularities exist. The definition of functional integral as a multidimensional integral on a lattice also evokes doubts: a lattice theory can be qualitatively different from the continuum theory, because renormalon contributions correspond to arbitrarily large momenta. This leads to a deadlock: an analysis of functional integrals is required to solve the renormalon problem, but the integrals remain ill defined until the renormalon problem is solved. The proposed scheme of self-consistent elimination of renormalon singularities appears to be the only remedy.

In this scheme, a continuum theory, by definition, is understood as the limit of lattice theories .[12]

## 6. PRACTICAL SUMMATION OF PERTURBATION SERIES

In this section, the practical summation of the following power series is discussed:

$$W(g) = \sum_{N=0}^{\infty} W_N (-g)^N, \qquad (6.1)$$

where the expansion coefficients have the asymptotic form $ca^N \Gamma(N + b)$ and their values are given numerically. The present analysis is restricted to alternating series. Accordingly, $(-1)^N$ is factored out, and $a = -1/S_0 > 0$, as in $\varphi^4$ theory.

### 6.1. Conformal–Borel Technique and Other Methods

Treating (6.1) as Borel's sum (see Section 3), consider the following modification of Borel transformation (5.9):

$$W(g) = \int_0^{\infty} dx\, e^{-x} x^{b_0 - 1} B(gx),$$

$$B(z) = \sum_{N=0}^{\infty} B_N(-z)^N, \quad B_N = \frac{W_N}{\Gamma(N + b_0)}, \qquad (6.2)$$

where an arbitrary parameter $b_0$ can be used to optimize the summation procedure [35]. Borel transform B(z) is assumed to have analyticity properties characteristic of $\varphi^4$ theory (see Section 5.2), i.e., it is analytic in the complex $z$ plane with a cut extending from $-1/a$ to $-\infty$ (see Fig. 8a). The series expansion of $B(z)$ is convergent in the circle $|z| < 1/a$. To calculate the integral in (6.2), B(z) should be analytically continued. When the expansion coefficients $W_N$ are given numerically, such continuation presents some problem. Its elegant solution proposed in [35] makes use of the conformal mapping $z = f(u)$ of the plane with a cut to the unit circle $|u| < 1$ (see Fig. 8b). Reexpanding $B(z)$ in powers of $u$,

$$B(z) = \sum_{N=0}^{\infty} B_N(-z)^N \bigg|_{z=f(u)} \quad (6.3)$$
$$\longrightarrow B(u) = \sum_{N=0}^{\infty} U_N u^N,$$

gives a series, convergent at any $z$. Indeed, the singular points of $B(z)$ ($P, Q, R, \ldots$) lie on the cut, while their images ($P, Q, Q', R, R', \ldots$) lie on the boundary of the circle $|u| = 1$. Thus, the latter series in (6.3) is convergent at $|u| < 1$. However, there is one-to-one correspondence between the interior $|u| < 1$ of the circle and the domain of analyticity in the $z$ plane. The conformal mapping is defined as

$$z = \frac{4}{a}\frac{u}{(1-u)^2} \quad \text{or} \quad u = \frac{(1+az)^{1/2}-1}{(1+az)^{1/2}+1}. \quad (6.4)$$

---

[12]This philosophy is hidden in the very concept of renormalizability. In fact, an effective theory is constructed for small momenta with a cutoff parameter $\Lambda$, and the same scale is supposed to be the upper boundary of the essential domain of integration. Renormalizability is understood as a possibility of taking the limit $\Lambda \longrightarrow \infty$ without destroying the structure of the theory. However, the actual contribution of momenta larger than $\Lambda$ can hardly be controlled within the scope of the effective theory. One can always imagine some "demon" that lurks in the large-momentum region, providing its essential contribution and running away with increasing $\Lambda$. The physical realization of such a demon was found in the theory of the Anderson transition: the role of a demon is played by contribution of the minimum of action associated with a lattice instanton [8, 36, 37].

Hence,

$$U_0 = B_0,$$
$$U_N = \sum_{K=1}^{N} B_K \left(-\frac{4}{a}\right)^K C_{N+K-1}^{N-K} \quad (N \geq 1), \quad (6.5)$$

where $C_N^K$ denotes binomial coefficients.

Since the $B(z)$ exhibits power-like behavior at infinity, the Borel integral in (6.2) is rapidly convergent, and its upper limit can be assigned a finite value in accordance with the required accuracy. Then, $u$ is bounded from above by $u_{max} < 1$, and the latter series in (6.3) is convergent. The substitution of $u = u(z)$ into the Borel integral (6.2) followed by integration, as done in [35], is somewhat risky, because permutation of summation and integration may lead to divergence of the algorithm. In fact, the scheme used in [35] is convergent, because the actual coefficients $U_N$ exhibit power-like asymptotic behavior (see [65, Section 2.1]).[13] As $b_0$ increases, oscillatory asymptotic behavior changes to monotonic [65], and this change was used in [35] as a basis for error estimation.

When the first $N_m$ coefficients in series (6.1) are known, formula (6.5) can be used to find the first $N_m$ coefficients of the convergent series in (6.3). If $g \sim 1$, then the dominant contribution corresponds to values of $u$ of the order a few tenths, which makes it possible to obtain accurate results even for small $N_m$. The summation error is estimated as

$$\delta W(g) \propto \exp\{-3(N_m^2/ag)^{1/3}\}, \quad ag \lesssim N_m^2. \quad (6.6)$$

Therefore, the highest value of $g$ for which satisfactory results can be obtained is of the order $N_m^2$. The described scheme was used [35] to calculate critical exponents in theory of phase transitions up to the third decimal place.

An alternative method of analytic continuation [66] consists in construction of the Padé approximants $[M/L]$, defined as the ratio $P_M(z)/Q_L(z)$ of polynomials of degrees $M$ and $L$, with the coefficients adjusted to reproduce for $B(z)$ its expansion (6.3) in the several lowest orders. Diagonal and quasi-diagonal approximants (with $M = L$ and $M \approx L$, respectively) are known to converge to the approximated function as $M \longrightarrow \infty$ for a broad class of functions, but the convergence rate is relatively low in the general case. When both $M$ and $L$ are finite, the approximate Borel transform generally exhibits incorrect behavior at infinity dictated by the particular Padé approximant $[M/L]$ employed. The corresponding incorrect behavior of $W(g)$ at $g \longrightarrow \infty$ gives rise to a certain error at $g \sim 1$ by continuity. Accordingly, different results are obtained by using different approximants, depending on the subjective choice of the user. The uniform convergence with respect to $g$ can be achieved by matching the behavior of Padé approximant at infinity with the asymptotic form of $B(z)$, if the asymptotic behavior of $W(g)$ in the strong-coupling limit is known. When the number of terms in the expansion is sufficiently large, the asymptotic behavior at strong coupling can be "probed" by analyzing the convergence rate, as done in [66]. Furthermore, information on the nature of singularity at $-1/a$ can be taken into account, and the approximation may be restricted by requiring that all poles of Padé approximants lie on the negative half-axis; then the information used in this method is the same as in the conformal–Borel technique described above. The results of the original calculations of critical exponents performed by this method in [66] were virtually identical with those obtained in [35]. The Padé–Borel technique is preferably to be used when the analyticity properties of $B(z)$ are not known: then information about the locations of the nearest singularities can be gained by constructing Padé approximants.

In multiple-charge models, Chisholm approximants (rational functions in many variables) can be used instead of Padé approximants [67]. A more efficient approach to problems of this type is based on the so-called resolvent expansion [68]: all charges are multiplied by an auxiliary parameter $\lambda$, Padé approximants in terms of this parameter are employed, and $\lambda$ is set to unity at the end of the calculations. In this method, the symmetry of the model is completely preserved, and projection onto any charge subspace of lower dimen-

---

[13]To determine the asymptotic form of $U_N$, the contributions found in [65] should be summed over all singular points, whose number is infinite. Since the sum is finite for any constant $N = N_0$, it can readily be shown that it is dominated by the term containing the highest power of $N$ as $N \longrightarrow \infty$.

1210sion does not lead to loss of information [69, 70]. A more complicated sequence of approximants can be constructed by using Winn's ε-algorithm [71] based on a "strong" Borel transform (see Section 7). The Sommerfeld–Watson summation scheme [13] makes use of the analyticity properties of the coefficient functions, instead of Borel transforms. A generalized conformal–Borel technique was employed in [12]. The last two methods make it possible to "guess" the strong-coupling asymptotics, analyzed systematically in the next section.

In another approach, variational perturbation theory [72, 73] is used to formulate a scheme of interpolation between the weak- and strong-coupling regions when some information about the latter is available. With regard to critical exponents, this information concerns behavior near the renormalization-group fixed point and can be expressed in terms of strong coupling in result of expansion in the bare charge. Thus, a divergent perturbation series is transformed into a convergent sequence of approximations, and an accuracy comparable to that of conformal–Borel technique is achieved [74]. Unfortunately, neither divergence of the series nor the Lipatov asymptotics is used in this approach explicitly. Information about the latter can be used only implicitly by interpolating the coefficient function. Since an attempt of this kind made in [75] did not result in any improvement in accuracy, Kleinert claimed that information about high-order terms is insignificant. It is obvious that such assertion is incorrect in the general case: since exact knowledge of the expansion coefficients is equivalent to exact knowledge of the function, appropriate use of any additional information should improve accuracy. The particular result obtained by Kleinert is simply related to the fact that variational interpolation is not less accurate than interpolation of the coefficient function.

The application of divergent series to calculation of critical exponents is based on the use of the Callan–Symanzik renormalization-group equation, which contains both Gell-Mann–Low function $\beta(g)$ and renormalization-group functions $\eta(g)$ and $\eta_4(g)$ (anomalous dimensions) [36, 76, 77]. If $g^*$ is a nontrivial zero of $\beta(g)$, then the critical exponents $\eta$ and $\nu$ can be expressed in terms of the anomalous dimensions at that point by the relations $\eta(g^*) = \eta$ and $\eta_4(g^*) = 1/\nu - 2 + \eta$, and the remaining exponents are determined by well-known relations [39]. The renormalization-group functions are calculated as series expansions in terms of $g$, which can be summed by methods mentioned above.

Based on this approach, substantial progress has been made in analysis of critical behavior of a various systems. The critical exponents for the $O(n)$-symmetric $\varphi^4$ theories with $n = 0, 1, 2,$ and 3 were originally calculated in [35, 66] for two- and three-dimensional spaces by using six- and four-loop expansions, respectively. Subsequently, these calculations were extended to larger $n$ and higher order coupling constants [78, 79]. In particular, this provided a basis for estimating the computational scope of the $1/n$ expansion. The seven-loop contributions to the renormalization-group functions for $d = 3$ in [75, 80] and five-loop ones for $d = 2$ in [81, 82] were found and used to refine the critical exponents. The latter studies revealed systematic deviation of the resummation results from the known exact values of critical exponents. However, their interpretation as a manifestation of nonanalytic contributions to renormalization-group functions (see [81, 82]) does not seem to be well grounded.

When cubic anisotropy is taken into account, a two-charge version of the $n$-component $\varphi^4$ theory is obtained, since the corresponding action contains two fourth-order invariants. The cases of $n = 3$ and $n = 0$ correspond to a cubic crystals and weakly disordered Ising ferromagnets, respectively. This can be shown by using the standard replica trick to average over the random impurity field [83]. The expansion coefficients and sums of renormalization-group series were calculated for these systems in the four-, five-, and six-loop approximations in [69, 84], [85], and [86, 87], respectively (see also [88, 89]).

Six-loop expansions have also been obtained for other two-charge field systems: the *mn* model describing certain magnetic and structural phase transitions (including the critical behavior of $n$-component disordered magnets in the case of $m = 0$ [87]) and the $O(m) \times O(n)$-symmetric model corresponding to the so-called chiral phase transitions [90]. The summation of the resulting series performed in [90–93] made it possible to elucidate the structure of the phase portraits of the renormalization-group equations and analyze the stability of nontrivial fixed points. Even more complicated (three-charge) versions of $\varphi^4$ theory arise in models of superconductors with nontrivial pairing, many-sublattice antiferromagnets, structural phase transitions, superfluid transition in neutron liquid, etc. Some of these have been analyzed in three-, four-, and six-loop approximations in [70, 94–96]. In recent studies, five-loop expansions were found and resummed for the two-dimensional chiral [97], cubic, and *mn* models [82]. Finally, summation of three- and four-loop expansions was used to analyze critical dynamics in pure and disordered Ising models [98, 99], as well as effects due to long-range interactions [100] and violation of replica symmetry [101].

Note that, instead of summing a series expansion in terms of the coupling constant in the space of physical dimension, one can sum up the divergent ε-expansions obtained for the formal problem of phase transition in the $(4 - \epsilon)$-dimensional space [77, 102, 103]. The four-loop expansions for the $O(n)$-symmetric theory [104] summed up in [105] were extended to the five-loop level [106] and summed in [80]. The five-loop expansions obtained in [107] for the cubic model were used in [108] as a basis for deriving five-loop $\sqrt{\epsilon}$-expansions of critical exponents for disordered Ising model, and their summation was discussed in [108–110].

Detailed discussion of the current status of theory of critical phenomena based on multiple-loop renormalization-group expansions, as well as extensive bibliography, can be found in recent reviews [89, 96, 111].

### 6.2. Summation in the Strong-Coupling Limit

The results obtained for increasingly stronger coupling are characterized by stronger dependence on the particular implementation of the summation procedure. To analyse the unsuing uncertainties, it is desirable to find the direct relation of asymptotic behavior of $W(g)$ for the strong coupling with the values of $W_N$. This problem is solved here by assuming that the asymptotic behavior can be represented by the power law

$$W(g) = W_\infty g^\alpha, \quad g \longrightarrow \infty, \quad (6.7)$$

which adequately represents all models that are amenable to analysis and close to realistic field-theoretical problems. This problem can be solved in the framework of the standard conformal–Borel technique (Sec.6.1) [65]. However, a more efficient algorithm can be developed by using a modified conformal mapping.

If $z = 0$ and $\infty$ are mapped to $u = 0$ and 1, respectively, then (6.4) is the only conformal mapping that can be used to find the analytic continuation of the Borel transform to arbitrary complex $z$. However, this is not necessary: the integral in (6.2) can be calculated if the analytic continuation to the positive half-axis is found. Therefore, it is possible to use a conformal mapping to any region for which $u = 1$ is the boundary point nearest to the origin (see Fig. 8c): under this condition, the latter series in (6.3) will be convergent if $|u| < 1$, and, in particular, on the interval $0 < u < 1$, which is the image of the positive half-axis.

One advantage of this conformal mapping is that the divergence of the re-expanded series in (6.3) is controlled by the nearest singular point $u = 1$, which is related to the singularity of $W(g)$ at $g \longrightarrow \infty$, so that the asymptotic form of $U_N$ is relared with the parameters of asymptotic formula (6.7). If $U_N$ is expressed in terms of $B(u)$ as

$$U_N = \oint_C \frac{du}{2\pi i} \frac{B(u)}{u^{N+1}}, \quad (6.8)$$

and the contour $C$ enclosing point $u = 0$ is deformed so that it goes around the cuts extending from the singular points to infinity, then it can easily be shown that the asymptotic form of $U_N$ for large $N$ is controlled by the nearest singular point $u = 1$. To reduce the contributions of the remaining singular points $P$, $Q$, $Q'$, ..., these points should be moved away as far as possible. As a result, we come to the extremal form of such transform, which is the mapping to the plane with the cut $(1, \infty)$ (see Fig. 8d),

$$z = \frac{u}{a(1-u)}, \quad (6.9)$$

for which

$$U_0 = B_0,$$
$$U_N = \sum_{K=1}^{N} \frac{B_K}{a^K}(-1)^K C_{N-1}^{K-1} \quad (N \geq 1). \quad (6.10)$$

The asymptotic form of $U_N$ for large $N$,

$$U_N = U_\infty N^{\alpha-1}, \quad N \longrightarrow \infty, \quad (6.11)$$

$$U_\infty = \frac{W_\infty}{a^\alpha \Gamma(\alpha)\Gamma(b_0+\alpha)}, \quad (6.12)$$

is determined by the parameters of asymptotic formula (6.7). Thus, a simple algorithm is proposed: use (6.2) to calculate the coefficients $B_N$ corresponding to the given $W_N$; substitute the results into (6.10) to find $U_N$; and find power-law fit (6.11) to determine the parameters $W_\infty$ and $\alpha$ in (6.7).

In practical applications of the algorithm, one has to deal with problems due to the growth of random errors The random error in $U_N$, corresponding to a relative computational or round-off error $\delta$ in $W_N$, is a rapidly increasing function of $N$:

$$\delta U_N \sim \delta \cdot 2^N. \quad (6.13)$$

In double-precision computations, when $\delta \sim 10^{-14}$, the value of $\delta U_N$ is comparable to unity if $N \approx 45$, and the corresponding error in the recovered asymptotic formula (6.7) is ~1%.

Fortunately, influence of smooth errors is essentially different and the algorithm is "superstable" in the sense that the output error is even smaller than the input error. Linear transformation (6.10) has a remarkable property:

$$\sum_{K=1}^{N} K^m (-1)^K C_{N-1}^{K-1} = 0 \quad (6.14)$$

for $m = 0, 1, ..., N-2$. Accordingly, the addition of an arbitrary polynomial $P_m(K)$ to $B_K/a^K$ (which behaves as a power of $K$) does not change the asymptotic form of $U_N$. An analogous property is valid for a broad class of smooth functions, which are accurately approximated by polynomials: in particular, when $B_K/a^K$ is replaced with $B_K/a^K + f(K)$, where $f(K)$ is an entire function with rapidly decreasing Taylor series expansion coefficients, the resulting change in $U_N$ is rapidly decreasing with increasing $N$. In practical problems, sev-eral low-order terms and asymptotic behavior of $W_N$ are known, while intermediate coefficients are found by interpolation. Since interpolation leads to smooth errors, they can be expected to play a minor role even if their magnitudes are large. Thus, the proposed algorithm is best suited to the most relevant formulation of the problem.

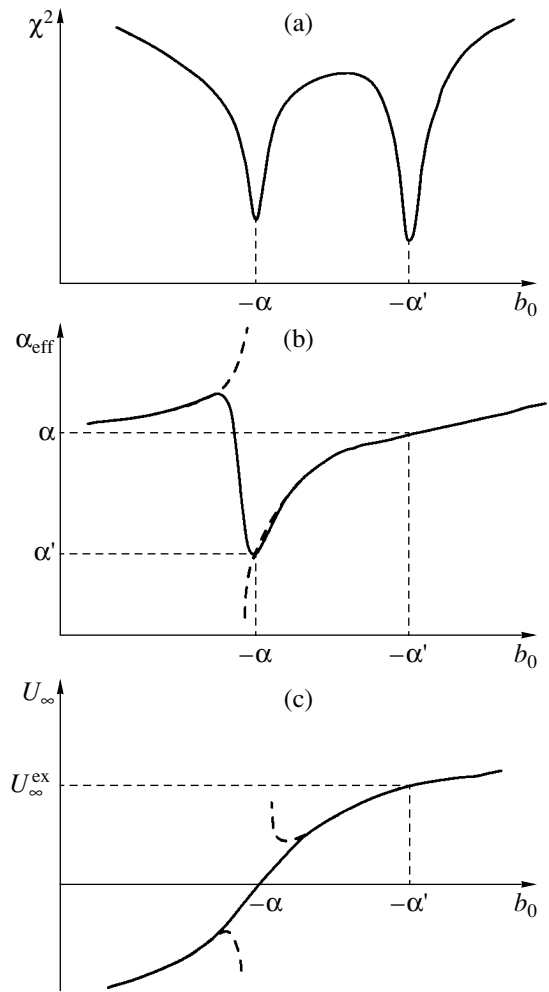

**Fig. 9.** Theoretical curves of $\chi^2$, $\alpha_{\text{eff}}$, and $U_\infty$ plotted versus $b_0$ with neglecting the corrections subsumed under the ellipsis in (6.16). Results obtained from log–log plots are frequently discontinuous (dashed curves) [65].

The approximation of $U_N$ by a power law can be performed by applying a standard $\chi^2$ minimization procedure [112], and the fitting interval $N_{\min} \leq N \leq N_{\max}$ can be chosen as follows. The upper limit $N_{\max}$ can be taken from the condition $\delta U_N \sim U_N$, since no additional information is contained in the coefficients with higher $N$. This condition is not restrictive, because points with large error are automatically discarded in $\chi^2$ minimization. The lower limit $N_{\min}$ is set by requiring that $\chi^2$ have "normal" values; it rules out the systematic error due to deviation of $U_N$ from asymptotic law (6.11).

The existence of an upper bound for $N$ entails strong dependence of the results on $b_0$ (see (6.2)), because this parameter determines the rate at which the asymptotic behavior is approached. To analyze this effect, suppose that asymptotic formula (6.7) is modified by adding power-law corrections:

$$W(g) = W_\infty g^\alpha + W'_\infty g^{\alpha'} + W''_\infty g^{\alpha''} + \ldots . \quad (6.15)$$

By analogy with (6.11) and (6.12), it follows that

$$U_N = \frac{W_\infty}{a^\alpha \Gamma(\alpha)\Gamma(b_0 + \alpha)} N^{\alpha - 1} \\ + \frac{W'_\infty}{a^{\alpha'} \Gamma(\alpha')\Gamma(b_0 + \alpha')} N^{\alpha' - 1} + \ldots . \quad (6.16)$$

When the corrections subsumed under the ellipsis in (6.16) are neglected, the formal approximation of (6.16) by power law (6.11) leads to satisfactory results, because the log–log plot of (6.16) is almost linear; however, the values of $\alpha$ and $U_\infty$ thus obtained should be interpreted as "effective" parameters.

Since the first and second terms in (6.16) vanish at the poles of the respective gamma functions, $U_N \propto N^{\alpha' - 1}$ and $U_N \propto N^{\alpha - 1}$ are obtained for $b_0 = -\alpha$ and $b_0 = -\alpha'$, respectively. The power-law fits are particularly accurate for these b_0 and the values of $\chi^2$ are low. The results expected by varying $b_0$ are illustrated by Fig. 9. The graph of $\chi^2$ has two sharp minima at $b_0 = -\alpha$ and $b_0 = -\alpha'$. The curve of $\alpha_{\text{eff}}$ drops to $\alpha'$ in the neighborhood of $b_0 = -\alpha$ and approaches $\alpha$ outside this neighborhood. At $b_0 = -\alpha'$, the exact equality $\alpha_{\text{eff}} = \alpha$ is reached and effective parameter $U_\infty$ exactly corresponds to $W_\infty$. In the neighborhood of $b_0 = -\alpha$, $U_\infty$ vanishes, while its linear slope

$$U_\infty \approx \frac{W_\infty}{a^\alpha \Gamma(\alpha)} (b_0 + \alpha) \quad (6.17)$$

yields an estimate for $W_\infty$ weakly sensitive to errors in $\alpha$. The effect of the terms discarded in (6.16) only slightly changes this pattern.

The analysis above suggests that independent estimates for the exponent $\alpha$ can be obtained by using: (1) the value of $\alpha_{\text{eff}}$ at the right-hand minimum of $\chi^2$, (2) the location of the left-hand minimum of $\chi^2$, (3,4) position of the zero for the function $U_\infty(b\_0)$, which can be found from log-log plot (3), or by power-law fitting with a fixed $\alpha$ taken from previous estimates (4).

Similarly, independent estimates for $W_\infty$ are obtained by using: (1) the value of $U_\infty$ at the right-hand minimum of $\chi^2$, and (2,3) the linear slope of $U_\infty(b_0)$ near its zero. In the latter estimate, the fixed value of $\alpha$ is used, which is varied within its uncertainty (followed from the previous estimares) to obtain upper and lower bounds for $W_\infty$.

The accuracy of the results can be estimated by using the fact that the discrepancies between different estimates for α and $W_\infty$ are comparable to their respective deviations from the exact value. Existence of several estimates makes the results more reliable: f.e. in the case of two estimates they may become close by accidental reasons and provide underestimation of the error, while accidental proximity of three or four independent estimates looks improbable.

As an example, consider the integral

$$W(g) = \int_0^\infty d\varphi \exp(-\varphi^2 - g\varphi^4), \qquad (6.18)$$

which can be interpreted as the zero-dimensional analog of the functional integral in $\varphi^4$ theory. Its asymptotic behavior is described by (6.7) with $\alpha = -1/4$ and $W_\infty = \Gamma(1/4)/4$, and the corrections can be represented as a power series in $g^{-1/2}$. The results obtained when $W_N$ are given with double-precision accuracy ($\delta \sim 10^{-14}$) are illusrated by Fig. 10.[14] The ensuing estimates,

$$\alpha = -0.235 \pm 0.025, \quad W_\infty = 0.908 \pm 0.025,$$
$$\alpha' = -0.75 \pm 0.08, \qquad (6.19)$$

are in good agreement with the respective exact values

$$\alpha = -0.25, \quad W_\infty = 0.9064\ldots,$$
$$\alpha' = -0.75. \qquad (6.20)$$

Stability of the algorithm with respect to interpolation can be checked by writing the expansion coefficients as[15]

$$W_N = ca^N \Gamma(N+b)$$
$$\times \left\{ 1 + \frac{A_1}{N} + \frac{A_2}{N^2} + \ldots + \frac{A_K}{N^K} + \ldots \right\}, \qquad (6.21)$$

since relative corrections to the Lipatov asymptotics have the form of a regular series expansion in terms of $1/N$. This representation can readily be used to interpolate the coefficient function: the series can be truncated, and the parameters $A_K$ can be found from correspondence with several low-order coefficients $W_N$. When interpolation is performed by using parameters of the Lipatov asymptotics, the lowest order correction $A_1/N$, and the single coefficient $W_1$ [65], the resulting values $\alpha = -0.245 \pm 0.027$ and $W_\infty = 0.899 \pm 0.014$ are almost equal to those in (6.20). The errors in these results are still determined by round-off error, even though the interpolation errors are greater by ten orders of magnitude.

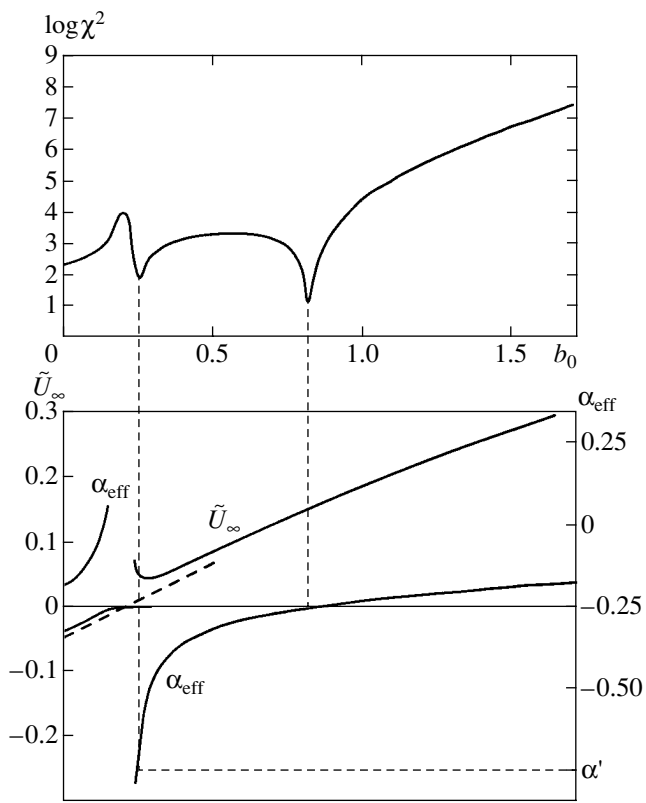

**Fig. 10.** Curves of $\chi^2$, $\alpha_\mathrm{eff}$, and $\tilde{U}_\infty$ calculated as functions of $b_0$ for integral (6.18) by using the interval $24 \le N \le 50$. Dashed curve represents $\tilde{U}_\infty(b_0)$ in the neighborhood of its zero obtained with fixed $\alpha = -0.25$.

---

[14] For technical reasons, the coefficients $\tilde{U}_N = U_N \Gamma(b_0 + N_0)$ are given below, which are normalized to have a constant limit at $b_0 \longrightarrow \infty$. Here, $N_0$ is the lower limit of summation in (6.5), which may differ from unity if the first terms in (6.1) vanish. Similarly, $\tilde{U}_\infty = U_\infty \Gamma(b_0 + N_0)$.

[15] Frequently arising questions concerning the analyticity of the coefficient function and its interpolation were discussed in [113] in the context of comments to [114].

Another example is the calculation of the ground-state energy $E_0(g)$ for anharmonic oscillator (1.1), which can be reduced to one-dimensional $\varphi^4$ theory. The parameters of the power law asymptotics (6.7) obtained by using the coefficients $W_N$ calculated in [6] up to $\delta \sim 10^{-12}$ [65],

$$\alpha = 0.317 \pm 0.032, \quad W_\infty = 0.74 \pm 0.14, \qquad (6.22)$$

agree with the exact values $\alpha = 0.3333\ldots$ and $W_\infty = 0.6679\ldots$ and demonstrate adequate estimation of errors.

The reliability of error estimation suggests a new approach to optimization of summation algorithms. On a conceptual level, optimization is performed by introducing a variation (characterized by a parameter λ) into the summation procedure and then fixing this parameter from the condition of the best convergence of the algorithm.

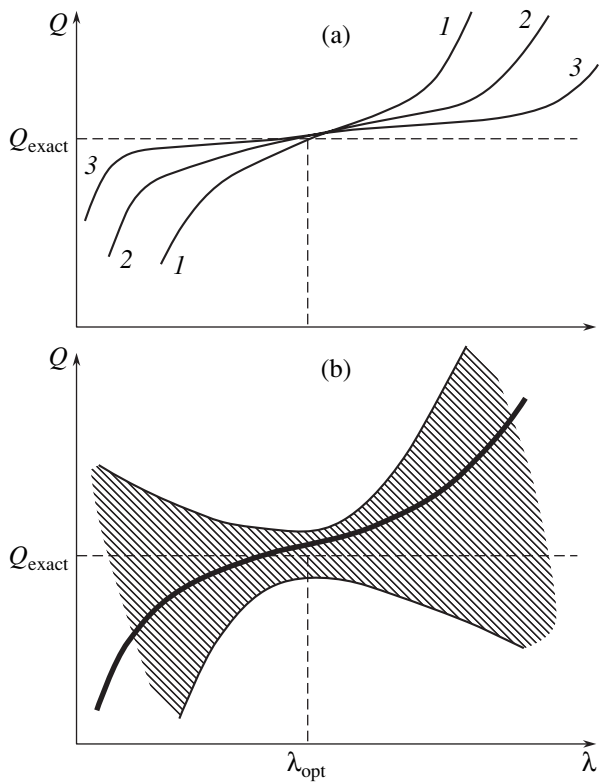

**Fig. 11.** (a) In theory, any quantity $Q$ obtained by summation of a series is independent of the optimization parameter $\lambda$. In practice, the dependence exists and changes from curve *1* to curves *2* and *3* with increasing amount of available information. The optimal value of $\lambda$ corresponds to the central plateau region. (b) The error of approximate calculation of $Q$ (hatched region) depends on $\lambda$. When the error is estimated correctly, the exact value $Q_{\text{exact}}$ is consistent with all data. In the "ideal" case illustrated here, the optimal value of $\lambda$ corresponds to the minimal error.

For example, series (6.1) can be raised to the power $\lambda$, and the summation algorithm can then be applied to the re-expanded series

$$W(g)^\lambda = \tilde{W}_0 - \tilde{W}_1 g + \tilde{W}_2 g^2 - \ldots \\ + \tilde{c} a^N \Gamma(N+b)(-g)^N + \ldots, \quad (6.23)$$

which has analogous properties (except for a different parameter $c$ in the Lipatov asymptotic form [8]). Theoretically, series (6.23) is equivalent to (6.1), and the value of any quantity $Q$ obtained as a result of summation should be independent of $\lambda$. When the available information about series (6.1) is incomplete, such dependence arises, but becomes weaker as the amount of information increases. In the general case, convergence is not uniform with respect to $\lambda$, and the approximate value of $Q$ is close to the exact one only within some plateau-like region (see Fig. 11a), while the error rapidly increases away from it. The plateau widens and flattens with increase in available information (e.g., see [115]). It is clear that the best convergence is achieved at the center of the plateau. However, the location of a "center" may not be easy to determine, since the plateau may be asymmetric or indistinct, its center may move in the course of convergence, etc. Therefore, the choice of $\lambda$ and the optimal value of $Q$, is a largely subjective one.

However, there exists an objective approach to optimization. Note that the $\lambda$ determines not only the approximate value of $Q$, but also the error of its calculation. If the error is estimated correctly, then the exact value $Q_{\text{exact}}$ should be consistent with the approximate results corresponding to any $\lambda$ (see Fig. 11b); i.e., spurious dependence of $Q$ on $\lambda$ is ruled out. If this "ideal" situation is attained, then optimization with respect to $\lambda$ reduces to minimization of error.

It is reasonable to perform optimization at the interpolation stage, because any uncertainty of results is ultimately due to imprecise knowledge of the coefficients $W_N$. If (6.21) is rewritten as

$$W_N = c a^N N^{\tilde{b}} \Gamma(N + b - \tilde{b}) \left\{ 1 + \frac{\tilde{A}_1}{N - \tilde{N}} \right. \\ \left. + \frac{\tilde{A}_2}{(N-\tilde{N})^2} + \ldots + \frac{\tilde{A}_K}{(N-\tilde{N})^K} + \ldots \right\} \quad (6.24)$$

and the interpolation is performed by truncating the series and determining the coefficients $\tilde{A}_K$, then the interpolation procedure can be parameterized by $\tilde{b}$ and $\tilde{N}$. Optimization with respect to $\tilde{b}$ can be made theoretically [65], and the optimal $\tilde{b} = b - 1/2$ corresponds to the Lipatov asymptotic form parameterized as $c a^N N^{b-1/2} \Gamma(N + 1/2)$. Optimization with respect to $\tilde{N}$ was demonstrated in [65] on the example of anharmonic oscillator, where interpolation was performed with the use of the first nine coefficients of the series. Coarse optimization of $\chi^2$ as a function of $\tilde{N}$ was performed for several constant values of $b_0$ having minima at $\tilde{N}$ between $-5.5$ and $-5.0$. This narrow interval determines the range of interpolations consistent with the power-law asymptotic behavior of $W(g)$. Next, a systematic procedure was executed to find $\alpha$ and $W_\infty$. The "ideal" situation illustrated by Fig. 11b was obtained by widening the error corridor for $\alpha$ by a factor of 1.3 and for $W_\infty$ by a factor of 1.1, which is admissible since the error is estimated up to order of magnitude. If the values of $\alpha$ and $W_\infty$ are chosen to be consistent with all data, and the one-sided error is minimized, the results are

$$\alpha = 0.38 \pm 0.05, \quad W_\infty = 0.52 \pm 0.12, \quad (6.25)$$

and their deviation from the exact results is adequately estimated by the respective errors.

If the available information concerning $W_N$ is sufficient to recover the asymptotic behavior of $W(g)$, then summing of series

(6.1) at arbitrary g presents no problem: calculating the lowest order coefficients $U_N$ using (6.10), we can continue them according to the asymptotics $U_\infty N^{\alpha-1}$; as a result, all coefficients in the convergent series (6.3) are found. The summation error is determined by the accuracy of determination of the asymptotic form of $U_N$, which is characterized by a quantity $\Delta$ assumed to be constant within a bounded interval (in fact, the actual error behaves logarithmically in $N$). If $N_c$ is the characteristic scale for which the relative error is of the order $\Delta$, then we can accept

$$\frac{\delta U_N}{U_N} = \begin{cases} 0, & N < N_c, \\ \Delta, & N \geq N_c, \end{cases} \quad (6.26)$$

and the summation error is

$$\frac{\delta W(g)}{W(g)} \sim \begin{cases} \Delta, & ag \gtrsim N_c, \\ \Delta \exp\{-2(N_c/ag)^{1/2}\}, & ag \lesssim N_c. \end{cases} \quad (6.27)$$

The error of a straightforward summation using $N_m$ known coefficients is given by (6.27) with $N_c = N_m$ and $\Delta \sim 1$ and is higher than estimated by (6.6). Nevertheless, the stability of the algorithm with respect to interpolation ensures that $\Delta \ll 1$ and $N_c \gg N_m$ even for small $N_m$ [65].

## 7. "NON-BOREL-SUMMABLE" SERIES

It is clear from Section 3 that a definition of the sum of a factorially divergent series cannot be essentially different from Borel's definition, i.e. it should coincide with Borel's definition or be equivalent to it. Otherwise, self-consistent manipulation of divergent series is impossible. Nevertheless, "non-Borel-summable" series are frequently discussed in the literature. This misleading concept is used in two situations.

In one of these situations, the coefficients of the series in question increase much faster than $N!$, and the standard transform defined by (3.11) is not effective. However, one can use the "strong" Borel transformation

$$F(g) = \sum_{N=0}^{\infty} F_N g^N$$
$$= \sum_{N=0}^{\infty} \frac{F_N}{\Gamma(kN+1)} \int_0^\infty dx\, x^{kN} e^{-x} g^N \quad (7.1)$$
$$= \int_0^\infty dx\, e^{-x} \sum_{N=0}^{\infty} \frac{F_N}{\Gamma(kN+1)} (gx^k)^N,$$

for summing series whose coefficients increase as $(N!)^k$ with arbitrary finite $k$.[16]

In the other situation, the factorial series in question have nonalternating coefficients. Analysis of the simple example

$$F(g) = \sum_{N=0}^{\infty} a^N N! g^N = \int_0^\infty dx \frac{e^{-x}}{1 - agx}, \quad (7.2)$$
$$ag > 0,$$

shows that the corresponding Borel image $B(z)$ has the singular points on the positive half-axis, which lie on the integration path in (3.12). Therefore, the Borel integral is ill defined and should be correctly interpreted. In particular, the contour of integration in (7.2) may lie above or below the singular point $x = 1/ag$, or the integral can be understood in the sense of the principal value.

To find all the set of the possible interpretations, let rewrite definition of the gamma function in the form

$$\Gamma(z) = \sum_i \gamma_i \int_{C_i} dx\, e^{-x} x^{z-1}, \quad \sum_i \gamma_i = 1, \quad (7.3)$$

where $C_1, C_2, \ldots$ are arbitrary contours extending from the origin to infinity in the right half-plane. Then, the Borel transformation leads to

$$F(g) = \sum_i \gamma_i \int_{C_i} dx\, e^{-x} x^{b_0-1} B(gx), \quad (7.4)$$

where the contours $C_i$ are not mutually equivalent because of the singularities of the Borel transform $B(z)$ and cannot be aligned with the positive half-axis as can be done in (7.3). The choice of interpretation is determined by the parameters $\gamma_i$ if the set $\{C_i\}$ contains all nonequivalent contours.

Correct interpretation of the Borel integral is impossible without additional information on the mathematical object represented by a divergent series. For this reason, current views on the prospects of recovering physical quantities from the corresponding perturbative expansions are largely pessimistic (see [116, 117]). However, the importance of additional information should not be overestimated. In our opinion, interpretation can be based on the principle of analyticity with respect to model parameters, which is valid in practically all physical applications.

---

[16]This leads one to the following question: why do we not use "strong" transforms of this kind in every case whatsoever? As far as exact calculations are concerned, the only criterion is analytical tractability: any transform is applicable if the required calculations can be carried through. In approximate calculations, "strong" summation methods are not as desirable as they seem to be. Mathematically, the nontriviality of a function is determined by the type and location of its singularities. Strong Borel transform (7.1) defines an entire function $B(z)$ having a complicated singular point at infinity. This very fact entails practical difficulties: the singular point is hardly amenable to analysis, whereas its impact is not any weaker. This explains why the "weakest" Borel transform is preferable: its singularities lie at finite points in the complex plane, and even their location provides essential information about the function (see Section 6).

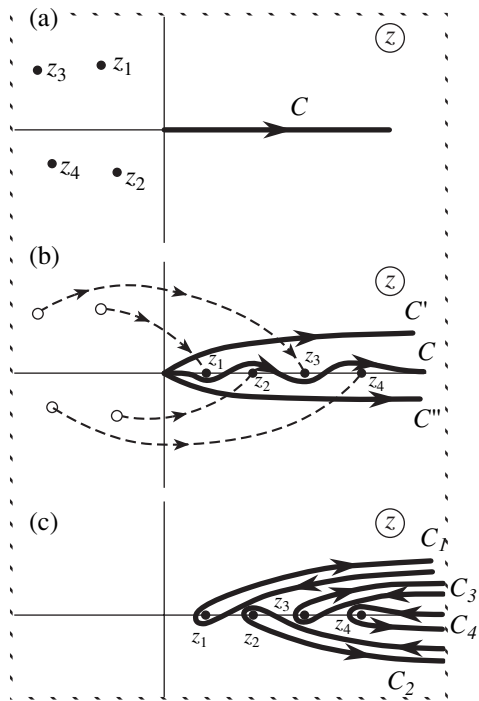

**Fig. 12.** Graphic illustration of non-Borel-summability.

If the values of the parameters are such that all singular points $z_i$ of the Borel transform lie in the left half-plane (see Fig. 12a), then Borel integral (3.12) for a factorial series with asymptotic expansion coefficients $ca^N\Gamma(N+b)$ is the only analytic function that satisfies the strong asymptotic condition within the sector $|\arg g| \leq \pi$, $|g| < g_0$ with an arbitrary $g_0$ (see [21, Section 8.1]), i.e., on the entire physical sheet of the Riemann surface. Therefore, the choice of a contour $C$ aligned with the positive half-axis (see Fig. 12a) is surely correct. Let the singular points of the Borel image move to the positive half-axis as the parameters are varied, and "non-Borel-summable" situation is reached. If the analyticity with respect to parameters of the model is to be preserved, then the moving singular points should not cross the contour $C$. Therefore, the contour should be deformed as shown in Fig. 12b. Accordingly, correct interpretation of the Borel integral requires that only one parameter $\gamma_i$ in (7.4) is not zero.

Interpretation in the sense of the principal value corresponds to the half-sum of integrals over contours $C'$ and $C''$. Its difference from the correct interpretation is determined by the half-sum of the integrals over contours $C_i$ going around the singular points $z_i$ (see Fig. 12c). The integral over a contour $C_i$ behaves as $\exp(-z_i/g)$. When interpretation in the sense of the principal value is used, nonperturbative contributions of such form should be explicitly added to the Borel integral. An expression of this kind was discussed in [118] with regard to the quantum mechanical problem of double-well potential.

Can one be sure that the same interpretation will be obtained by analytic continuation of the Borel integral with respect to different model parameters? This question is nontrivial, since it cannot be answered positively in the case of analytic continuation with respect to the coupling constant. Indeed, the action $S\{g,\varphi\}$ (in $\varphi^4$ and related theories) can be transformed into $S\{\phi\}/g$, and the change $g \longrightarrow g e^{i\psi}$ is equivalent to $S\{\phi\} \longrightarrow S\{\phi\}e^{-i\psi}$. Thus, the pattern of singularities of the Borel transform in the complex plane is rotated by an angle $\psi$. The singularities move from the positive half-axis into the left half-plane if $\psi > \pi/2$, which is impossible if the convergence of the functional integral is preserved.

In realistic field theories, the set of parameters is strongly restricted. Due to translational invariance and other symmetries the action can contain only the corresponding invariants, while the renormalizability condition requires that only low powers of fields and their gradients should be included in the action. Since the coefficients of the highest powers of fields are generally associated with the coupling constant, they are not amenable to analytic continuation. The coefficients of the terms that are quadratic in fields cannot be used either, because their variation may lead to vacuum instability and corresponding phase transitions. The remaining possibilities include the coefficients of the intermediate powers of fields (such as $\varphi^3$ in $\varphi^4$ theory) and the cross terms representing interactions between different fields. Analyticity with respect to these coefficients is preserved in any part of the complex plane by virtue of (a) convergence of the functional integrals defined on a finite-size lattice; (b) possibility of taking infinite-volume limit everywhere except for phase-transition points, because the system can be partitioned into quasi-independent subsystems owing to the finite correlation length; and (c) possibility of elimination of ultraviolet cutoff due to renormalizability. These considerations are illustrated here by several examples.

### 7.1. Zero-Dimensional Model

Consider the integral

$$W(g) = \int_{-\infty}^{\infty} d\varphi \exp\{-\varphi^2 - 2\gamma\sqrt{g}\varphi^3 - g\varphi^4\}$$

$$= g^{-1/2} \int_{-\infty}^{\infty} d\phi \exp\{-S\{\phi\}/g\}, \quad (7.5)$$

as a zero-dimensional analog of the functional integral in $\varphi^4$ theories containing cubic terms. According to [50], the singular points in the Borel plane correspond to the extrema of the dimensionless action $S\{\phi\}$:

$$S\{\phi_c\} = \frac{1}{32}[-27\gamma^4 + 36\gamma^2 - 8 \pm \gamma(9\gamma^2 - 8)^{3/2}]. \quad (7.6)$$

If $\gamma = 0$, then there exist two saddle points with the same value of action, $\phi_c = \pm i/\sqrt{2}$, and there is a singularity at $z = S\{\phi_c\} = -1/4$ on the negative half-axis. If $0 < \gamma < \gamma_c$, where $\gamma_c = (8/9)^{1/2} \approx 0.942$, there exist two complex conjugate saddle points. If $\gamma > \gamma_c$, then these two points lie on the positive half-axis (Fig. 13a), and situation becomes "non-Borel-summable." When $\gamma$ equals $\gamma_c$, another minimum of $S\{\phi\}$ appears on the real axis.

When $\gamma = 0$, the contour $C$ in (5.9) is aligned with the positive half-axis. This choice of $C$ can obviously be retained for $\gamma < \gamma_c$ (Fig. 13b). The configuration obtained by making cuts from the singular points in Fig. 13b to infinity along the rays emanating from the origin can be considered as the quadrilateral $A_1A_2A_3A_4$ with vertices $A_2$ and $A_4$ at infinity, that can be mapped to the unit circle by the Christoffel–Schwarz integral [51, 119] and then to the plane with a cut (Fig. 8d). The latter mapping is defined as

$$z = p\frac{u}{(1-u)^\beta}, \quad p = \frac{\beta^\beta(1-\beta)^{1-\beta}}{a}, \quad (7.7)$$

where $2\pi\beta$ is the angle between the cuts and $1/a$ is the distance from the singular points to the origin. Note that (6.4) and (6.9) are special cases of (7.7) corresponding to $\beta = 2$ and $\beta = 1$. The coefficients $U_N$ of the resummed series in (6.3) are expressed as

$$U_0 = B_0,$$

$$U_N = \sum_{K=1}^{N} B_K p^K \frac{\Gamma(N-K+\beta K)}{\Gamma(N-K+1)\Gamma(\beta K)}, \quad N \geq 0. \quad (7.8)$$

Their asymptotic form in the limit of $N \longrightarrow \infty$,

$$U_N = U_\infty N^{-1+\alpha\beta}, \quad U_\infty = \frac{W_\infty}{\Gamma(\alpha\beta)\Gamma(\alpha+b_0)}, \quad (7.9)$$

determines values of parameters in the asymptotics of $W(g)$ in the strong-coupling limit. When $0 < \gamma < \gamma_c$, the pattern of minima of $\chi^2$ is analogous to that corresponding to $\gamma = 0$ (see Fig. 10). The summation results obtained for $\gamma = 0.25$ and $0.75$ are presented in Tables 1 and 2, respectively. As in the case of $\gamma = 0$, the accuracy of summation depends on the error of reconstructing the asymptotics, which increases as $\gamma$ approaches $\gamma_c$ (cf.

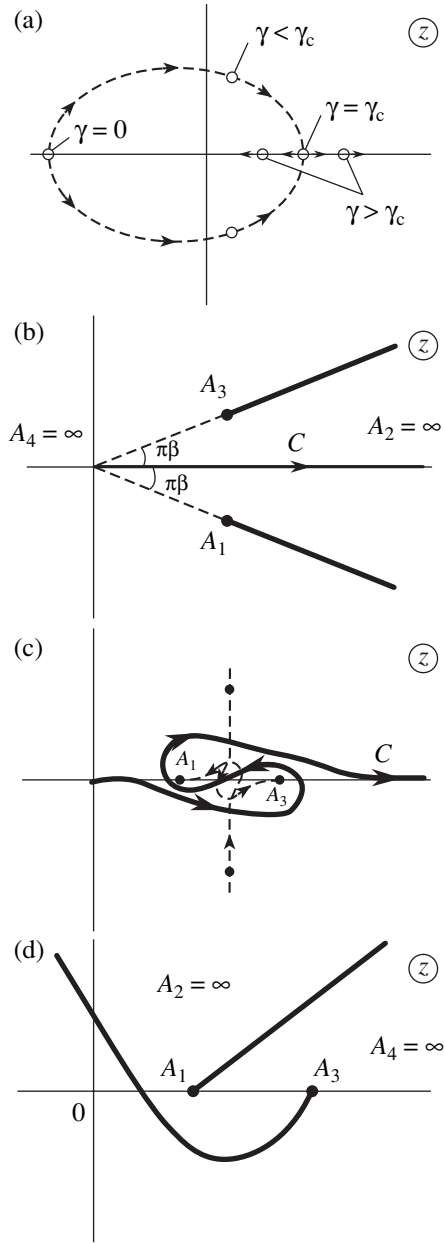

**Fig. 13.** (a) Singularities in Borel plane for different $\gamma$ in integral (7.5). Contour of integration: (b) $\gamma < \gamma_c$; (c) $\gamma > \gamma_c$. (d) Point $A_1$ remains on the physical sheet of the Riemann surface if the cut emanating from $A_3$ is parabolic.

Tables 1 and 2). Indeed, the leading correction to (7.9) has the order $N^{-1+\alpha'\beta}$, and the actual large parameter in the expansion is $N^\beta$ (since $\alpha - \alpha' \sim 1$). This conclusion is confirmed by an estimate for summation error analogous to (6.27):

$$\frac{\delta W(g)}{W(g)} \sim \begin{cases} \Delta, & ag \gtrsim N_c^\beta, \\ \Delta\exp\{-(1+\beta)[(1-\beta)^{1-\beta}N_c^\beta/ag]^{1/(1+\beta)}\}, & ag \lesssim N_c^\beta. \end{cases} \quad (7.10)$$

**Table 1.** Sum of the series for integral (7.5) with $\gamma = 0.25$

| g | W(g) | |
|---|---|---|
| | exact value | resummed value |
| 0.0625 | 1.718915 | 1.718915 |
| 0.125 | 1.674422 | 1.674422 |
| 0.25 | 1.604821 | 1.604821 |
| 0.50 | 1.508008 | 1.508008 |
| 1 | 1.387746 | 1.387745 |
| 2 | 1.252226 | 1.252220 |
| 4 | 1.110955 | 1.11093 |
| 8 | 0.972181 | 0.97212 |
| 32 | 0.722937 | 0.72272 |
| $g \longrightarrow \infty$ | $1.812g^{-0.25}$ | $1.835g^{-0.252}$ |

**Table 2.** Sum of the series for integral (7.5) with $\gamma = 0.75$

| g | W(g) | |
|---|---|---|
| | exact value | resummed value |
| 0.0625 | 1.902930 | 1.902928 |
| 0.125 | 1.937627 | 1.93755 |
| 0.25 | 1.903621 | 1.90300 |
| 0.50 | 1.787743 | 1.7851 |
| 1 | 1.615170 | 1.608 |
| 2 | 1.419861 | 1.406 |
| 4 | 1.226524 | 1.205 |
| 8 | 1.048303 | 1.020 |
| 32 | 0.753306 | 0.714 |
| $g \longrightarrow \infty$ | $1.812g^{-0.25}$ | $1.885g^{-0.275}$ |

The parameter $N^\beta$ decreases as $\gamma \longrightarrow \gamma_c$, since $\beta \longrightarrow 0$ and the value of $N$ cannot be increased indefinitely for technical reasons. Thus, the algorithm formulated above cannot be used to deal with the case of $\gamma = \gamma_c$, but there are no principal restrictions for approaching this case arbitrarily closely.

To analyze the case $\gamma > \gamma_c$, the degeneracy of the singular points at $\gamma = \gamma_c$ is eliminated by adding a small imaginary constant $i\delta$ to $\gamma$. When $\gamma \approx \gamma_c$, saddle-point action (7.6) contains the singular contribution $(\gamma - \gamma_c)^{3/2}$. As $\gamma$ increases, vertical displacement of the singular points is followed by horizontal after the rotation by an angle of $3\pi/2$ is performed, and the contour $C$ folds (Fig. 13c). If the cut emanating from $A_3$ is rotated so as to coincide with the positive half-axis, point $A_1$ appears on another sheet of the Riemann surface and does not contribute to the divergence of the perturbation series, as it can be verified directly by calculating the expansion coefficients. This agrees with the fact that the formally calculated contribution of $A_1$ to the coefficients is purely imaginary. To hold the singular point $A_1$ on the physical sheet of the Riemann surface, the cut emanating from $A_3$ must be curved. If the cut is parabolic (see Fig. 13d), the following constructive algorithm can be used: the mapping $w = \sqrt{z - A_1}$ transforms the domain in Fig. 13d into a plane with a straight cut, which is mapped to a unit circle by the Christoffel–Schwarz integral; then any desired domain (see Fig. 8c) can be obtained.

### 7.2. Double-Well Potential

Consider the ground state of a quantum particle in the potential

$$U(x) = \frac{1}{2}x^2 - \gamma\sqrt{g}x^3 + \frac{1}{2}gx^4, \quad (7.11)$$

which reduces to anharmonic oscillator (1.1) when $\gamma = 0$ and becomes a double-well potential with symmetric minima when $\gamma = 1$. The latter model is of interest as a typical case of two degenerate vacuums: according to [116] the problems of this kind cannot be solved by summation of perturbation series in principle. Model (7.11) reduces to a one-dimensional field theory, which has two comlex conjugate instantons for $0 < \gamma < 1$, while the corresponding dimensionless action is

$$S\{\phi_c\} = -\frac{2}{3} + \gamma^2 - \frac{1}{2}\gamma(\gamma^2 - 1)\left[\ln\frac{1+\gamma}{1-\gamma} \pm \pi i\right]. \quad (7.12)$$

If $m$-instanton configurations are taken into account, then the singularities of the Borel transform lie at the points $z_m = -(2/3)m$ on the negative half-axis when $\gamma = 0$ and on two rays emanating from the origin when $0 < \gamma < 1$ (see Fig. 14a). In the latter case, the perturbation series can be resummed by using the conformal mapping defined by (7.7). The value $\gamma = 1$ corresponds to the critical case approached as $\beta \longrightarrow 0$. This situation is unreachable in a rigorous sense, but there is no principal restrictions that forbids approaching it to an arbitrarily small distance. This can be done without using perturbation series in terms of an arbitrary $\gamma$. It will suffice to analyze the change in the Lipatov asymptotics caused by a small deviation of $\gamma$ from unity [43].

### 7.3. Yang–Mills Theory

In the Yang–Mills theories, the Borel integral can be interpreted by using a procedure that resembles analytical continuation with respect to the coupling between fields, but preserves gauge invariance. This is facilitated by invoking the results obtained in [18], where an SU(2) Yang–Mills field $A_\nu^a$ coupled to a complex scalar field $\varphi$ was considered,

$$S\{A, \varphi\} = \int d^4x \left\{ \frac{1}{4}(F_{\mu\nu}^a)^2 + |(\partial_\mu - ig\tau^a A_\mu^a)\varphi|^2 \right.$$
$$\left. + \frac{1}{2}\lambda^2 |\varphi|^4 \right\}, \quad (7.13)$$

$$F_{\mu\nu}^a = \partial_\mu A_\nu^a - \partial_\nu A_\mu^a + g\epsilon_{abc} A_\mu^b A_\nu^c,$$

with $\tau^a = \sigma^a/2$ and $\sigma^a$ denoting Pauli matrices. After changing to new field variables, $A \longrightarrow B/g$, $\varphi \longrightarrow \phi/\lambda$, action (7.13) is represented as

$$S\{A, \varphi\} = \frac{S_0\{B\}}{g^2} + \frac{S_1\{B, \phi\}}{\lambda^2} \equiv \frac{S\{B, \phi\}}{g^2}, \quad (7.14)$$

where the last equality is written by introducing $\chi = \lambda^2/g^2$. In this theory, an arbitrary quantity $Z(g^2, \lambda^2)$ can be represented as a double series in powers of $g^2$ and $\lambda^2$,

$$Z(g^2, \lambda^2) = \sum_{K, M} Z_{K, M} g^{2M} \lambda^{2K}, \quad (7.15)$$

with coefficients $Z_{K, M}$ determined by the saddle-point configurations of functional (7.14) modified by adding $-M\ln g^2$ and $-K\ln\lambda^2$. The saddle-point values of $g^2$ and $\lambda^2$ are

$$g_c^2 = \frac{S_0\{B_c\}}{M}, \quad \lambda_c^2 = \frac{S_1\{B_c, \phi_c\}}{K}, \quad (7.16)$$

while the saddle-point field configuration is given by

$$B_\mu^a(x) = 4\eta_{\mu\nu}^a x_\nu \frac{\rho^4 - 1}{(x^2 + \rho^2)(\rho^2 x^2 + 1)},$$

$$\phi(x) = \pm i\sqrt{\chi} U \frac{4\sqrt{3}}{[(x^2 + \rho^2)(\rho^2 x^2 + 1)]^{1/2}}, \quad (7.17)$$

$$\rho^4 = 12\chi - 1,$$

where $\eta_{\mu\nu}^a$ denotes 't Hooft matrices, $U$ is a constant spinor ($UU^* = 1$). The saddle-point action is expressed as

$$S\{B_c, \phi_c\} = 16\pi^2 \left[ -2 + \frac{3(\sinh 4\xi_0 - 4\xi_0)}{2\sinh^2 2\xi_0} \right],$$

$$S_1\{B_c, \phi_c\} \quad (7.18)$$

$$= 16\pi^2 \chi(-6\cosh 2\xi_0) e^{-2\xi_0} \frac{2\xi_0 \cosh 2\xi_0 - \sinh 2\xi_0}{\sinh^3 2\xi_0},$$

$$e^{2\xi_0} = \rho^2,$$

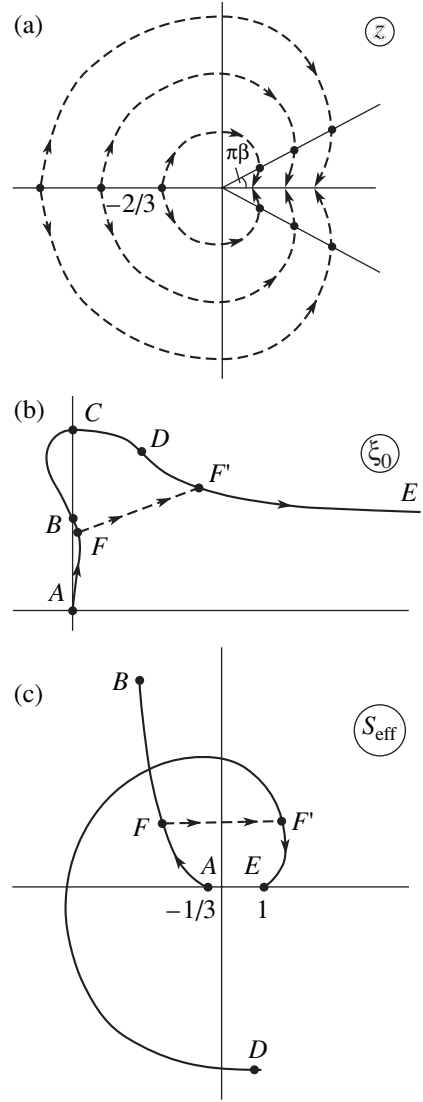

**Fig. 14.** (a) Singularities in the Borel plane corresponding to different $\gamma$ in potential (7.11). Interpretation of the Borel integral in the Yang–Mills theory: (b) curve *ABCDE* in the $\xi_0$ plane defined by the condition $\mathrm{Im}f(\xi_0) = 0$ and (c) the corresponding action $S_{\mathrm{eff}}$ (in units of $16\pi^2$).

while the corresponding value of \chi is

$$\chi = \frac{\lambda_c^2}{g_c^2} = \frac{M}{K} \frac{S_1\{B_c, \phi_c\}}{S_0\{B_c\}}$$

or $\dfrac{\chi S\{B_c, \phi_c\}}{S_1\{B_c, \phi_c\}} = \dfrac{M + K}{K}.$ (7.19)

If $K$ and $M$ in (7.15) are in a constant ratio to N,

$$K = \gamma N, \quad M = \delta N, \quad \gamma + \delta = 1, \quad (7.20)$$

then the asymptotic behavior of the expansion coeffi-cients is described by the formula

$$Z_{K,M} = Z_{\gamma N, \delta N} \propto N! \operatorname{Re}[S_{\text{eff}}(\xi_0)]^{-N}, \quad (7.21)$$

where

$$S_{\text{eff}}(\xi_0) = 16\pi^2 \left[ -2 + \frac{3(\sinh 4\xi_0 - 4\xi_0)}{2\sinh^2 2\xi_0} \right] \\ \times \left( \frac{e^{4\xi_0} + 1}{12} \right)^{\gamma}, \quad (7.22)$$

and $\xi_0$ is found by solving the second equation in (7.19) rewritten by using (7.18) as

$$f(\xi_0) = e^{2\xi_0} \tanh 2\xi_0 \\ \times \frac{2\xi_0 + \frac{1}{3}\sinh^2 2\xi_0 - \frac{1}{2}\sinh 2\xi_0 \cosh 2\xi_0}{2\xi_0 \cosh 2\xi_0 - \sinh 2\xi_0} = \frac{1}{\gamma}. \quad (7.23)$$

Equation (7.23) has a pair of complex conjugate solutions, and the symbol Re in (7.21) is a result of summation over these solutions. Figure 14b shows the curve $ABCDE$ defined by the condition $\operatorname{Im} f(\xi_0) = 0$. On this curve, $\operatorname{Re} f(\xi_0)$ varies from 1 to $\infty$ along segment $AB$, from $-\infty$ to 0 along $BC$, and from 0 to $\infty$ along $CE$. Physical values of $\gamma$ ($0 \leq \gamma \leq 1$) correspond to segments $AB$ and $DE$, where $|S_{\text{eff}}(\xi_0)|/16\pi^2$ varies from 1/3 to 5.4 and from 4.2 to 1, respectively (see Fig. 14c). Asymptotic form (7.21) is determined by the saddle point with the minimum value $|S_{\text{eff}}(\xi_0)|$, and the variation of $\gamma$ from 1 to 0 corresponds to the movement along the trajectory $AFF'E$ with a jump between points $F$ and $F'$, which are associated with equal values of $\gamma$ and $|S_{\text{eff}}(\xi_0)|$. The jump in action (see Fig. 14c) can be eliminated by moving via complex values of $\gamma$ defined by relation (7.23) on the segment $FF'$ in the complex $\xi_0$ plane (see Fig. 14b).

Point $A$ corresponds to the value $\rho^2 = 1$ for which the Yang–Mills field vanishes (see (7.17)), and $S_{\text{eff}}(\xi_0)$ corresponds to the saddle-point action in $\varphi^4$ theory. On the other hand, the parameter $\chi$ increases indefinitely at the right endpoint of the curve ($\xi_0 \longrightarrow \infty + i\pi/4$), the field $\varphi$ vanishes accordingly, and $S_{\text{eff}}(\xi_0)$ corresponds to the value of action for an instanton–anti-instanton pair for the pure Yang–Mills theory. If coefficients (7.21) formally define the series

$$Z(\tilde{g}^2) = \sum_N Z_{\gamma N, \delta N} \tilde{g}^N, \quad \tilde{g} = g^{2\gamma} \lambda^{2\delta}, \quad (7.24)$$

then $\gamma$ can be varied to perform a continuous change from the series for $\varphi^4$ theory to the series for the Yang–Mills theory and to monitor the evolution of the singularities of the Borel transform at $z = S_{\text{eff}}(\xi_0)$ and $z =$ $S_{\text{eff}}^*(\xi_0)$. If the $m$-instanton configurations are taken into account, then the result is similar to that obtained for the double-well potential (Fig. 14a). Therefore, the summation of the series for the Yang–Mills theory should make use of conformal mapping (7.7) with a sufficiently small parameter $\beta$.

## 8. GELL-MANN–LOW FUNCTIONS IN BASIC FIELD THEORIES

This section presents a scheme for finding the Gell-Mann–Low functions in basic field theories with arbitrary coupling constants (see Fig. 15).

### 8.1. $\varphi^4$ Theory

The first attempt to recover the Gell-Mann–Low function in the four-dimensional scalar $\varphi^4$ theory was made in [11]. The analysis of the strong-coupling limit presented in [12] predicted the asymptotic behavior $0.9g^2$, which differs from the one-loop result $1.5g^2$ valid for $g \longrightarrow 0$ only by a numerical factor. Similar asymptotic behavior, $1.06g^{1.9}$, was obtained in [13]. The variational perturbation theory developed in [120] predicts $2.99g^{1.5}$. All of these results indicate that $\varphi^4$ theory is internally consistent (or "trivial"), which contradicts the absence of renormalon singularities established in Section 5. An additional argument follows from the fact that $\varphi^4$ theory can be rigorously derived from a reasonable model of a disordered system [103, 121–123], which is well defined in the continuum limit.

The Gell-Mann–Low function can be found by means of the algorithm described in Section 6.2 with $\beta(g)$ playing the role of $W(g)$ [65]. The input data used here are the same as in [12]: the values of the first four coefficients of the $\beta$ function expansion in the subtraction scheme [124, 125],

$$\beta(g) = \frac{3}{2}g^2 - \frac{17}{6}g^3 + \frac{154.14}{8}g^4 - \frac{2338}{16}g^5 + \ldots, \quad (8.1)$$

and their high-order asymptotics [7] with the first correction to it calculated in [126],

$$\beta_N = \frac{1.096}{16\pi^2} N^{7/2} N! \left\{ 1 - \frac{4.7}{N} + \ldots \right\}. \quad (8.2)$$

This asymptotic expression is determined by the expansion coefficients for the invariant charge, which corresponds here to the vertex with $M = 4$ (cf. (4.1.18)). The "natural" normalization is used for charge $g$, with the parameter $a$ in (2.5) set equal to unity. In this case, the nearest singular point of the Borel transform is sepa-rated by unit distance from the origin, and the characteristic scale of variation of $\beta$ function is $g \sim 1$.

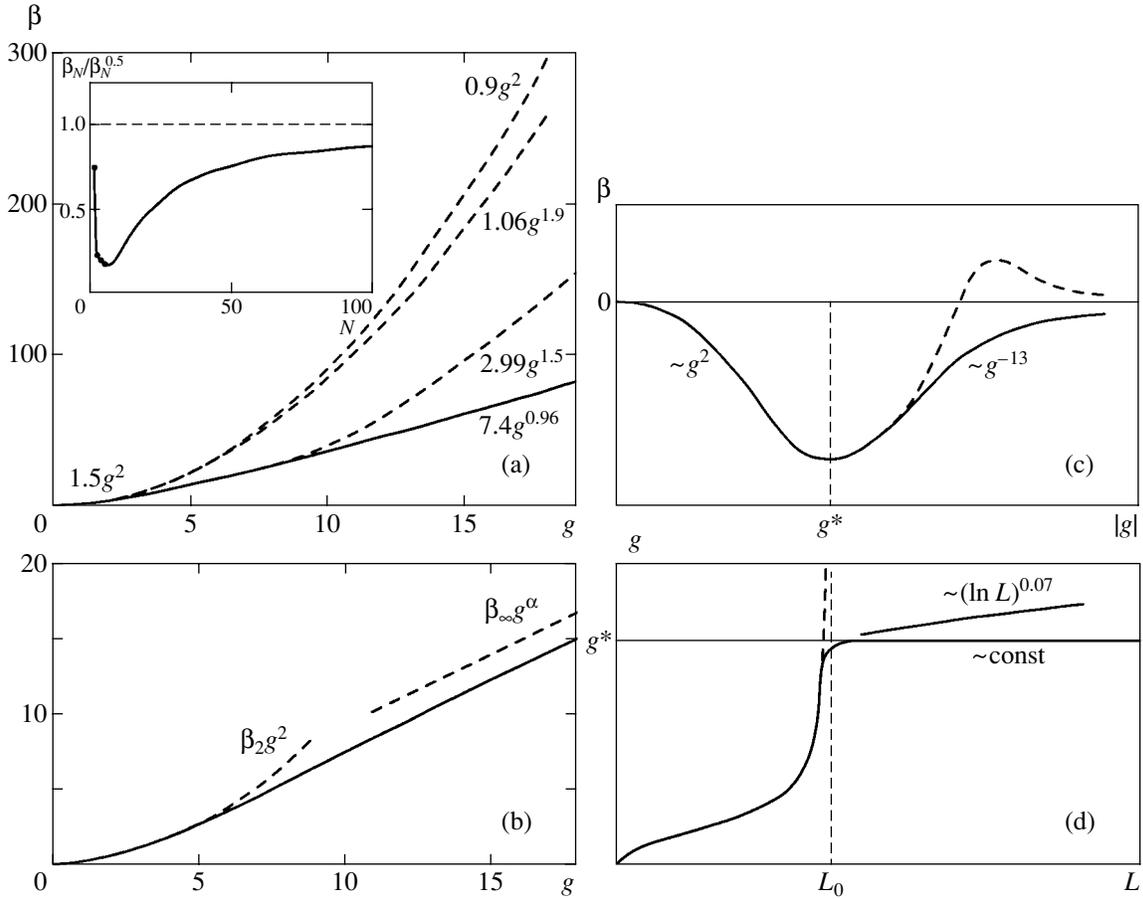

**Fig. 15.** Gell-Mann–Low functions for (a) $\varphi^4$ theory, (b) QED, and (c) QCD. (d) Effective coupling in QCD.

Interpolation was performed using (6.24) with $\tilde{b} = 4$ as an optimal value. Coarse optimization of $\chi^2$ as a function of $\tilde{N}$ was performed for several constant values of $b_0$ [65] to determine the range of interpolations ($\tilde{N}$ between –0.5 and 0.5) consistent with the power-law asymptotic behavior of $W(g)$. Figure 16 shows the behavior of $U_N$ for a nearly optimal interpolation with $\tilde{N} = 0$. Since all curves except for those corresponding to $b_0 \gg 1$ and $b_0 \approx -2$ (whose approach to their asymptotica is dragged out) tend to constants at large $N$, the value of $\alpha$ is close to unity. This result agrees with the value of $\alpha_{\text{eff}}$ at the right-hand minimum of $\chi^2$, the location of the left-hand minimum of $\chi^2$, and the behavior of $U_\infty$ near its zero (see Fig. 17). Figure 18 illustrates the dependence of the results on $\tilde{N}$. The behavior of $\alpha$ corresponding to the "ideal" situation shown in Fig. 11b is obtained by widening the error corridor by a factor of 2 (short-dashed curves in Fig. 18a). The resulting value $\alpha = 0.96$ is consistent with all results obtained for various $\tilde{N}$. The "ideal" situation for $W_\infty$ is obtained immediately (Fig. 18b), and the corresponding value $W_\infty = 7.4$ is consistent with all results. Thus,

$$\alpha = 0.96 \pm 0.01, \quad W_\infty = 7.4 \pm 0.4. \tag{8.3}$$

A similar pattern is observed when $\tilde{b}$ is varied in (6.24) [65].

Figure 15a compares the $\beta$ function obtained for $g \leq 20$ by series summation (solid curve) with results obtained in [12, 13, 120] (upper, middle, and lower dashed curves, respectively). The asymptotic form of $\beta(g)$ found in [12, 13] corresponds to the stable line segment $\tilde{U}_N \approx 1.1N$ at $N \lesssim 10$ in Fig. 16, which is inevitably interpreted as the true asymptotics if it is calculated by using only the known expansion coefficients. Actually, this segment is associated with a dip in the reduced coefficient function \beta_N/\beta^as_N at $N \lesssim 10$ (see insert to Fig. 15a). This dip has also manifestation in the $\beta$ function, resulting to its one-loop behavior being extended[17] to $g \sim 10$ [65]. Thus, the results obtained in [12, 13] reflect the actual properties of the $\beta$ function and should not be considered as essentially incorrect (see detailed discussion in [65, Section 8.3]). The variational perturbation theory developed in [120] provides a somewhat better description of the region of $g \lesssim 10$ in Fig. 15a, but does not guarantee correct results in the strong-coupling limit even theoretically.

---

[17]Being more pronounced for the Borel transform, this behavior is less obvious for the $\beta$ function because of the integration in (6.2).

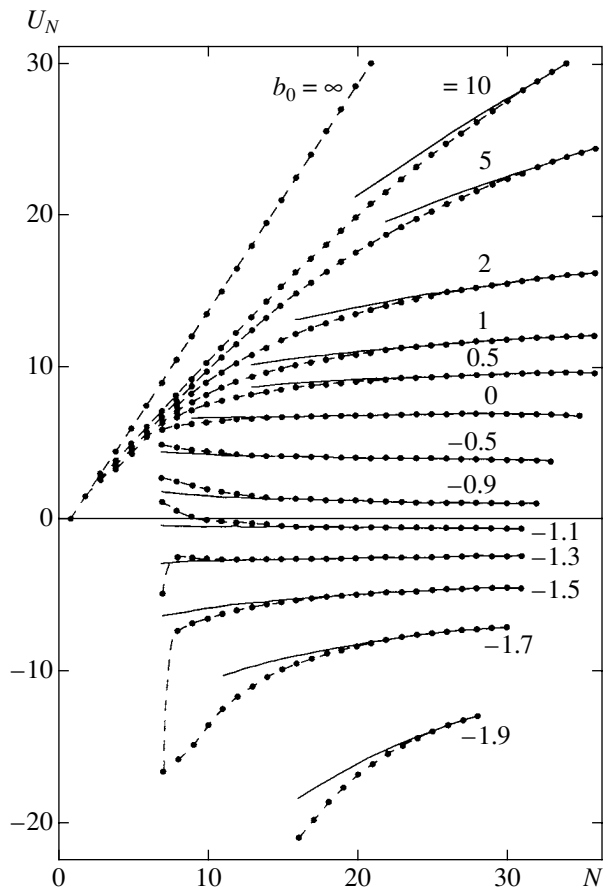

**Fig. 16.** Coefficients $\tilde{U}_N$ versus $N$ for different $b_0$ (symbols) and corresponding approximate power laws (solid curves) obtained for $\varphi^4$ theory by nearly optimal interpolation with $\tilde{b} = 4$ and $\tilde{N} = 0$.

The value of $\alpha$ obtained is close to unity. Even though the deviation from unity exceeds the error, the exact equality $\alpha = 1$ cannot be ruled out, because asymptotic expansion (6.7) may contain logarithmic corrections,

$$W(g) = W_\infty g^\alpha (\ln g)^{-\gamma}, \quad g \longrightarrow \infty, \qquad (8.4)$$

which may be interpreted as a slight decrease in $\alpha$ if $\gamma > 0$. In this case, expansion (6.11) contains the factor $(\ln N)^{-\gamma}$, while $U_\infty$ does not change, and the resulting $U_N$ can be fitted by using (8.4) with

$$\alpha = 1, \quad \gamma \approx 0.14, \quad W_\infty \approx 7.7 \qquad (8.5)$$

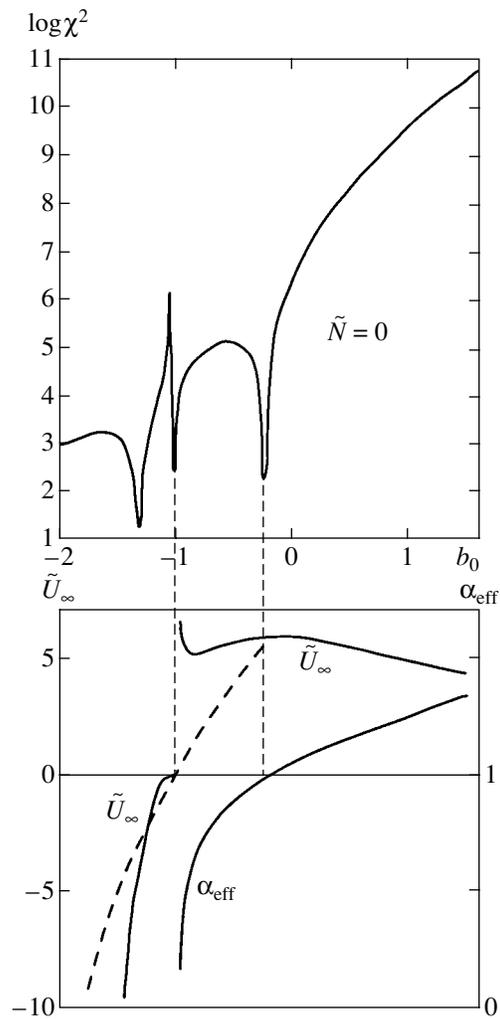

**Fig. 17.** Pattern of $\chi^2$ minima obtained for $\varphi^4$ theory by using the interval $20 \leq N \leq 40$ and curves of $\alpha_{\text{eff}}$ and $\tilde{U}_\infty$ versus $b_0$ for nearly optimal interpolation $\tilde{N} = 0$. Dashed curve is $U_\infty(b_0)$ corresponding to fitting with fixed $\alpha = 1$.

without any increase in $\chi^2$. Logarithmic branching appears to be quite plausible for the following reasons.

1. Logarithmic branching is inevitable when the exact equality $\alpha = 1$ holds. Indeed, series (6.1) can be represented as the Sommerfeld–Watson integral [7, 13]

$$W(g) = \sum_{N = N_0}^{\infty} W_N(-g)^N$$
$$= -\frac{1}{2i}\oint_C dz \frac{\mathcal{W}(z)}{\sin \pi z} g^z, \qquad (8.6)$$

where $\mathcal{W}(z)$ is the analytic continuation of $W_N$ to the complex plane ($\mathcal{W}(N) = W_N$), $C$ is a contour encompassing the points $N_0, N_0 + 1, N_0 + 2, \ldots$. When $g$ is large, the contour $C$ can be extended and shifted left-wards until the rightmost singular point of $\mathcal{W}(z)/\sin\pi z$ at $z = \alpha$ is reached. This singularity determines the behavior of $W(g)$ as $g \longrightarrow \infty$. Power law (6.7) and asymptotic formula (8.4) correspond to the existence of a simple pole at $z = \alpha$ and a singularity of the form $(z - \alpha)^{\gamma - 1}$, respectively.

The term $\beta_0$ in expansion (8.1) vanishes by definition. However, the zero value of the coefficient $\beta_1$ is accidental: f.e. in the $(4 - \epsilon)$-dimensional $\varphi^4$ theory, it has a finite value of the order $\epsilon$, and $\mathcal{W}(1) \sim \epsilon$ accordingly. As $\epsilon \longrightarrow 0$, the four-dimensional value $\mathcal{W}(1) = 0$ is obtained, and there is no simple pole if $\alpha = 1$. If zero is approached according to the law $\mathcal{W}(z) = \omega_0(z - 1)^\gamma$, then

$$\beta(g) = \frac{\omega_0}{\Gamma(1 - \gamma)} g (\ln g)^{-\gamma}, \quad g \longrightarrow \infty, \qquad (8.7)$$

and the positive value of $\gamma$ has a natural explanation.

2. The class of field theories with the interaction $\varphi^n$ (generalizations of $\varphi^4$ theory) was analyzed in [127] for space of dimension $d = 2n/(n - 2)$, for which logarthmic situation takes place. In theories of this type, the coefficient $\beta_1$ vanishes, but becomes finite as $d$ decreases. Therefore, $\mathcal{W}(1) = 0$ as shown above. The Gell-Mann–Low function can be calculated exactly as $n \longrightarrow \infty$ [127], and the rightmost singularity of $\mathcal{W}(z)$ has the form $(z - 1)^{3/2}$, which corresponds to the asymptotic behavior $\beta(g) \propto g(\ln g)^{-3/2}$. By continuity, nonanalyticity of the type $(z - 1)^\gamma$ should hold for finite $n$, and the singularity at $z = 1$ should remain rightmost. Therefore, asymptotic behavior (8.7) is natural for field theories of this kind, and it is no surprise that it holds even for $n = 4$. Note that $W_\infty$ is negative as $n \longrightarrow \infty$, and the Gell-Mann–Low function has a zero. A similar conclusion can be drawn for $\varphi^4$ theory by straightforward extrapolation to $n = 4$ [127]. Actually, one should have in mind in this extrapolation that the exponent $\gamma$ varies from 3/2 to small values like (8.5). Accordingly, the asymptotics (8.7) obviously changes sign for $\gamma = 1$ due to gamma-function. ($\omega_0$ is positive since $\mathcal{W}(2) \sim \omega_0$ and $\beta_2$ is positive [127]).

Thus, one has to choose between two possibilities: power law (6.7) with $\alpha$ slightly below unity and asymptotic expression (8.7) with $\gamma > 0$. In either case, $\varphi^4$ theory is self-consistent, which contradicts the widespread view that $\varphi^4$ theory is trivial. Let us discuss the origin of this belief (for a more detailed discussion, see [65, Section 8.4]).

It has been rigorously proved that $\varphi^4$ theory is trivial for $d > 4$ and nontrivial for $d < 4$ [128, 129]. The inequalities obtained for $d = 4$ are "just a bit" insufficient for proving triviality [130, Section 14]. For mathematicians, it looks as an annoying minor problem, and the triviality of $\varphi^4$ theory is commonly regarded as "almost proved." For physicists, there is no reason to be so optimistic about it: from the modern perspective, the afore-mentioned results obtained for $d \neq 4$ are nothing more than elementary corollaries of renormalizability theory and one-loop renormalization group. As for the situation for $d = 4$, it is extremely complicated from viewpoint of its physics, and no analytical approach to the problem has been found to this day.

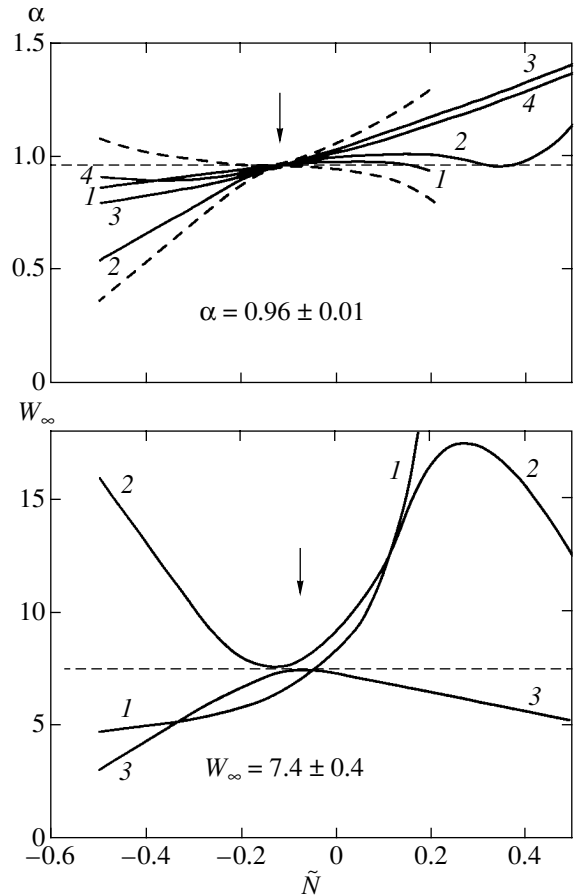

**Fig. 18.** Curves of $\alpha$ and $W_\infty$ obtained for $\varphi^4$ theory: numbers at curves correspond to numbering of estimates in Section 6.2; short-dashed curves illustrate the widening of the error corridor for $\alpha$ by a factor of 2.

It is generally believed that the triviality of $\varphi^4$ theory is convincingly demonstrated by numerical experiments on lattices. However, most of them reveal only a decrease in the effective charge $g(L)$ with increasing $L$, which is quite natural because the $\beta$ function has no zero, whereas convincing evidence of "zero charge" can hardly be obtained on any finite-size lattice. There is considerable misunderstanding with regard to charge normalization. Even under the "natural normalization" used here, a one-loop quadratic law extends till $g \sim 10$. For the conventional normalizations, it holds on wider intervals, for example, till $g \sim 2000$ when the interaction term is written as $g\varphi^4/4!$. Accordingly, behavior of any variable is impossible to distinguish from "trivial" over a wide range of parameter values. The very concept of triviality is frequently misunderstood. Many authors identify it to the mean-field values of critical exponents in the four-dimensional theory of phase transitions; but this indisputable fact is due to the mere absence of nontrivial zero of the β function.

Issues related to triviality were analyzed by Agodi, Consoli, and others in a recent series of publications (e.g., see [131, 132]). An unconventional scenario for continuum limit in $\varphi^4$ theory was proposed and claimed to be logically consistent. The validity of the conventional perturbation theory was basically denied, which seems to be a premature conclusion. Since the numerical lattice results, used as supportive evidence, were obtained in the weak-coupling region, they cannot provide any information about triviality. The analyses presented in [131, 132] were performed to resolve the difficulties arising in the Higgs sector of the Standard Model in view of the triviality of $\varphi^4$ theory. No difficulties of this kind arise when the theory is internally-consistent.

### 8.2. Quantum Electrodynamics

In QED, four terms of the expansion of the β function are known in the MOM scheme [133]:

$$\beta(g) = \frac{4}{3}g^2 + 4g^3 + \left[\frac{64}{3}\zeta(3) - \frac{202}{9}\right]g^4 \\ + \left[186 + \frac{256}{3}\zeta(3) - \frac{1280}{3}\zeta(5)\right]g^5 + \ldots, \quad (8.8)$$

and the corresponding asymptotic expression is

$$\beta_N^{as} = \text{const} \times 4.886^{-N}\Gamma\left(\frac{N+12}{2}\right), \quad N \longrightarrow \infty. \quad (8.9)$$

It is identical, up to a constant factor, to the asymptotic behavior of coefficients for the invariant charge [7], which is determined by $gD$ in QED, where $D$ is the photon propagator (see (4.2.11) for $M = 2$ and $L = 0$).

The summation procedure for this series should be modified in comparison with Section 6, because the Lipatov asymptotic form is $ca^N\Gamma(N/2 + b)$ rather than $ca^N\Gamma(N + b)$. The Borel transformation gives

$$\beta(g) = \int_0^\infty dx\, e^{-x} x^{b_0-1} B(ag\sqrt{x}),$$

$$B(z) = \sum_{N=0}^\infty B_N(-z)^N, \quad (8.10)$$

$$B_N = \frac{\beta_N}{a^N \Gamma(N/2 + b_0)},$$

where $b_0$ is an arbitrary parameter. The conformal mapping $z = u/(1-u)$ is applied to obtain a convergent series in $u$ for the Borel transform, with coefficients

$$U_N = \sum_{K=1}^N B_K (-1)^K C_{N-1}^{K-1} \quad (N \geq 1),$$

$$U_0 = B_0, \quad (8.11)$$

whose behavior at large $N$,

$$U_N = U_\infty N^{\alpha-1}, \quad U_\infty = \frac{\beta_\infty}{a^\alpha \Gamma(\alpha) \Gamma(b_0 + \alpha/2)}, \quad (8.12)$$

determines the parameters of the asymptotic expression $\beta(g) = \beta_\infty g^\alpha$ as $g \longrightarrow \infty$.

Interpolation was performed by using (6.24) with $\tilde{b} = b - 1/2 = 5.5$ [65]. In contrast to $\varphi^4$ theory, the constant factor in (8.9) is not known. Technically, this is not a problem, because the constant $c$ can be factored into the curly brackets in (6.24) to replace 1 with a parameter $\tilde{A}_0$ treated as unknown and determined by interpolation. However, this leads to a much higher uncertainty in the reduced coefficient function $F_N = \beta_N/\beta_N^{as}$: its values $F_2 = 63.1$, $F_3 = -7.02$, $F_4 = 0.34$, and $F_5 = 1.23$ (measured in units of $10^{-3}$) exhibit only weak convergence to a constant, and the predicted $\tilde{A}_0 = \lim_{N \to \infty} F_N$ varies by orders of magnitude as a function of $\tilde{N}$. Nevertheless, the "superstability" of the algorithm mentioned above (see Section 6.2) suggests that reasonable results can be obtained even in this situation. To verify this possibility, a test experiment was performed for $\varphi^4$ theory. The complete input data (including four coefficients $\beta_2, \beta_3, \beta_4, \beta_5$ and parameters $\tilde{A}_0$ and $\tilde{A}_1$) resulted in values $\alpha = 0.96 \pm 0.01$ and $\beta_\infty = 7.4 \pm 0.4$ (recall Section 8.1). Similar procedure performed without using $\tilde{A}_0$ and $\tilde{A}_1$ resulted in $\alpha = 1.02 \pm 0.03$ and $\beta_\infty = 1.7 \pm 0.3$. Since the uncertainty in the coefficient function (estimated by varying $\tilde{N}$ within unity about its optimal value) is a few percent in the former case and more than an order of magnitude in the latter, this robustness of results is rather satisfactory.[18] Of course, the results presented below should be considered as a zeroth approximation.

Coarse optimization of $\chi^2$ as a function of $\tilde{N}$ was performed to determine the range of interpolations $(-0.5 \lesssim \tilde{N} \lesssim 1.0)$ for which $U_N$ may exhibit power-law behavior. The dependence of $\chi^2$, $U_\infty$ and $\alpha$ on $b_0$ illustrated by Fig. 19 implies that $\alpha \approx 1$. Indeed, $U_\infty$ changes sign when $\beta_0 = -\alpha/2 \approx -0.5$ (see (8.12)). The same value of $b_0$ corresponds to the minimum of $\chi^2$ due to zero value of the leading contribution to the asymptotic $U_\infty N^{\alpha-1}$. The values of $\alpha_{\text{eff}}$ corresponding to the minima of $\chi^2$ at $b_0 = -\alpha'/2, -\alpha''/2, \ldots$, where the corresponding corrections to (8.12) vanish, are closest to the exact value $\alpha \approx 1$.[19]

---

[18]The shift in $\beta_\infty$ is not controlled by error estimation. This can be explained by the fact that the procedure of error estimation validated in [65] is justified only when the discrepancy with the exact result is sufficiently small and all deviations can be linearized.

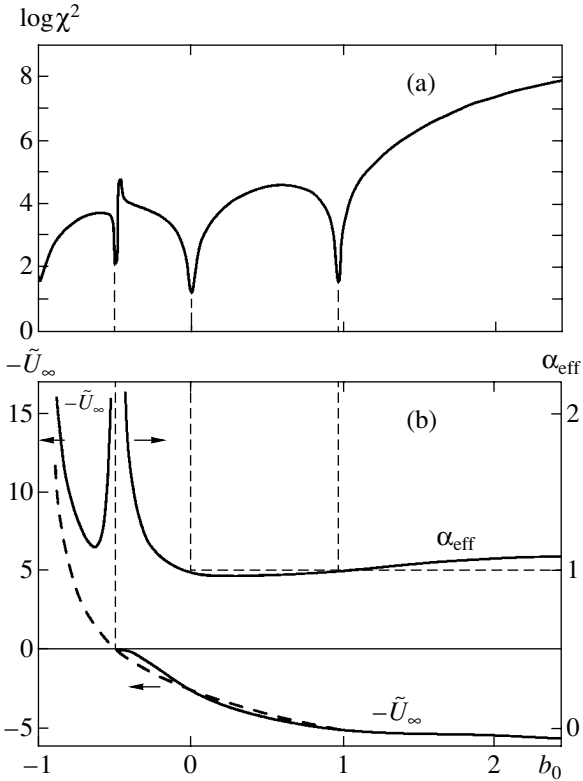

**Fig. 19.** Pattern of $\chi^2$ minima and curves of $\alpha_{\text{eff}}$ and $\tilde{U}_\infty$ versus $b_0$ obtained for quantum electrodynamics by using the interval $20 \leq N \leq 40$ (notation as in Fig. 17).

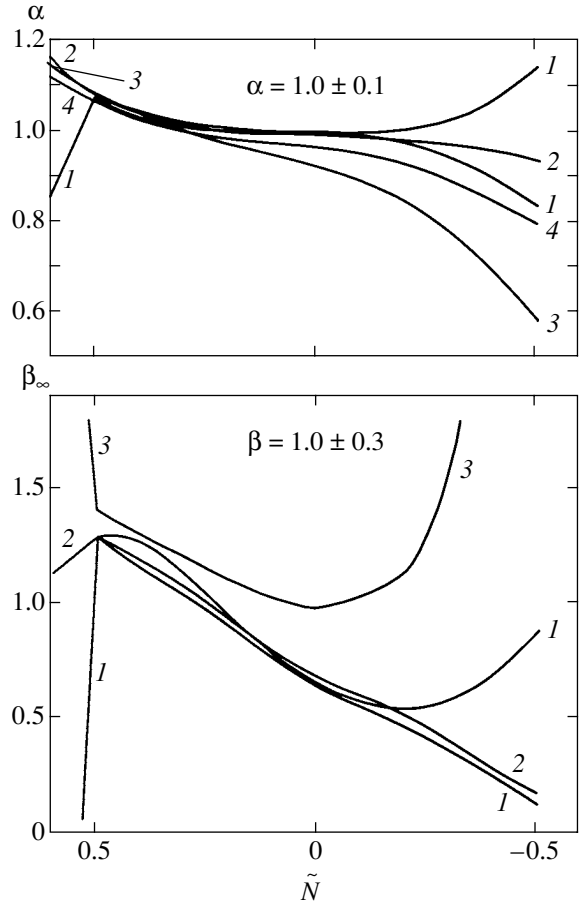

**Fig. 20.** Curves of $\alpha$ and $W_\infty$ estimated for QED: numbers at curves correspond to estimates in Section 6.2.

Figures 20a and 20b show different estimates for $\alpha$ and $\beta_\infty$ as functions of $\tilde{N}$. The values of $\alpha$ obtained for $\tilde{N} \leq 0.25$ are consistent with a value slightly below unity. The systematic growth to 1.08 observed at $\tilde{N} > 0.25$ is not controlled by error estimation, but the corresponding minima of $\chi^2$ are weak and unstable. Similar behavior is characteristic of $\beta_\infty$. The results obtained for the central part of the examined interval of $\tilde{N}$ are accepted as more reliable, with a conservative error estimate including systematic variations:

$$\alpha = 1.0 \pm 0.1, \quad \beta_\infty = 1.0 \pm 0.3. \quad (8.13)$$

In view of the above remarks concerning errors, even this estimate is somewhat unreliable.

Figure 15b shows the results obtained by summing the series for $\tilde{N} = 0.2$ and $b_0 = 0$. The one-loop law $\beta_2 g^2$ is matched with the asymptotic $\beta_\infty g^\alpha$ at $g \sim 10$. The difference between $\beta(g)$ and the one-loop result is negligible at $g < 5$. The asymptotic $\beta(g)$ agrees with the upper bound in the inequality $0 \leq \beta(g) < g$, derived in [134] from a spectral representations, within uncertainty. If $\alpha = 1$ and $\beta_\infty = 1$, then the fine structure constant in pure electrodynamics increases as $L^{-2}$ in the small-length limit.

Results obtained in lattice QED [135, 136] are differently interpreted by specialists. Overall, these results point to the triviality in Wilson's sense: the $\beta$ function does not have a nontrivial zero, and phase transitions are characterized by mean-field critical exponents. This conclusion agrees with the results presented above.

---

[19] Usually, only the minima of $\chi^2$ corresponding to $\alpha$ and $\alpha'$ (recall Section 6.2) are observed in test examples. Additional minima may appear when certain relations between the coefficients $W'_\infty$, $W''_\infty$, … are satisfied. Probably, this occurs in the cases with small amount of available information. Such additional minima were also observed in the test experiment for \phi^4 theory described above.

### 8.3. QCD

In QCD, the first four terms of the expansion of the Gell-Mann–Low function are known in the MS scheme [137]:

$$\beta(g) = -\sum_{N=0}^{\infty} \beta_N g^N = -\beta_2 g^2 - \beta_3 g^3 - \beta_4 g^4 - \ldots,$$

(8.14)

$$g = \frac{\bar{g}^2}{16\pi^2},$$

$$\beta_2 = 11 - \frac{2}{3}N_f, \quad \beta_3 = 102 - \frac{38}{3}N_f,$$

$$\beta_4 = \frac{2857}{2} - \frac{5033}{18}N_f + \frac{325}{54}N_f^2,$$

(8.15)

$$\beta_5 = \left[\frac{149753}{6} + 3564\zeta(3)\right]$$

$$- \left[\frac{1078361}{162} + \frac{6508}{27}\zeta(3)\right]N_f$$

$$+ \left[\frac{50065}{162} + \frac{6472}{81}\zeta(3)\right]N_f^2 + \frac{1093}{729}N_f^3,$$

where $\bar{g}$ is the coupling constant in QCD Lagrangian

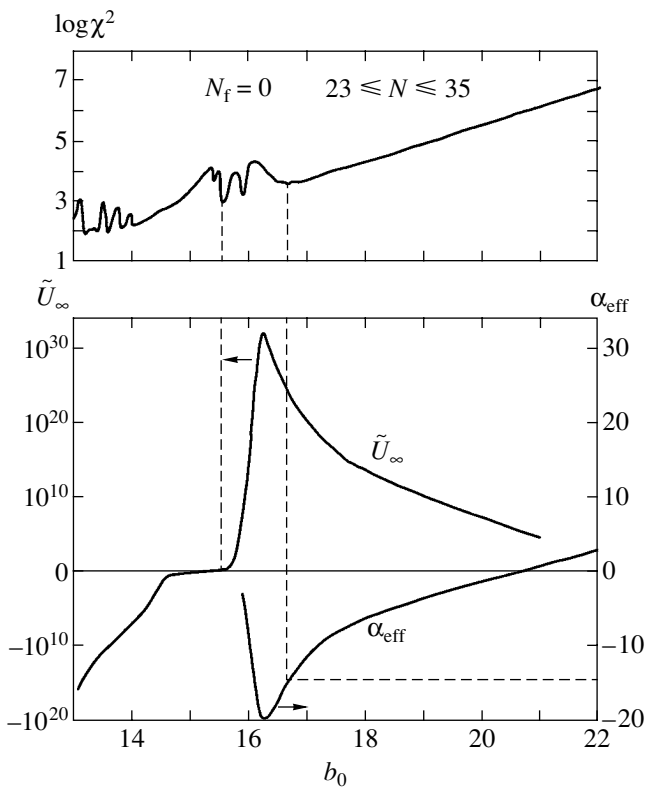

**Fig. 21.** Curves of $\chi^2$, $\alpha_{\text{eff}}$, and $\tilde{U}_\infty$ versus $b_0$ obtained for QCD. The minima at $b_0 = 15.4$ and $15.9$ are interpreted as satellites moving with the main minimum at $b_0 = 15.5$.

(4.5.7). The asymptotic form of the coefficients in series (8.14)

$$\beta_N = \text{const}\,\Gamma\left(N + 4N_c + \frac{11(N_c - N_f)}{6}\right)$$

(8.16)

is determined by the expansion of the invariant charge, which can be found by using any vertex in view of the generalized Ward identities [47]. Formula (8.16) for $N_c = 2$ and $N_f = 0$ agree with the result obtained in [17].

Since (8.14) is a nonalternating series, its summation should be performed by the method described in Section 7. However, a simpler procedure [47] can be applied by assuming that $B(z) \sim z^\alpha$ at infinity. Irrespective of interpretation of Borel integral (7.4), the result is

$$\beta(g) = \beta_\infty g^\alpha, \quad g \longrightarrow \infty,$$

$$\beta(g) = \bar{\beta}_\infty |g|^\alpha, \quad g \longrightarrow -\infty,$$

(8.17)

where the exact relation between $\beta_\infty$ and $\bar{\beta}_\infty$ depends on $\gamma_i$ and $C_i$, but generally $\beta_\infty \sim \bar{\beta}_\infty$. Consequently, the summation of series (8.14) for negative $g$ can be used to determine the exponent $\alpha$ and estimate $\beta_\infty$.

Interpolation of the coefficient function is performed by using (6.24) with $\tilde{b} = b - 1/2$. As in QED, the parameter $c$ in the Lipatov asymptotic form is not known. In Section 8.2, it was calculated in the course of interpolation. In the present case, the results of an analogous procedure are characterized by considerable uncertainties, which cannot be reduced by optimization. For this reason, interpolation was performed for a trial value of $c$ which was varying between $10^{-5}$ and $1$.[20] The change in the results due to this variation was negligible as compared to other uncertainties. The results presented below were obtained for $N_c = 3$, $N_f = 0$, and $c = 10^{-5}$.

By finding a power-law fit for $U_N$ and analyzing $\chi^2$ as a function of $\tilde{N}$ [47], it was found that the minimal values of $\chi^2$ correspond to $0.5 \lesssim \tilde{N} \lesssim 2.0$. Thus, the range of interpolations consistent with the power-law behavior of $U_N$ was determined. The typical curves of $\chi^2$ and effective $U_\infty$ and $\alpha$ plotted versus $b_0$ in Fig. 21 demonstrate that $\alpha \approx -15$. Indeed, $U_\infty$ changes sign (see (6.12)) at $b_0 = -\alpha \approx 15.5$, and the left-hand minimum of $\chi^2$ is located at the same point. A similar estimate, $\alpha \approx -15$, is obtained by using the value of $\alpha_{\text{eff}}$ at the right-hand minimum of $\chi^2$. The values of $\alpha$ estimated by these methods agree only for $\tilde{N}$ close to the optimal value $\tilde{N} = 1.58$ (Fig. 21) and tend to disagree as the difference between $\tilde{N}$ and this value increases.

---

[20] The parameter $c$ is estimated as the product of the squared 't Hooft constant $c_H$ in one-instanton contribution (4.5.6) ($c_H^2 \sim 10^{-5}$ and $10^{-4}$ for $N_f = 0$ and 3, respectively) with the dimensionless integral of instanton configuration. The latter factor is relatively large (its characteristic scale is $16\pi^2$).

The resulting value of $\alpha$ cannot be accepted as final. First, a large value of $\alpha$ may be indicative of exponential behavior. Second, since $\Gamma(\alpha)$ has poles at $\alpha = 0, -1, -2, \ldots$ (see (6.12)), the leading contribution to the asymptotic behavior of $U_N$ may vanish, and the result may correspond, for example, to $\alpha'$ in (6.15). For this reason, the function $W(g) = g^{n_s}\beta(g)$ is introduced, and the integer parameter $n_s$ is increased until the exponent $\alpha_W = \alpha + n_s$ becomes positive. The results obtained by this method (Fig. 22a) demonstrate that the true behavior is a power law with a large noninteger negative exponent rather than an exponential (if $\alpha = -n$, the exponent would behave as illustrated by the inset.) Each point in Fig. 22a is obtained by independent optimization in $\tilde{N}$. The optimal $\tilde{N}$ decreases monotonically with increasing $n_s$. The uncertainty of the results is primarily due to their dependence on the lower limit of the averaging interval $N_{\min} \le N \le N_{\max}$. The higher lying data points in Fig. 22a correspond to small $N_{\min}$ and minimum values of $\chi^2$ of the order $10^6$. As $N_{\min}$ increases, $\alpha$ decreases monotonically until $\chi^2$ reaches values of the order $10^3$ (lower lying data points). With a further increase in $N_{\min}$, the pattern of $\chi^2$ minima becomes indistinct and the uncertainty of the results sharply increases. The value of $\alpha$ is then allowed to decrease further until $\chi^2 \sim 10$ is reached as required, and this is taken into account in error estimation. Even though the uncertainty in $\bar{\beta}_\infty$ amounts to several orders of magnitude (Fig. 22b), the value of the order $10^5$ is consistent with most data and looks to be most probable. Thus,

$$\alpha = -13 \pm 2, \quad \bar{\beta}_\infty \sim 10^5. \qquad (8.18)$$

for $N_f = 0$. For $N_f = 3$, the result is $\alpha = -12 \pm 3$, and the same most probable value is obtained for $\bar{\beta}_\infty$ (though it is scattered between 1 and $10^7$). The consistency of results with different choice of summation procedure means that their uncertainty has been adequately estimated.

The large uncertainty in $\bar{\beta}_\infty$ corresponds to relatively small uncertainty in the $\beta$ function itself: the one-loop law $\beta_2 g^2$ is matched with asymptotic expression (8.17) at $g^* \sim 2$, and $\bar{\beta}_\infty$ changes by four orders of magnitude as $g^*$ changes by a factor of two. When $\alpha_W$ is negative, the sign of $\bar{\beta}_\infty$ is indeterminate, because the error in $\alpha$ is large and the factor $\Gamma(\alpha)$ in Eq. (6.12) is alternating, but this sign is definitely negative for positive $\alpha_W$ (large $n_s$). Figure 15c illustrates the behavior of $\beta$-function at $g < 0$ (solid curve). Its analytic continuation to positive $g$ is expected to exhibit qualitatively similar behavior, but the sign of (8.17) may change (dashed curve).[21] Nevertheless, the behavior of the effective coupling as a function of the length scale $L$ (Fig. 15d) appears to be rather definite. In the one-loop approximation, $g(L)$ has a pole at $L = L_0 = 1/\Lambda_{QCD}$ (dashed curve). For the obtained $\beta$ function, $g(L)$ increases in the neighborhood of $L_0$ until a value of the order $g^*$ is reached (see Fig. 15c) and then either becomes constant (if $\beta_\infty > 0$) or nearly constant, increasing as $(\ln L)^{0.07}$ (if $\beta_\infty < 0$).

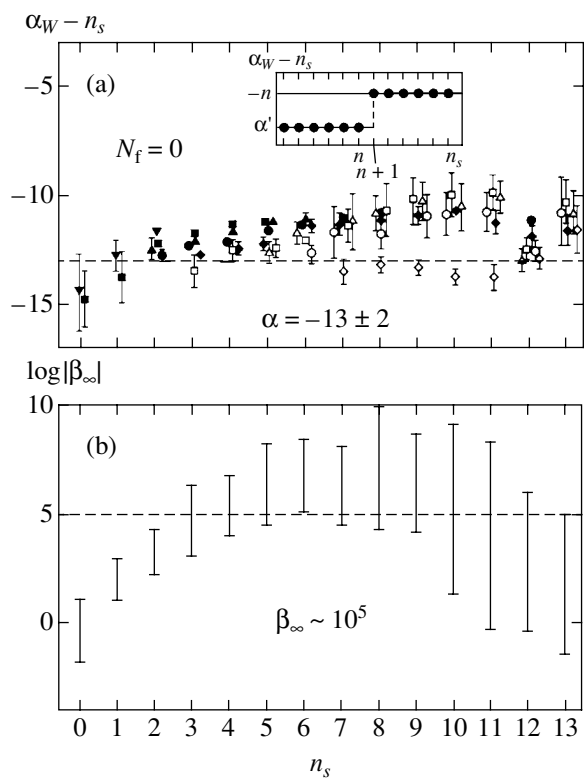

**Fig. 22.** (a) Exponent $\alpha_W$ for QCD obtained by summation of series for $W(g) = g^{n_s}\beta(g)$ versus $n_s$ for different intervals $N_{\min} \le N \le N_{\max}$: ▼ corresponds to $N_{\min} = 22 + n_s$, $N_{\max} = 35 + n_s$; ■, ▲, ●, □, △, ○, and ◇ correspond to $N_{\min}$ increased in unit steps. (b) $\bar{\beta}_\infty$ versus $n_s$.

In the weak-coupling region, the quark–quark interaction potential $V(L)$ is described by the modified Coulomb law $\bar{g}^2(L)/L$, and the sharp increase in $\bar{g}(L)$ in the neighborhood of $L = L_0$ points to a tendency to confinement. In the strong-coupling region, the relation between $V(L)$ and $\bar{g}(L)$ is not known. However, the

---

[21]In particular, $\beta_\infty = \bar{\beta}_\infty \cos\pi\alpha$, if the Borel integral is interpreted in the sense of the principal value.

close in spirit result was obtained by Wilson [138] for the lattice version of QCD:

$$V(L) = \frac{\ln 3\bar{g}^2(a)}{a^2}L, \quad \bar{g}(a) \gg 1, \qquad (8.19)$$

where $a$ is the lattice constant. Since the result should be independent of $a$, the β function in the strong-coupling region may be estimated as $\beta(g) \sim g\ln g$ [139], which is, however, incorrect. The transverse size of the string estimated for $a \gg 1/\Lambda_{QCD}$ is of the order $a$, which is much larger than its actual physical size ($\sim 1/\Lambda_{QCD}$). This means that lattice effects are rather strong and there is no reason to expect that the result is independent of $a$. With regard to $a \ll 1/\Lambda_{QCD}$, there is a reason to these expectations, but Eq. (8.19) does not apply since the coupling constant $\bar{g}(a)$ is small. Thus, Eq. (8.19) may be valid only for $a \sim 1/\Lambda_{QCD}$. In the plateau region, $\bar{g}(L) \sim \sqrt{2 \cdot 16\pi^2} \sim 20$, and the sharp increase in $\bar{g}(L)$ in the neighborhood of $L = L_0$ (Fig. 15d) implies that the conditions $a \sim 1/\Lambda_{QCD}$ and $\bar{g}(a) \gg 1$ are compatible; probably, it is sufficient to justify applicability of lattice formula (8.19) to actual QCD.

## 9. HIGH-ORDER CORRECTIONS TO THE LIPATOV ASYMPTOTICS

As noted above, corrections to Lipatov asymptotic form (2.5) can be represented by a regular expansion in terms of $1/N$:

$$W_N = ca^N \Gamma(N+b)$$
$$\times \left\{ 1 + \frac{A_1}{N} + \frac{A_2}{N^2} + \ldots + \frac{A_K}{N^K} + \ldots \right\}. \qquad (9.1)$$

Knowledge of all coefficients $A_K$ is equivalent to knowledge of the exact coefficient function $W_N$, and their calculation offers an alternative to direct calculation of low-order diagrams [81, 106, 125, 133, 137]. Currently, the lowest order corrections are known in $\varphi^4$ theory [126] and a number of quantum-mechanical problems [6, 140].

It was shown in [141] that series (9.1) is factorially divergent, and high-order expansion coefficients can be calculated by using a procedure analogous to Lipatov's method: an exact expression for the $K$th coefficient can be written as a functional integral and found by the saddle-point method for large $K$. Typically, $A_K$ has the asymptotic form

$$A_K = \tilde{c} \left( \ln \frac{S_1}{S_0} \right)^{-K} \Gamma\left(K + \frac{r'-r}{2}\right), \qquad (9.2)$$

where $S_0$ and $S_1$ are the values of action for the first and second instantons in the field theory under analysis, and $r$ and $r'$ denote the corresponding number of zero modes. The instantons are enumerated in the order of increasing of their action.

Detailed calculations of the asymptotic form of $A_K$ for the $n$-component $\varphi^4$ theory were presented in [33]. Available information about higher-lying instantons in $\varphi^4$ theory is incomplete. However, the most probable candidate for the role of the second instanton is a combination of two elementary instantons [33, 142]. Then, expression (9.2) should be modified, because it is correct only when the equipartition law is valid (see Section 4.1), i.e., when all fluctuational modes can be distinctly divided to zero and oscillatory ones. For two-instanton configurations, there always exist a soft mode that corresponds to variation of the distance between the elementary instantons and it can be reduced to oscillation in a potential well with nonanalytic minimum. Accordingly, logarithmic corrections appear in (9.2) if $d = 1, 2, 3$ and even the power law ones for $d = 4$.

If $d = 1$, then the asymptotic form of the coefficients $A_K$ corresponding to the $M$-point Green function $G_M(g)$ is

$$A_K = -\frac{2^{-M/2}}{(\pi/2)\Gamma(n/2)} \left(\frac{3}{2\ln 2}\right)^{n/2}$$
$$\times \Gamma\left(K + \frac{n}{2}\right)(\ln 2)^{-K}[\ln K + C], \qquad (9.3)$$

$$C = C_E + \ln\left(\frac{6}{\ln 2}\right) + \frac{\psi(1/2) - \psi(n/2)}{2},$$

where $C_E$ is Euler's constant and $\psi(x)$ is the logarithmic derivative of the gamma function. If $d = 2$, then

$$A_K = -\frac{2^{-M/2}}{19.7} \frac{(0.702)^n}{\Gamma(n/2)}$$
$$\times \Gamma\left(K + \frac{n+1}{2}\right)(\ln 2)^{-K}\ln^2 K \qquad (9.4)$$

to logarithmic accuracy. Similarly,

$$A_K = -\frac{2^{-M/2}}{2.12} \frac{(0.704)^n}{\Gamma(n/2)}$$
$$\times \Gamma\left(K + \frac{n+2}{2}\right)(\ln 2)^{-K}\ln^3 K \qquad (9.5)$$

for $d = 3$. The results obtained for $d = 4$ depend on the coordinates entering to the Green functions and have cumbersome expressions [33]. They can be simplified by passing to the momentum representation and choosing momenta $p_i$ corresponding to the symmetric point ($p_i \sim p$):

$$A_K = Be^{\nu \ln(\mu/p)}\Gamma\left(K + \frac{n+4}{2} + \nu\right)(\ln 2)^{-K}, \qquad (9.6)$$

where $\mu$ is a point of the charge normalization, $\nu = (n + 8)/3$, and the values of $B$ are listed in Table 3. In the scalar theory ($n = 1$), the leading contribution to the asymptotic expression vanishes, and the asymptotic behavior is expected to be determined by the next-order term in $1/K$:

$$A_K = \text{const}\, e^{\nu \ln(\mu/p)} \times \Gamma\left(K + \frac{n+4}{2} + \nu - 1\right)(\ln 2)^{-K}. \quad (9.7)$$

The results for the logarithm of the vacuum integral $Z_0(g)$ are formally obtained by setting $M = 0$ and introducing a factor of $1/2$ in (9.3)–(9.6). In particular, the following result is obtained for the ground-state energy of the anharmonic oscillator ($d = 1$, $n = 1$):

$$A_K = -\frac{\ln K + 2.74}{3.78}\Gamma\left(K + \frac{1}{2}\right)(\ln 2)^{-K}. \quad (9.8)$$

Figure 23 compares this prediction with numerical results obtained in [6].

If $d = 1$, then the entire instanton spectrum can be represented by combinations of elementary instantons. If $d \geq 2$, then there may exist a nonspherically symmetric instanton with action lower than $2S_0$. In this case, there are no soft modes, and formula (9.2) with $r' - r = d(d-1)/2$ is valid, because an asymmetric instanton is associated with $d(d-1)/2$ additional modes corresponding to rotations in the coordinate space. Since modes of this kind have never been considered, the calculation of the constant $\tilde{c}$ in (9.2) is a technically nontrivial problem. The technique of integration over these modes developed in [33] should be instrumental in quantum electrodynamics, where even the first instanton is asymmetric [23].

## 10. OUTLOOK

Finally, let us discuss the most promising lines of further research.

### 10.1. Calculation of c in the Lipatov Asymptotics

Complete Lipatov asymptotic forms are known only in $\varphi^4$ theory and a number of quantum-mechanical problems. In other models, the common factor $c$ has yet to be calculated. For the Gell-Mann–Low function in QCD, the factor $c$ has been calculated only in the case of $SU(2)$ symmetry [17]. However, this calculation is based on unconventional definition of the $\beta$ function, and consistency of the asymptotic form of $\beta_N$ with the renormalization schemes used in actual diagrammatic calculations remains an open question. The factor $c$ has been formally calculated for the quark–quark correlation function in QCD [24]. However, the procedure used to eliminate divergences in (4.5.13) evokes doubts [48], and even the general form of the result may be revised.

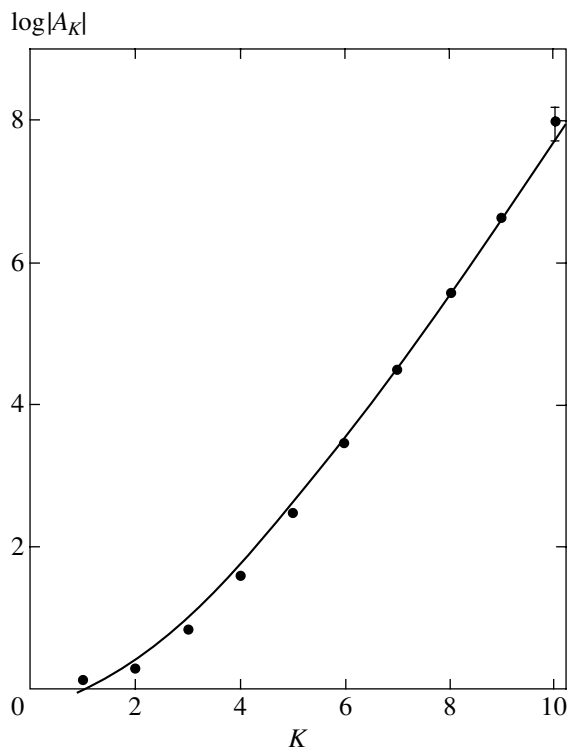

**Fig. 23.** Predictions of asymptotic formula (9.8) (curve) compared with coefficients $A_K$ calculated numerically in [6] (symbols).

### 10.2. A Priori Proofs of Absence of Renormalons

A proof of this kind exists only for $\varphi^4$ theory [48]. The constructive scheme proposed in Section 5.3 for eliminating renormalon singularities in QED and QCD is substantiated by results presented in Section 8. However, it relies on approximate determination of the Gell-Mann–Low function, which may seem questionable to a skeptical reader. Therefore, extension of the analysis presented in [48] to other field theories is highly desirable.

**Table 3.** Parameter $B$ in formula (9.6)

| $n$ | $B \times 10^4$ | |
|---|---|---|
| | $M = 2$ | $M = 4$ |
| 0 | –9.05 | –8.72 |
| 1 | 0 | 0 |
| 2 | 3.25 | 1.45 |
| 3 | 4.55 | 1.50 |

### 10.3. Development and Application of Highly Accurate Summation Methods

In the conventional conformal–Borel technique presented in Section 6.1, the cut in the Borel plane extends from –∞ to the nearest instanton singularity $S_0$. However, the cut can be extended to point $S^*$ such that $S_0 < S^* < 0$. In this case, all singularities of the Borel transform remain on the boundary of a unit circle in the $u$ plane, and the resummed series is convergent at every point in the Borel plane that does not lie on the cut. If $S_0 < S^* < 0$, then the results are independent of $S^*$. This conclusion, empirically confirmed in [115], implies that knowedge of exact value of $S_0$ is not necessary. Thus, the conformal-Borel technique [35] in fact does not use any quantitative information about the Lipatov asymptotics. In our opinion, interpolation of the coefficient function and explicit use of the asymptotic behavior at strong coupling (Sec.6.2) will substantially improve the accuracy of evaluation of critical exponents already for available information. This example illustrates the inefficiency of the current use of information that requires enormous labor resources to be acquired.

Additional improvement of efficiency can be achieved by using information concerning high-order corrections to the Lipatov asymptotic form. The scheme described in Section 9 facilitates the calculation of several parameters characterizing the coefficient function. In terms of efficiency, this is equivalent to advancing by several orders in perturbation theory, whereas advancement to the next order in diagrammatic calculations requires about ten years.

The method for finding strong-coupling asymptotics described in Section 6.2 is effective when information is scarce, but cannot be classified as a highly accurate one. When more information is available, construction of Padé approximants for the coefficient function [13] looks as a more effective tool. Preliminary studies show that this method can be combined with some strategy for selecting most suitable Padé approximants.

### 10.4. Summation of Nonalternating Series

In essence, the analysis presented in Section 7 solves the problem of non-Borel-summability for the most interesting problems. However, the summation schemes formulated therein are insufficiently effective, and improved methods should be developed.

The summation of QCD perturbation series is performed in Section 8.3 without invoking the technique developed in Section 7. A certain trick is used to circumvent the problem, such as the exponent $\alpha$ is correct, whereas $\beta_\infty$ is determined only up to an order of magnitude. Currently, this rough approximation is acceptable in view of large uncertainty in $\beta_\infty$ (see Fig. 22b). Nevertheless, it is quite desirable to make attempt of the proper summation "by following all the rules" and to analyze the arising uncertainties.

In relation with the confinement problem, summation of series for anomalous dimensions is desirable. The formation of a string-like "flux tube" between quarks is not controlled by the $\beta$ function, being determined by properties of correlation functions, which depend on the values of anomalous dimensions in the "plateau" region of the coupling constant (see Fig 15d).

### 10.5. Analytical Methods for Strong-Coupling Problems

The exponent $\alpha$ is close to unity in both QED and $\varphi^4$ theory. Moreover, there are reasons to believe that its exact value is $\alpha = 1$. Simple results of this kind should be obtainable by analytical methods. Since it is always easier to substantiate a known result than to obtain it a priory, there are grounds for an optimistic outlook. Once the equality $\alpha = 1$ is proved, the accuracy of analysis of strong-coupling asymptotics will substantially improve, because the number of parameters to be determined will reduce from two to one. It is obvious that progress in this area will be stimulated by acquiring additional "experimental" information concerning strong-coupling asymptotica.

### 10.6. Applications to the Theory of Disordered Systems

The theory of disordered systems is unique in that high-order contributions are essential even in the weak-coupling region. Description of a particle moving in a Gaussian random field can be rigorously reformulated as a $\varphi^4$ theory with "incorrect" sign of the coupling constant [103, 121–123]. In formally unstable field theories of this kind, nonperturbative contributions of the form $\exp(-a/g)$ play an important role and can be found by summing perturbation series. One example is the fluctuation tail of density of states [143], which is directly related to the Lipatov asymptotic form [8, 144]. Combination of instanton calculations with parquet approximation was used in [8] to develop a complete theory of density of states for a disordered system in the $(4 - \epsilon)$-dimensional space. Next in order is the development of an analogous theory for calculating transport properties of disordered systems, which requires an analysis of a $\varphi^4$-type theory with two vector fields [122, 123]. The qualitative importance of high-order perturbative contributions for such an analysis is due to the purely non-perturbative nature of the diffusion pole in the localized phase [145, 146]. If all characteristics of the pole can be elucidated in the framework of the instanton method, then an explanation can be found for the "simple" critical exponents obtained in the symmetry-based approach to the theory of the Anderson transition [147].

### ACKNOWLEDGMENTS

I thank M.V. Sadovskii and A.I. Sokolov, who read a preliminary version of the manuscript and made important remarks; L.N. Lipatov for stimulating discussions; K.B. Varnashev for help in selecting referenced papers; and participants of seminars at the Kapitza Institute for Physical Problems, Lebedev Physical Institute, Institute of Theoretical and Experimental Physics, and Konstantinov Institute of Nuclear Physics for interest in this study and numerous discussions.

This work was supported by the Russian Foundation for Basic Research, project no. 03-02-17519.

*Translated by A. Betev*